\documentclass[aps,pra,twocolumn,showpacs]{revtex4-1}

\usepackage{amsmath,amssymb}
\usepackage[pdftex]{graphicx}

\newcommand{\ii}{\mathrm{i}}
\newcommand{\dd}{\mathrm{d}}
\newcommand{\ad}{a^\dagger}

\let \Re \relax
\DeclareMathOperator{\Re}{Re}
\let \Im \relax
\DeclareMathOperator{\Im}{Im}

\begin{document}

\title{Dynamics of the Dicke model close to the classical limit}

\author{L. Bakemeier}
\email{bakemeier@physik.uni-greifswald.de}
\author{A. Alvermann}
\author{H. Fehske}
\affiliation{Institut f\"ur Physik, Ernst-Moritz-Arndt-Universit\"at, 17487 Greifswald, Germany}

\begin{abstract}
We study the dynamical properties of the Dicke model for increasing spin length,
as the system approaches the limit of a classical spin.
First, we describe the emergence of collective excitations above the groundstate
that converge to the coupled spin-oscillator oscillations  found in the classical limit. The corresponding Green functions reveal quantum dynamical signatures
close to the superradiant quantum phase transition.
Second, we identify signatures of classical quasi-periodic orbits in the quantum time evolution using numerical time-propagation of the wave function.
The resulting phase space plots are compared to the classical 
trajectories.
We complete our study with the analysis of individual eigenstates close to the quasi-periodic orbits.
\end{abstract}


\maketitle

\section{Introduction}

The relation between quantum dynamical systems and their classical counterparts is of fundamental interest, but also important for the understanding of the quantum dynamics itself.
Specific questions concern the construction of and convergence to the classical limit~\cite{Yaff82, DGS00},
the relation between classical and quantum chaos~\cite{Gutz90, Haak10}, or between quantum chaos and thermalization~\cite{Sred94,AH12,AH12a}.
This includes the identification of specific signatures of the classical dynamics, in particular of stable or unstable periodic orbits characteristic for regular or chaotic motion, in the eigenstates and quantum phase space dynamics.

A paradigmatic example studied intensively in this context is the Dicke model~\cite{Dick53} of quantum optics.
The Dicke model, with Hamilton operator
\begin{equation}\label{Ham}
 H  = \Delta J_z + \lambda (a^\dagger + a) J_x + \Omega a^\dagger a \;,
\end{equation}
describes a spin (with operators $J_x$, $J_z$) of length $j$ coupled to a harmonic oscillator (with bosonic operators $a^{(\dagger)}$).
While the Dicke model acquires non-trivial behavior through the coupling of the spin to the oscillator, it remains accessible to analytical studies in the classical spin limit $j \to \infty$.
In this limit, the Dicke model shows a quantum phase transition at the critical coupling $\lambda^2 = (\Delta \Omega)/2j$, from a ground state with zero bosonic expectation value ($\langle a \rangle = 0$) to a ``superradiant'' ground state with finite bosonic expectation value ($\langle a \rangle \ne 0$)~\cite{HL73,WH73}.
This superradiant quantum phase transition (QPT) is accompanied by a divergence of spin-oscillator entanglement~\cite{LEB04,LEB05,VD06}.
This is in contrast to the QPT in the ``static'' oscillator limit $\Omega \to 0$, which occurs already for finite spin length and shows no divergence of entanglement~\cite{BAF12}.

The Dicke model gives also an example for quantum chaotic behavior as seen in the level statistics~\cite{Kus85,GH86,LNMPW91,EB03b}.
The quantum chaos is accompanied by classical chaos in the corresponding semi-classical (SC) equations of motion for spin and oscillator expectation values~\cite{AFLN92}.
It was further shown that classical chaos strongly influences the dynamics of entanglement~\cite{FNP98, SMYW12} and spin squeezing~\cite{SYMW09}.
The build-up and decay of entanglement is closely linked with the collapse and revival dynamics at finite $j$~\cite{RWK87,BSMDHRH96,ABF12}.

In this paper we study the dynamical properties of the Dicke model as the classical limit is approached.
Our goal is to compare the quantum dynamics at large $j$ with the SC dynamics in the limit $j\to\infty$.
Our comparison includes the linearized dynamics around the groundstate,
seen as the collective response to a weak perturbation,
and the full non-linear dynamics in the entire phase space.
With modern numerical tools, in particular Chebyshev algorithms for the computation of spectral functions~\cite{WWAF06} and time-propagation~\cite{TK84}, we can produce unbiased numerical results for large $j$ (up to $j=400$).
This allows for a direct analysis of the emergence of ``classical'' behavior as the $j \to \infty$ limit is approached.

The paper is organized as follows.
In Sec.~\ref{sec:EOM} we discuss the SC equations of motions that hold in the limit $j \to \infty$.
In Sec.~\ref{sec:CollModes} we compute the classical modes in the vicinity of the stationary state(s), and compare to the quantum mechanical excitation spectrum that is given by a spin-spin Green function.
In Sec.~\ref{sec:Chaos} we address the quantum dynamics at higher energies.
Convergence towards the classical dynamics is studied with the spin Husimi (phase space) function, both for individual eigenstates and the time evolution of initial coherent states.
We finally conclude in Sec.~\ref{sec:Conc}.
The appendices give details for the derivation of the SC equations of motion from the Dirac-Frenkel variation principle (App.~\ref{app:DF}),
for the computation of the classical collective modes (App.~\ref{app:EOM}),
and for the numerical computation of the time averaged Husimi function through a modification of Chebyshev time propagation (App.~\ref{app:Husimi}).

\section{The semi-classical equations of motion}
\label{sec:EOM}

We first derive the SC equations of motion for the spin and oscillator expectation values.
They are only an approximation to the true dynamics for finite $j$, but become exact in the limit $j \to \infty$~\cite{GH84}.

To obtain the SC equations of motion we can start with the Ehrenfest equations of motion $d \langle A \rangle/dt = \ii \langle [H,A] \rangle$
for the spin ($J_x, J_y, J_z$) and oscillator ($a^{(\dagger)}$) observables,
e.g. $(d/dt) \langle J_y \rangle = \Delta \langle J_x \rangle - \lambda \langle (a^\dagger+a) J_z \rangle$.

The SC approximation consists in neglecting spin-oscillator correlations~\cite{GH84}, replacing a mixed operator product $\langle A B \rangle$ by $\langle A \rangle \langle B \rangle$ , e.g.
$\langle (a^\dagger+a) J_z \rangle \mapsto \langle a^\dagger +a \rangle \langle J_z \rangle$ in the equation of motion for $\langle J_y \rangle$.
This results in the SC equations of motion
\begin{equation} \label{SC1}
\frac{d}{dt} \begin{pmatrix} \langle J_x \rangle \\  
  \langle J_y \rangle \\ \langle J_z \rangle \end{pmatrix} = \begin{pmatrix} 2 \lambda \Re \langle a \rangle \\ 0 \\ \Delta \end{pmatrix} \times \begin{pmatrix} \langle J_x \rangle \\  
  \langle J_y \rangle \\ \langle J_z \rangle \end{pmatrix} 
\end{equation}
for the spin observables and 
\begin{equation}\label{SC2}
 \ii \frac{d}{dt} \langle a \rangle = \Omega \langle a \rangle + \lambda \langle J_x \rangle
\end{equation}
for the oscillator observables.
Intuitively, the spin moves in the magnetic field generated by the oscillator,
and the oscillator moves in the constant force exerted upon it by the spin.
In this sense, the SC approximation gives a mean field description of the system dynamics.

Eqs.~\eqref{SC1},~\eqref{SC2} describe a five dimensional dynamical system in the real variables $\langle J_{x,y,z}\rangle$, $\Re \langle a \rangle$, $\Im \langle a \rangle$ with two conserved quantities, energy 
\begin{equation}
E = \Delta \langle J_z \rangle + 2 \lambda \Re \langle a \rangle \langle J_x \rangle + \Omega |\langle a \rangle|^2
\end{equation}
and spin length
\begin{equation}\label{eq:SpinLength}
j^2 = \langle J_x\rangle^2 + \langle J_y\rangle^2 + \langle J_z\rangle^2 \;.
\end{equation}
Note that the latter equation coincides with $\langle J^2 \rangle = j(j+1)$ only in  the limit $j \to \infty$.

To eliminate one degree of freedom, using the conservation of $j^2$, we switch to planar coordinates for the spin~\cite{ZFG90}.
With spherical coordinates $\theta, \phi$ and
\begin{equation}\label{ZSpin1}
 \begin{pmatrix} \langle J_x \rangle \\  
  \langle J_y \rangle \\ \langle J_z \rangle \end{pmatrix} = 
\begin{pmatrix}
 j \cos \phi \sin \theta \\
 j \sin \phi \sin \theta \\
 - j \cos \theta 
\end{pmatrix}
 \;,
\end{equation}
the complex variable
\begin{equation}\label{ZSpin2}
z = e^{-\ii\phi}\tan(\theta/2)
\end{equation}
gives a mapping of the Bloch sphere onto the complex plane.
We note $\langle J_x \rangle = 2 j \Re z /(1+|z|^2)$,
$\langle J_y \rangle = - 2 j \Im z /(1+|z|^2)$, and
$\langle J_z \rangle = j (|z|^2-1)/(1+|z|^2)$.

For the oscillator we introduce the complex variable
\begin{equation}\label{Alpha}
\bar \alpha = \frac{\Omega}{j \lambda} \langle a \rangle \;.
\end{equation}
The prefactor guarantees a well-defined limit $j \to \infty$.
We can identify $\bar \alpha$ with (the suitably scaled) position and momentum of the harmonic oscillator:
\begin{equation}\label{AlphaQP}
 Q = \Re \bar \alpha \;, \quad P = \Im \bar \alpha \,.
\end{equation}

Expressed in $z$, $\bar \alpha$, Eqs.~\eqref{SC1},~\eqref{SC2} become
\begin{align}
\label{SCEom}
\begin{split}
\ii\dot{ \bar \alpha} &= \Omega \Big( \bar\alpha + \frac{2 \Re z}{1+|z|^2} \Big) \;, \\
\ii\dot{z} &= \Delta \Big( z + \frac{\kappa}{2} (1-z^2) \Re \bar\alpha \Big) \;.
\end{split}
\end{align}
Here, we introduced the dimensionless coupling constant
\begin{equation}
 \kappa = \frac{2 j \lambda^2}{\Delta \Omega} \;.
\end{equation}
The quantum phase transition occurs at $\kappa=1$.
Conservation of spin length is imminent, and only four real dynamical variables remain. 
Note that $j$ does not appear in the equations.
Rescaling of the time variable $t$ would further allow the elimination of either $\Omega$ or $\Delta$.
The energy is given by
\begin{equation} \label{EZAlpha}
 E(z,\bar \alpha)/(j\Delta) = \frac{|z|^2-1}{|z|^2+1} + 2\kappa\frac{\Re z \Re \bar \alpha }{1+|z|^2} + \dfrac{\kappa}{2}|\bar \alpha|^2 \;.
\end{equation}
We note the parity symmetry $z \mapsto -z$, $\bar \alpha \mapsto -\bar \alpha$ of Eqs.~\eqref{SCEom},~\eqref{EZAlpha}.

To shed further light on the meaning of the SC approximation,
we stress that the SC equations of motion can also be derived from a time-dependent variational ansatz
\begin{equation}
\label{ProdState}
|\psi_\text{SC}(t)\rangle = |\alpha(t)\rangle \otimes |z(t)\rangle 
\end{equation}
for the wave function.
Here,
\begin{equation}
|\alpha\rangle = e^{-|\alpha|^2/2}e^{\alpha \ad}|0\rangle
\end{equation}
with $a |\alpha \rangle = \alpha |\alpha\rangle$ for $\alpha \in \mathbb C$
and
\begin{equation}
|z\rangle = (1+|z|^2)^{-j}e^{z J_+}|j,-j\rangle 
\end{equation}
denote oscillator and spin coherent states, respectively~\cite{ZFG90}.
The relation between $z$ and the spin observables is as in Eqs.~\eqref{ZSpin1},~\eqref{ZSpin2}, the relation between $\alpha=\langle a \rangle$ and $\bar \alpha$ as in Eq.~\eqref{Alpha}.

The time-dependence of $|\psi_\text{SC}\rangle$ now follows from the Dirac-Frenkel time-dependent variational principle \cite{Di30,Fren34}.
 The equation of motion is
\begin{equation}
\label{eq:DFEOM}
\dfrac{d}{dt}|\psi_\text{SC}\rangle= \mathcal P \, \dfrac{1}{\ii}H |\psi_\text{SC}\rangle \;,
\end{equation}
where $\mathcal P$ is the orthogonal projection onto the tangent space of the manifold of $|\psi_\text{SC}\rangle$ states.
Evaluation of the projection (see App.~\ref{app:DF}) recovers the equations of motion~\eqref{SCEom}.
The SC approximation is thus equivalent to the assumption that the system stays in a coherent product state as in Eq.~\eqref{ProdState} during time evolution.
This explains the restrictions of the SC approximation,
and hence part of the deviations between classical and quantum dynamics to be observed later.

\section{Classical and quantum collective modes}
\label{sec:CollModes}

We now consider the classical dynamics in the vicinity of the stationary solutions of Eq.~\eqref{SCEom}, and compare to the collective response of the Dicke model for small perturbations of the groundstate.

\subsection{Classical collective modes}
\label{ssec:CollModes}

Depending on the value of $\kappa$, Eq.~\eqref{SCEom} has one or two stable stationary solutions, which give the groundstate of the Dicke model at zero temperature and in the limit $j \to \infty$.
For $\kappa < 1$, the only stationary solution is 
$z= \bar \alpha = 0$.
For $\kappa > 1$,
this solution becomes unstable and the two stable solutions 
\begin{equation}\label{StatSol}
z_\pm=\pm\sqrt{\dfrac{\kappa-1}{\kappa+1}} \;, \hspace*{0.3cm} \bar \alpha_\pm=\mp \dfrac{\sqrt{\kappa^2-1}}{\kappa}
\end{equation}
appear.
Upon change of the value of $\kappa$,
Eq.~\eqref{SCEom} thus displays a (supercritical) pitchfork bifurcation~\cite{AFN91}.
The appearance of stable solutions with $\langle a \rangle \ne 0$, which break the parity symmetry, signals the 
superradiant quantum phase transition at the critical coupling $\kappa = 1$.

For small oscillations 
$z = z_s + \delta z$, $\bar \alpha = \bar \alpha_s + \delta \bar \alpha$
around a stationary solution $z_s, \bar \alpha_s$, linearization of Eq.~\eqref{SCEom} gives 
\begin{equation}\label{LinEOM}
\begin{split}
 \ii \dot {\delta \bar \alpha} &= \Omega \Big( \delta \bar \alpha + \frac{2(1-z_s^2)}{(1+z_s^2)^2} \Re \delta z \Big) \;, \\
 \ii \dot {\delta z} &= \Delta \Big( \big(1- \kappa \bar \alpha_s z_s \big) \delta z + \frac{\kappa}{2} \big(1-z_s^2 \big) \Re \delta \bar \alpha \Big) \;.
\end{split}
\end{equation}

\begin{figure}
\includegraphics[width=0.32\linewidth]{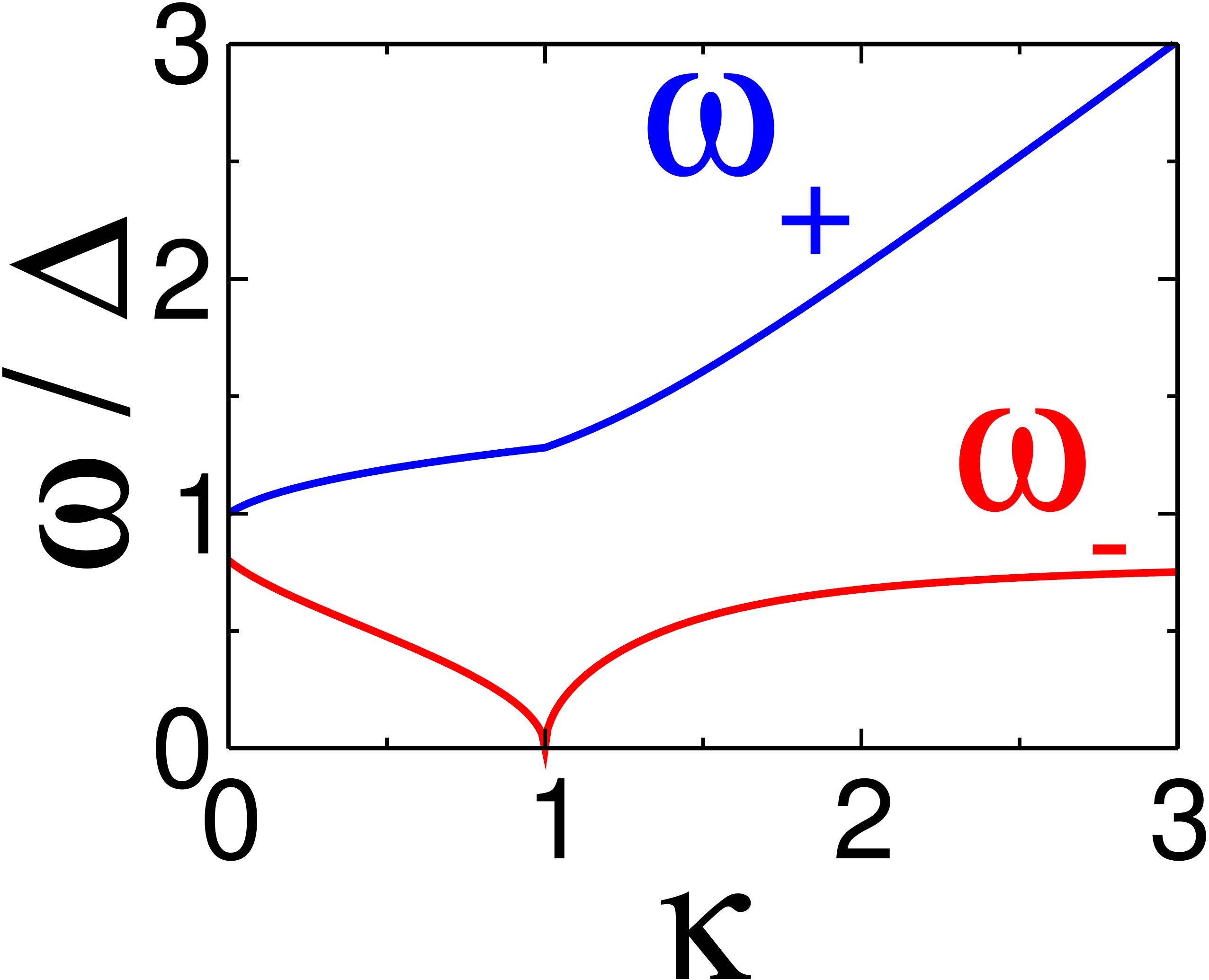}\hfill
\includegraphics[width=0.32\linewidth]{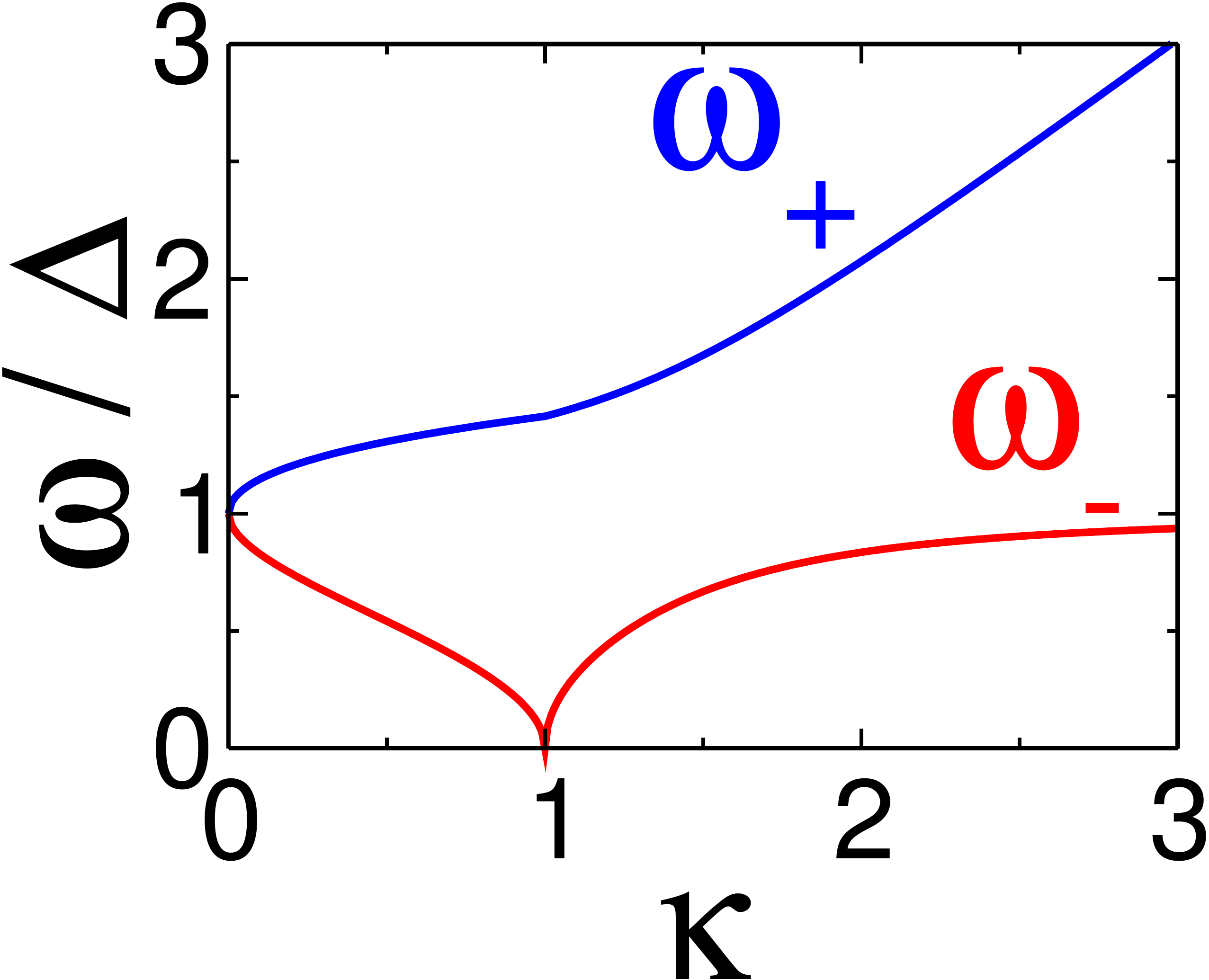}\hfill
\includegraphics[width=0.32\linewidth]{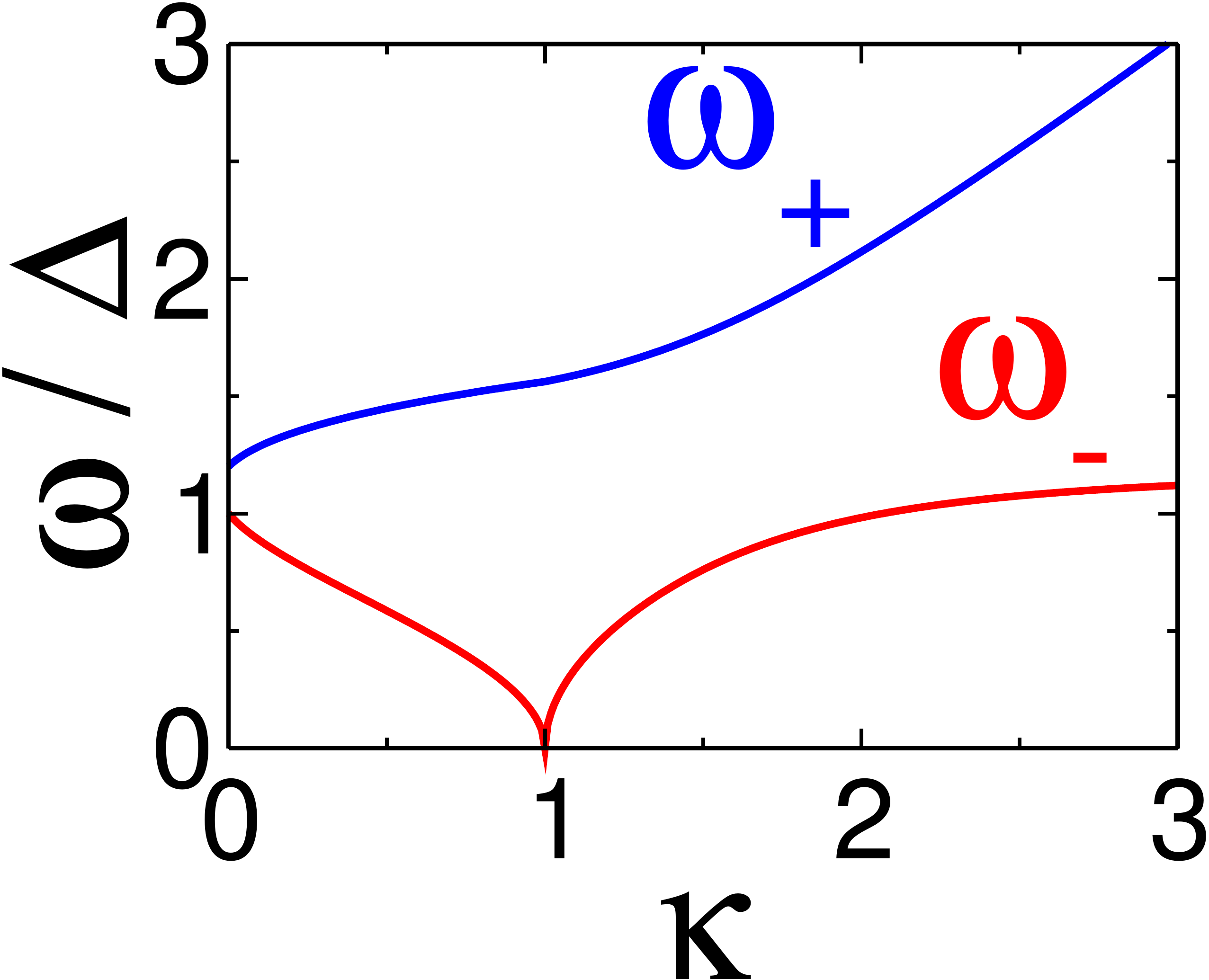} \\
\includegraphics[width=0.32\linewidth]{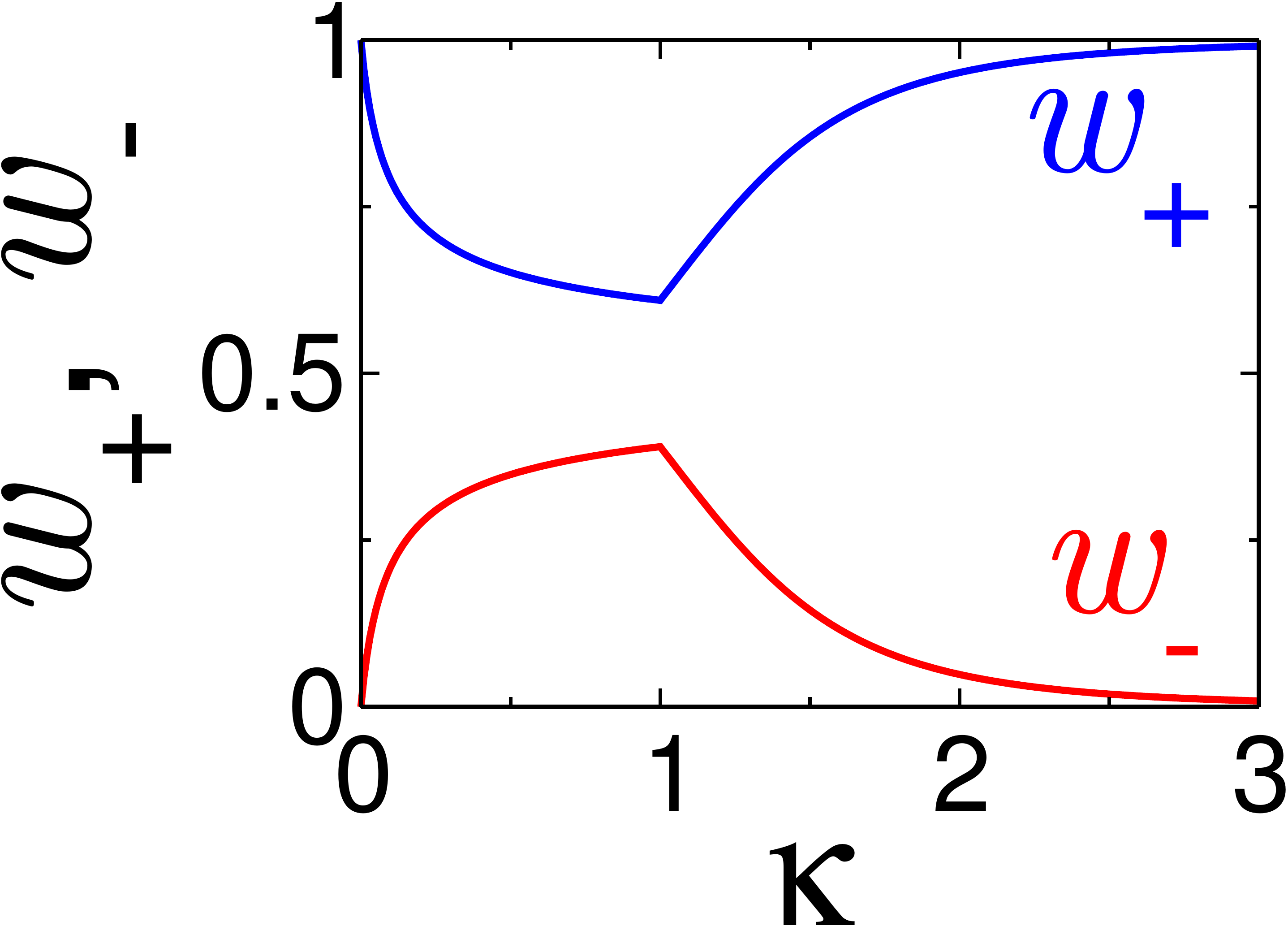}\hfill
\includegraphics[width=0.32\linewidth]{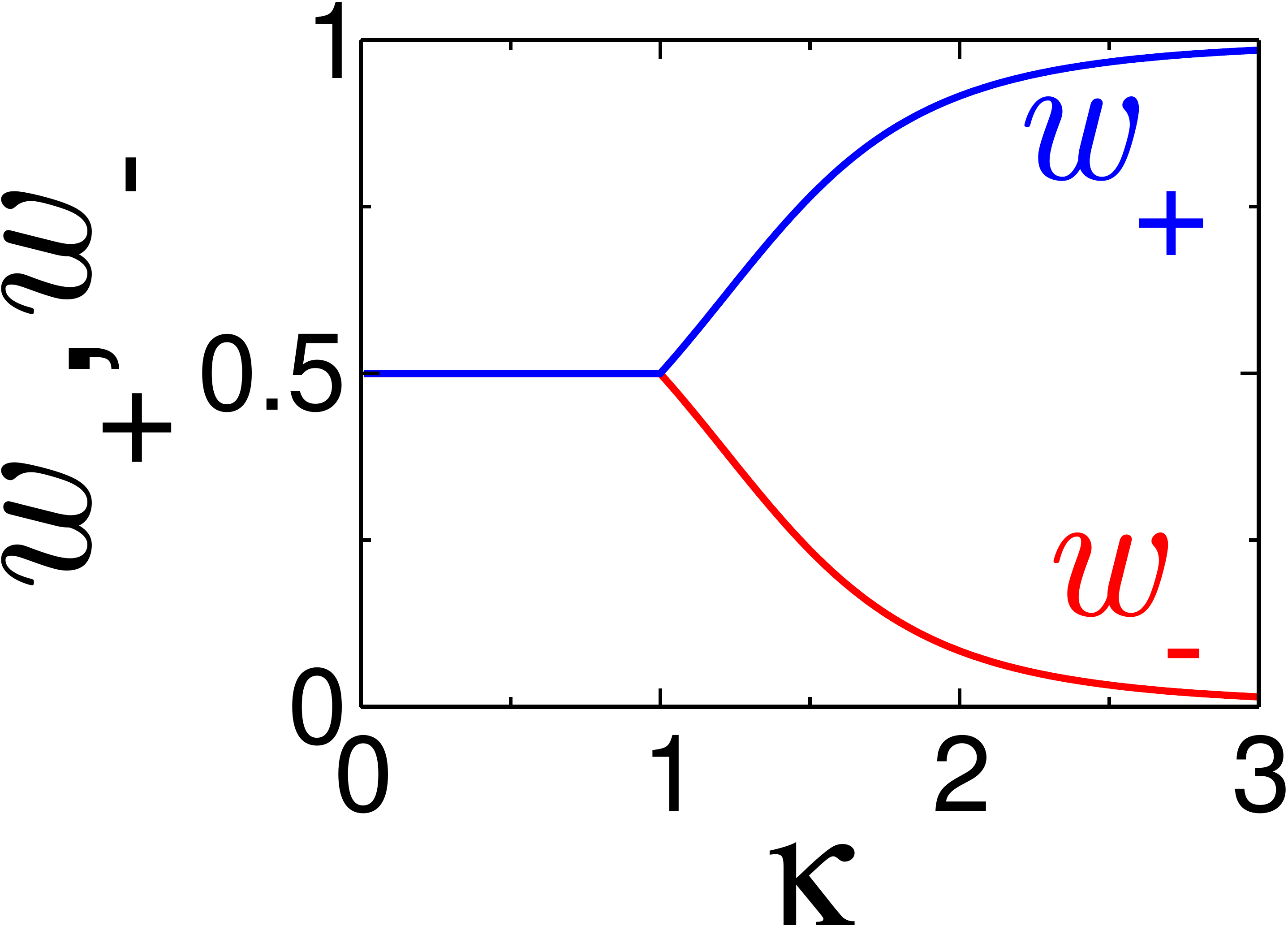}\hfill
\includegraphics[width=0.32\linewidth]{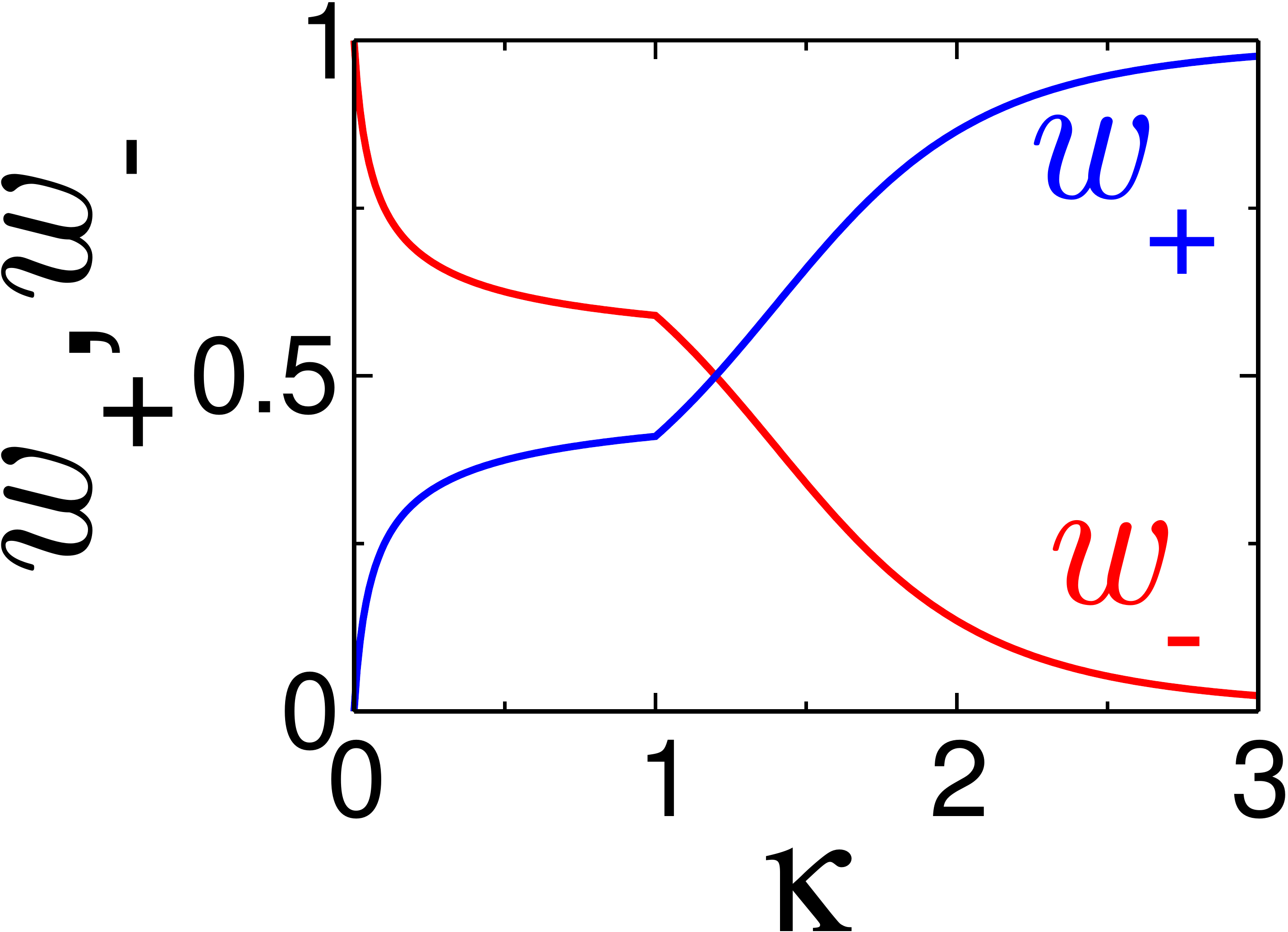} \\
\caption{(Color online) Collective mode frequencies $\omega_\pm$ from Eqs.~\eqref{FreqClass1},~\eqref{FreqClass2} (upper row) and weights $w_\pm$ from Eqs.~\eqref{ClassWeight},~\eqref{WeightBelow},~\eqref{WeightAbove} (lower row)  as a function of $\kappa$ for $\Omega/\Delta = 0.8, 1.0, 1.2$ from left to right.
}
\label{fig:collmodes}
\end{figure}

Eq.~\eqref{LinEOM} is a linear equation of motion for 
the deviations $\delta \bar \alpha$, $\delta z$, with purely imaginary eigenvalues (see App.~\ref{app:EOM}).
They give the frequencies of small oscillations around the groundstate of the Dicke model in the $j \to \infty$ limit.
Two different modes exist, with frequencies 
\begin{equation}\label{FreqClass1}
\omega_{\pm}^2 = \dfrac{\Omega^2+\Delta^2}{2}\pm\sqrt{\left(\dfrac{\Omega^2-\Delta^2}{2}\right)^2+(\Delta\Omega)^2 \kappa}
\end{equation}
for $\kappa < 1$, and
\begin{equation}\label{FreqClass2}
\omega_{\pm}^2 = \dfrac{\Omega^2+(\Delta\kappa)^2}{2}\pm\sqrt{\left(\dfrac{\Omega^2-(\Delta\kappa)^2}{2}\right)^2 +(\Delta\Omega)^2}
\end{equation}
for $\kappa > 1$.
The frequencies are plotted in Fig.~\ref{fig:collmodes}.
The appearance 
of a ``soft mode'' with $\omega_{-} =0$ at $\kappa = 1$ signals the second order QPT. 
The frequencies obtained here directly from the SC equations of motion agree with the result obtained with a Holstein-Primakoff-transformation of the spin operators in Ref.~\cite{EB03b}.
Both approaches are mathematically identical because they give, implicitly, the same linearized equations of motion around the stationary solutions of Eq.~\eqref{SCEom}.

From the eigenvectors of Eq.~\eqref{LinEOM} the full dynamical response can be determined (see App.~\ref{app:EOM}).
Let us exemplarily focus on the 
response to a small rotation of the spin around the $y$-axis, such that $J_x \mapsto J_x + \delta J_x$.
Within the linear approximation of Eq.~\eqref{LinEOM}
it is
\begin{equation}\label{JxSC1}
 \frac{\delta J_x(t)}{\delta J_x(0)} = w_- \cos \omega_- t + w_+ \cos \omega_+ t \;,
\end{equation}
with Fourier transform 
\begin{equation}\label{JxSC2}
\begin{split}
 \frac{\delta J_x (\omega)}{\delta J_x(0)} &= \frac{1}{\delta J_x(0)} \int_{-\infty}^\infty \delta J_x(t) e^{+\ii \omega t} dt \\
&= 
\pi w_- \big(\delta(\omega-\omega_-) + \delta (\omega+\omega_-) \big) \\
 & \; \;
+ \pi w_+ \big(\delta(\omega-\omega_+) + \delta (\omega+\omega_+) \big) \;.
\end{split}
\end{equation}

The Fourier transform has four $\delta$-peaks at frequencies $\pm \omega_\pm$.
The weight $w_\pm$ of the peaks, as shown in Fig.~\ref{fig:collmodes}, is given by 
\begin{equation}\label{ClassWeight}
 w_- = \cos^2 \beta \;, \quad w_+ = \sin^2 \beta \;,
\end{equation}
with
\begin{equation}\label{WeightBelow}
 \tan 2\beta = \frac{2 \Omega \Delta \sqrt{\kappa}}{\Omega^2 - \Delta^2}
\end{equation}
for $\kappa < 1$ and
\begin{equation}\label{WeightAbove}
 \tan 2\beta = \frac{2 \Omega \Delta }{\Omega^2 - (\Delta\kappa)^2}
\end{equation}
for $\kappa > 1$.
Note that these equations determine the angle $\beta$ only up to multiples of $\pi/2$. The correct choice is $0 \le \beta \le \pi/2$ for $\Omega \ge \Delta$ 
(with $\omega_- \to \Delta$, $w_- \to 1$ for $\kappa \to 0$) and
$\pi/2 \le \beta \le \pi$ for $\Omega < \Delta$
(with $\omega_+ \to \Delta$, $w_+ \to 1$ for $\kappa \to 0$).
At resonance $\Omega=\Delta$, it is
 $w_- = w_+ = 1/2$ for $\kappa < 1$ below the critical coupling.
For $\kappa > 1$, the weight $w_+$ of the high frequency peak grows,
and $w_+ \to 1$, $w_- \to 0$ for $\kappa \to \infty$ (cf. Fig.~\ref{fig:collmodes}).

\subsection{Quantum collective modes}
\label{sec:QuantExc}

\begin{figure}
\includegraphics[width=0.49\linewidth]{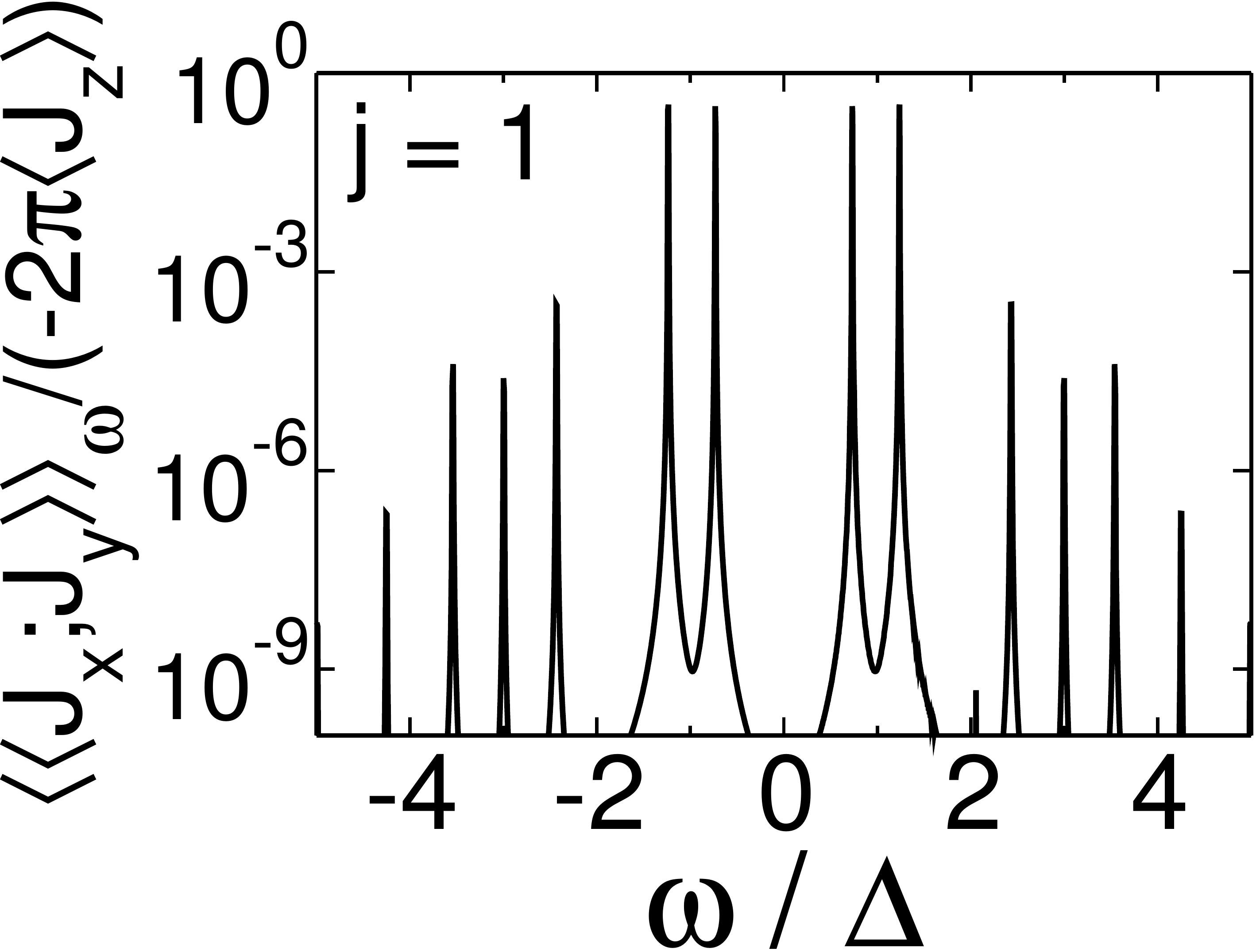}
\includegraphics[width=0.49\linewidth]{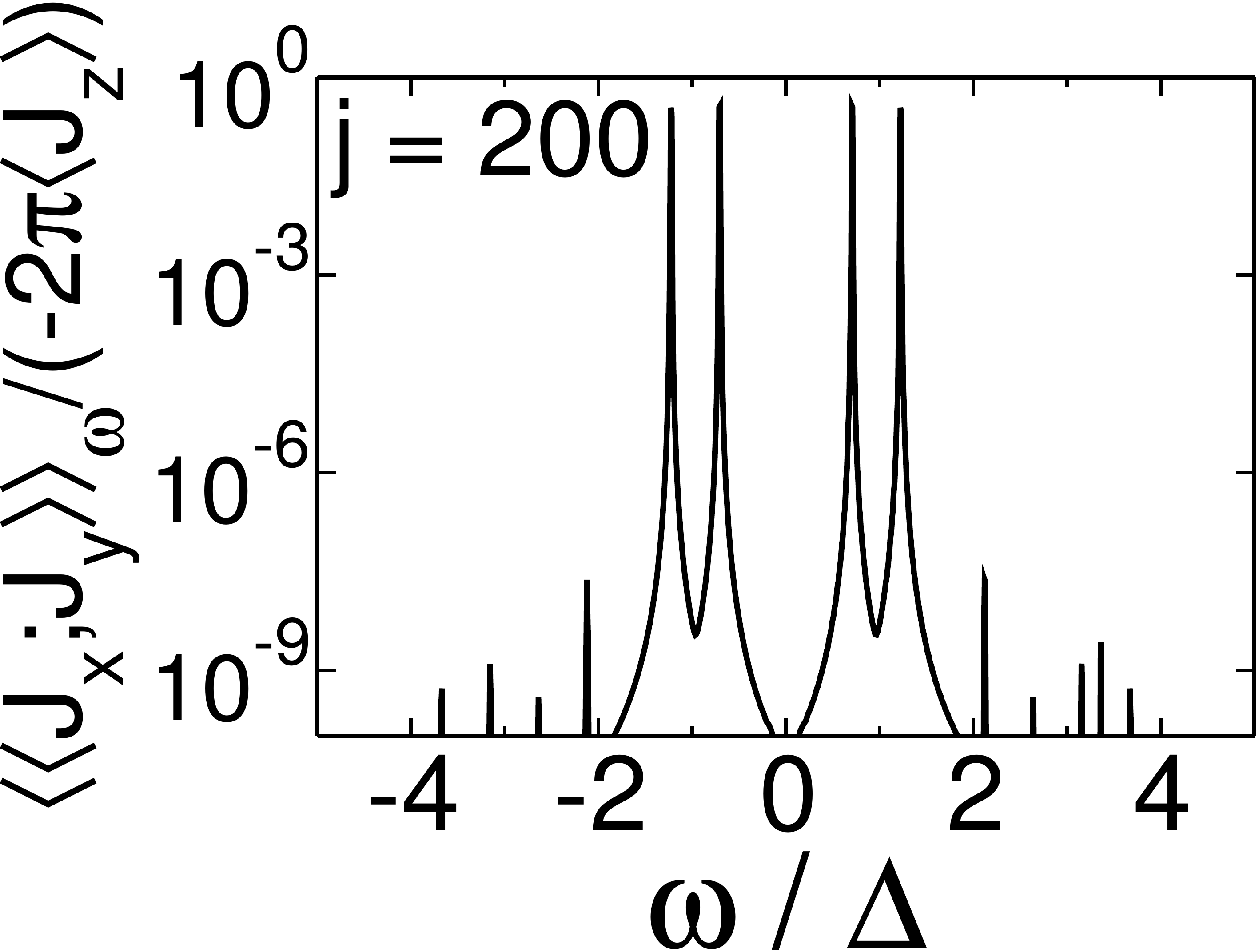}
\caption{Frequency spectrum of the normalized Green function as defined in Eq.~\eqref{eq:CorrFunc}, for
 $\Omega/\Delta=1$, $\kappa=0.95$ and $j=1$ (left panel), $j=100$ (right panel). The peaks are plotted with a finite width acquired from convolution with a narrow Gaussian.
}
\label{fig:freqspec}
\end{figure}

For a comparison of the quantum dynamics with the classical collective oscillations, we slighty disturb the groundstate and then determine the time evolution of the wave function.
With the operator for a spin rotation around the $y$-axis
\begin{equation}
S(\theta)=e^{i\theta J_y} \,,
\end{equation}
the initial state is given by
\begin{equation}\label{eq:distGS}
 |\psi_\delta \rangle = S(\delta \theta) |\psi_0\rangle
\end{equation}
for small $\delta \theta \ll 1$.
The expectation value of $J_x$ then is
\begin{equation}
J_x(t) = \langle \psi_\delta(t) | J_x | \psi_\delta (t) \rangle = \langle \psi_0 | S(-\delta \theta) J_x(t) S(\delta \theta) |\psi_0 \rangle \;.
\end{equation}
Linearization for small $\delta \theta$ gives
\begin{equation}
\begin{split}
J_x(t) & = \langle \psi_0 | (1 - \ii \delta \theta J_y)  J_x(t) + J_x(t) (1+ \ii \delta \theta J_y) | \psi_0 \rangle \\
& = \langle \psi_0 |J_x |\psi_0 \rangle + \ii \langle \psi_0 | [J_x(t), J_y] | \psi_0 \rangle \delta \theta \;.
\end{split}
\end{equation}
The relevant quantity for comparison with the \mbox{SC} result in Eqs.~\eqref{JxSC1},~\eqref{JxSC2} thus is the commutator Green function
\begin{equation}
 \langle\langle J_x (t); J_y \rangle\rangle = \ii \langle \psi_0 | [J_x(t), J_y] |\psi_0 \rangle \;,
\end{equation}
with Fourier transform
\begin{equation}
\label{eq:CorrFunc}
\begin{split}
\langle\langle J_x; J_y \rangle\rangle_\omega =& \int_{-\infty}^\infty \langle \langle J_x(t) ; J_y \rangle \rangle e^{i\omega t}  dt \\
=&  2 \pi \ii \langle \psi_0 | J_x \delta[ \omega - (H-E_0) ] J_y |\psi_0 \rangle \\[0.5ex]
 & -2\pi \ii \langle \psi_0| J_y \delta[ \omega + (H-E_0) ] J_x |\psi_0 \rangle
\;.
\end{split}
\end{equation}
We note that $\langle\langle J_x (t); J_y \rangle\rangle \in \mathbb R$,
hence $\langle\langle J_x; J_y \rangle\rangle_\omega = \langle\langle J_x; J_y \rangle\rangle_{-\omega}^*$,
and have the sum rule
\begin{equation}
\int_{-\infty}^\infty \langle\langle J_x; J_y \rangle\rangle_\omega \, d\omega  = - 2\pi \langle J_z \rangle \,.
\end{equation}
It is $\langle J_z \rangle < 0$ for $\Delta > 0$, as chosen here.
For a real Hamiltonian such as for the Dicke model, time-reversal symmetry  $\langle\langle J_x (-t); J_y \rangle\rangle = \langle\langle J_x (t); J_y \rangle\rangle^*$ holds, and $\langle\langle J_x; J_y \rangle\rangle_\omega~\in~\mathbb R$.

The Green function is computed with the kernel polynomial method (KPM)~\cite{WWAF06}, which allows us to treat large $j$.
According to Eqs.~\eqref{Alpha},~\eqref{StatSol} the average number of bosons in the groundstate scales as $j\Delta/(2\Omega)(\kappa^2-1)/\kappa$ for $\kappa > 1$, in addition to significant bosonic fluctuations at the QPT~\cite{BAF12}.
Therefore, up to $10^3$ bosons are kept in the calculation to ensure a negligible error from truncation of the infinite-dimensional
Hilbert space.
The spectral resolution of $\langle\langle J_x; J_y \rangle\rangle_\omega$ can be arbitrarily refined by increasing the number of Chebyshev moments.

For $j \to \infty$, the Green function should converge to the classical result from Eq.~\eqref{JxSC2}.
Some care has to be taken about the correct normalization of $\langle\langle J_x; J_y \rangle\rangle_\omega$ in comparison to Eq.~\eqref{JxSC2},
because the relation between $\delta J_x(0)$ and $\delta \theta$ depends on the value of the stationary solution $z_s$.
According to Eq.~\eqref{ZSpin1} it is $\delta J_x(0) = -\langle J_z \rangle \delta \theta$, which is just the factor from the sum rule for $\langle\langle J_x; J_y \rangle\rangle_\omega$. Therefore, we can use the normalized Green function $\langle\langle J_x; J_y \rangle\rangle_\omega/(-2 \pi \langle J_z \rangle)$.

We show $\langle\langle J_x; J_y \rangle\rangle_\omega$ in Fig.~\ref{fig:freqspec} for small and large $j$. The function consists of several peaks, but a (pair of) two peaks close to the classical frequencies $\pm \omega_\pm$ from Eqs.~\eqref{FreqClass1},~\eqref{FreqClass2} dominate the spectrum already at $j=1$.

\begin{figure}
\includegraphics[width=0.47\linewidth]{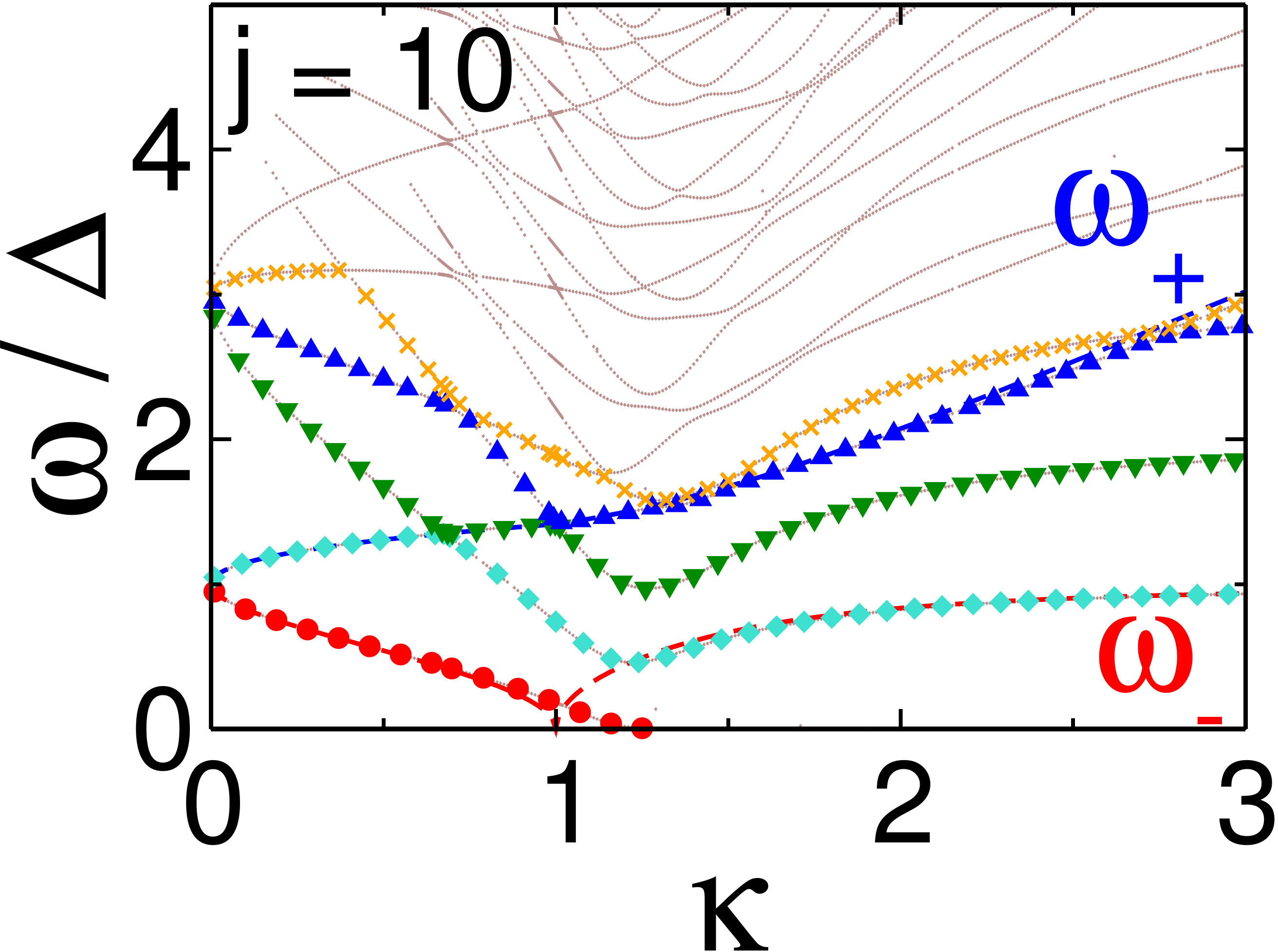}
\includegraphics[width=0.49\linewidth]{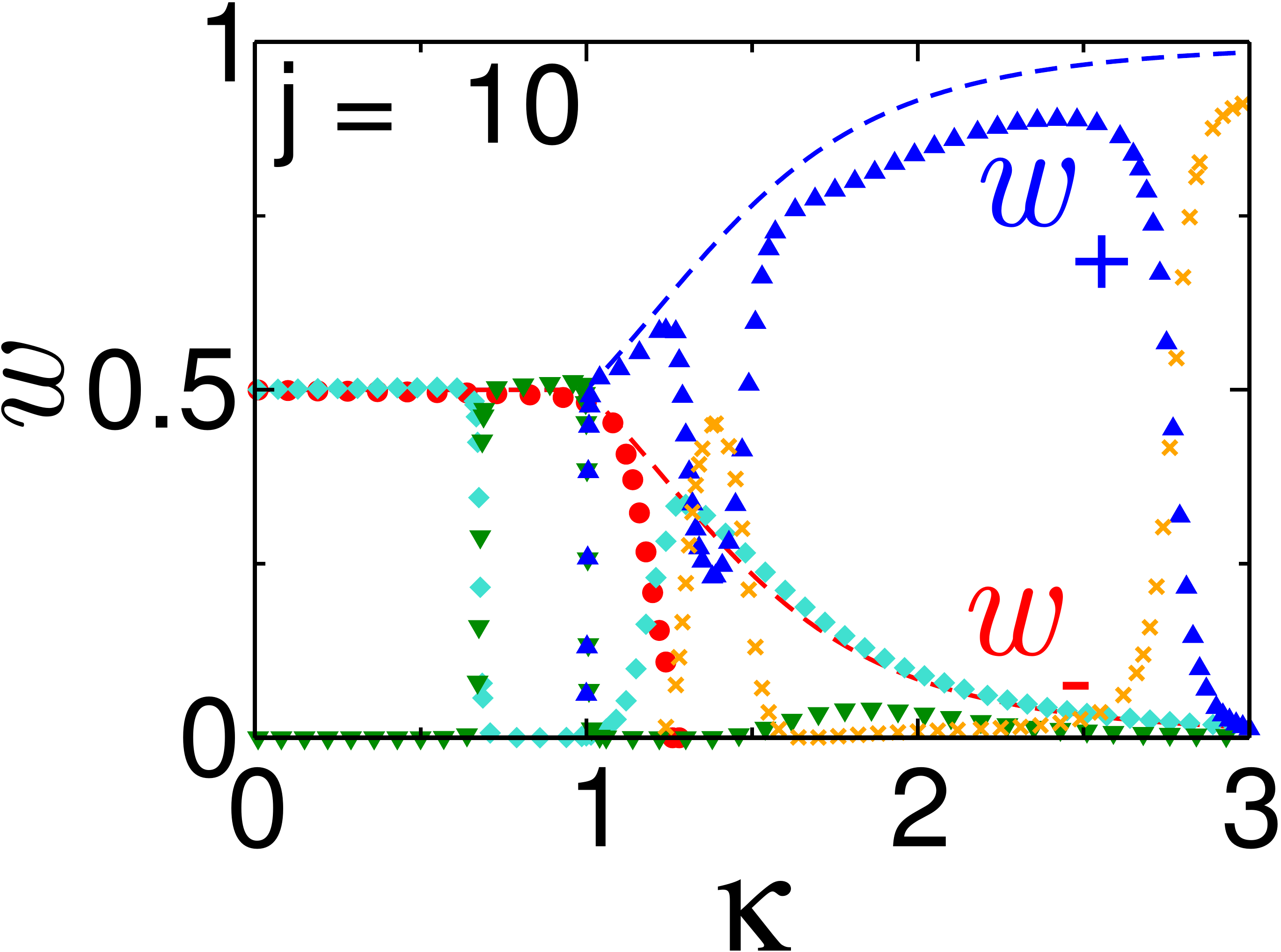}
\includegraphics[width=0.47\linewidth]{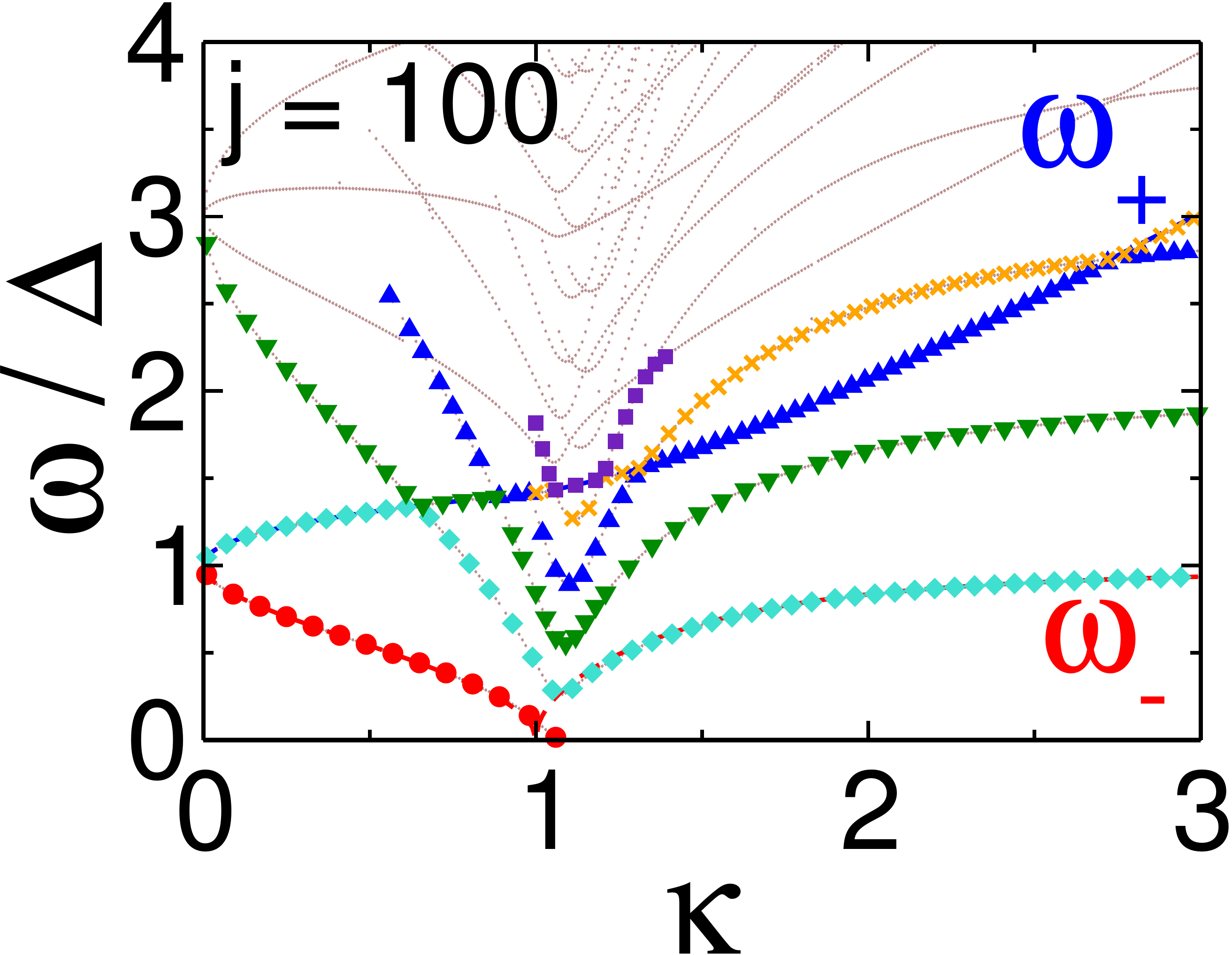}
\includegraphics[width=0.49\linewidth]{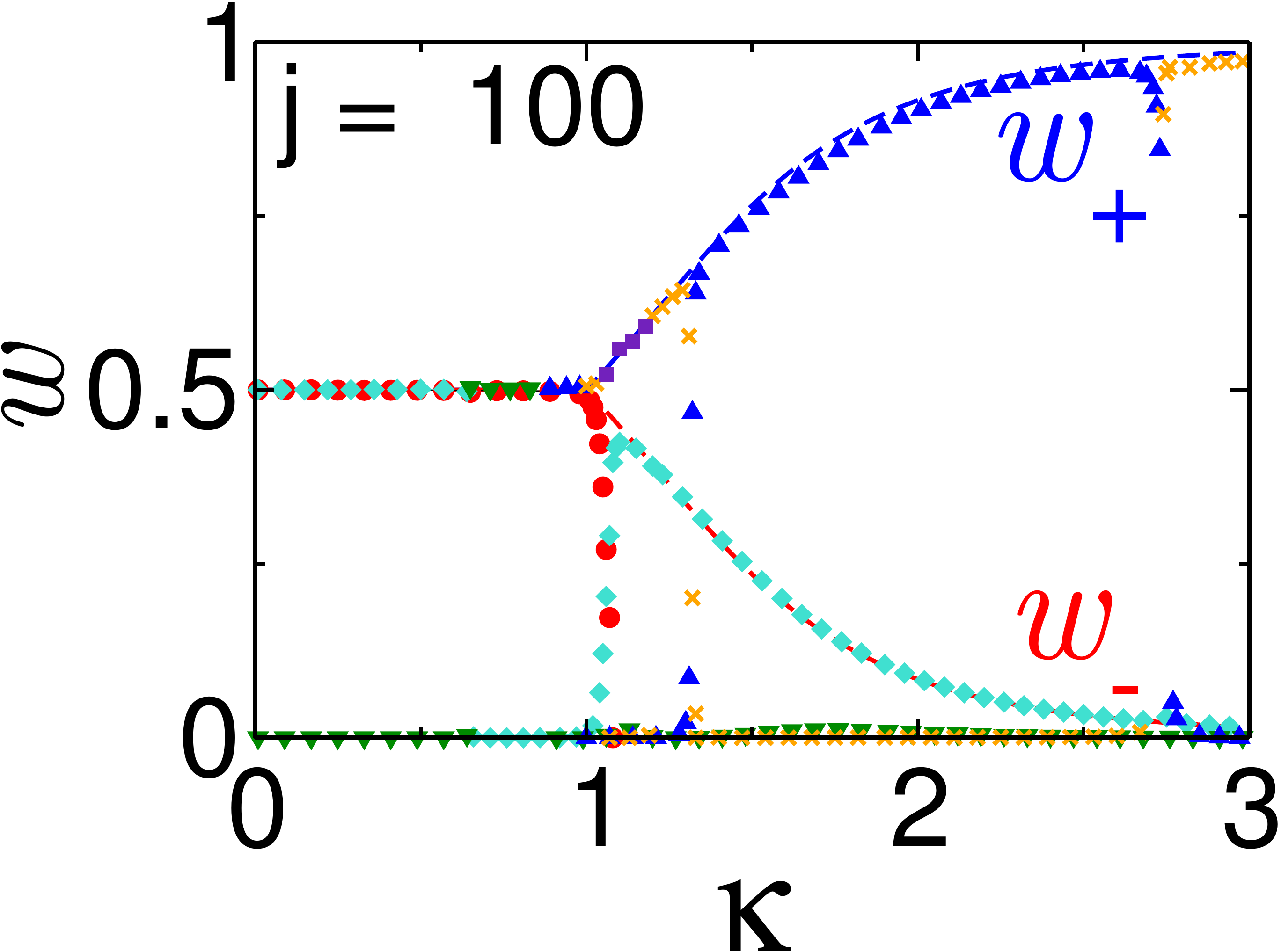}
\includegraphics[width=0.47\linewidth]{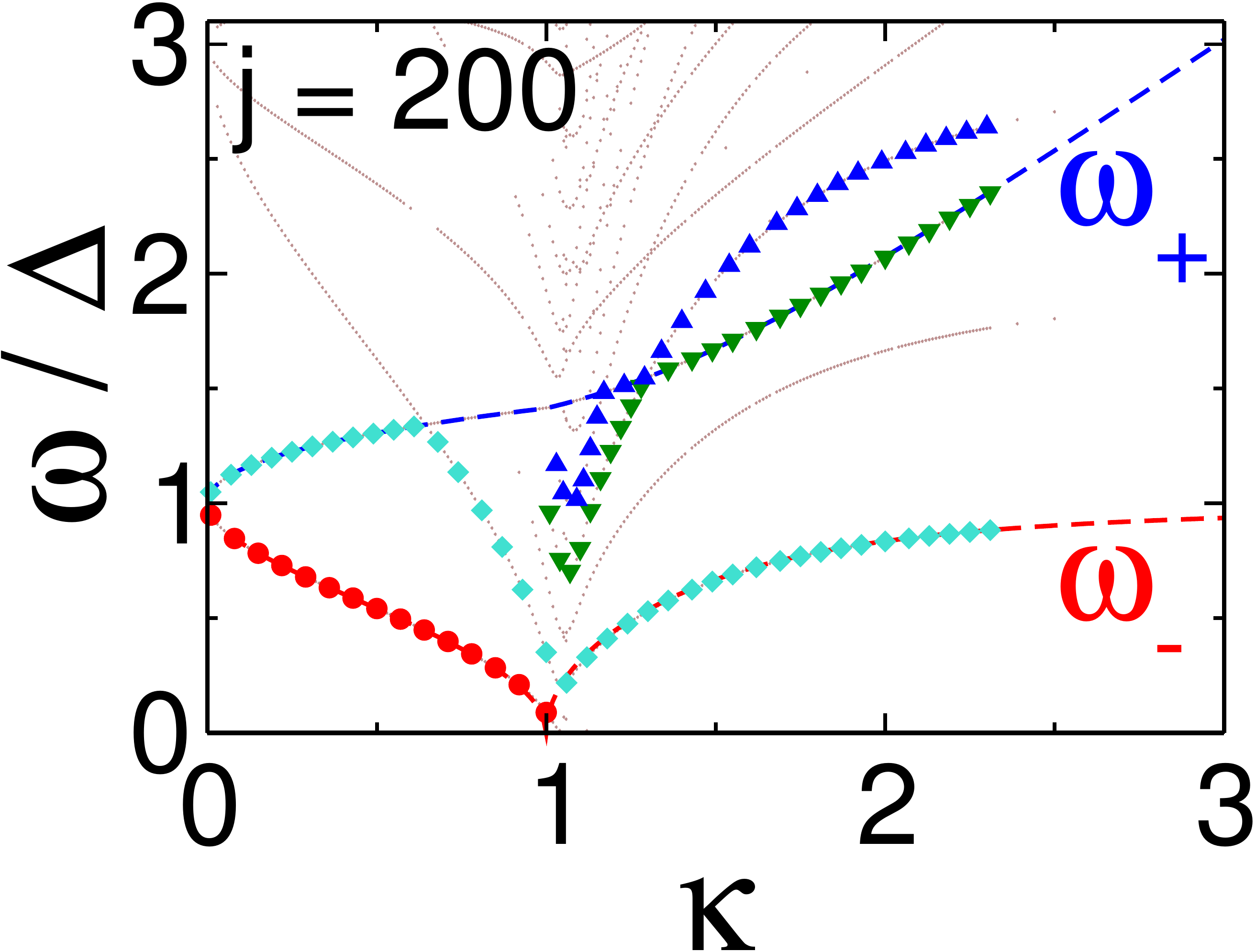}
\includegraphics[width=0.49\linewidth]{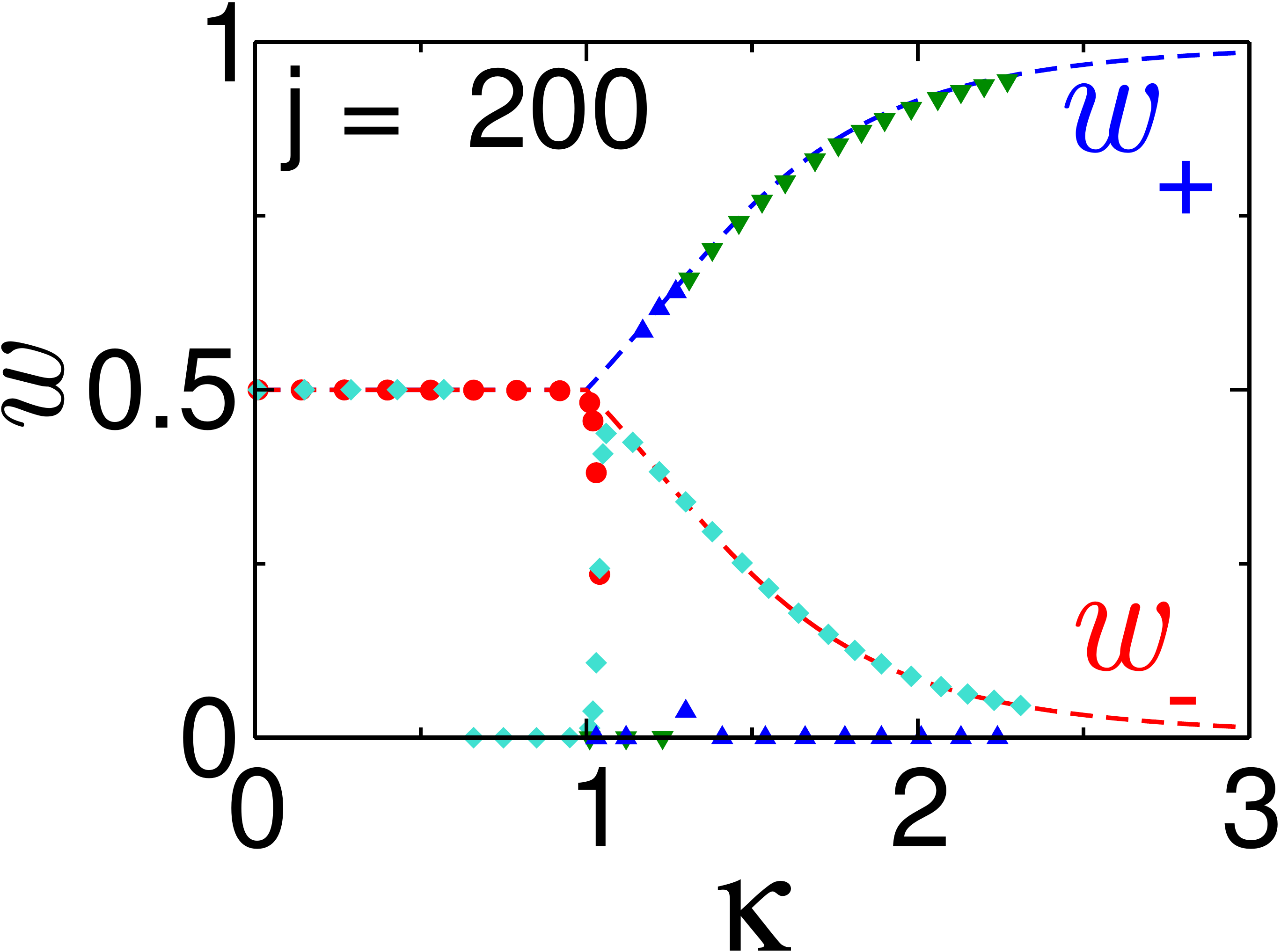}
\caption{(Color online) Position (left panel) and weight (right panel) of the peaks in the normalized Green function $\langle\langle J_x; J_y \rangle\rangle_\omega/(-2 \pi \langle J_z \rangle)$,
for $\Omega/\Delta=1$ and
$j=10, 100, 200$ from top to bottom.
The dashed lines show the classical frequencies $\omega_\pm$ and weights $w_\pm$ from Eqs.~\eqref{FreqClass1}---\eqref{WeightAbove}.
Those branches of the quantum excitation spectrum which gain significant weight are accentuated by colored symbols.
}
\label{fig:peakposgew}
\end{figure}

For a quantitative comparison with the classical limit, we show in Fig.~\ref{fig:peakposgew} the peak positions and weights as extracted from $\langle\langle J_x; J_y \rangle\rangle_\omega/(-2 \pi \langle J_z \rangle)$.
We see that with increasing $j$, the quantum mechanical Green function
indeed converges to the result in the classic limit (Eq.~\eqref{JxSC2}),
in the sense that the weight of the dominant peaks increases and their position shifts towards the frequencies $\pm \omega_\pm$ of the classical collective modes.
Since there is no QPT for finite $j$, convergence is slowest for $\kappa$ in the vicinity of the critical $\kappa=1$.
For example for $j = 100$, the peaks at $\pm \omega_\pm$ contribute $96 \%$ of the spectral weight for all $|\kappa -1 | > 0.4$, but only $ 79 \%$ for $\kappa \approx 1.06$.
In particular, precursors of the ``soft mode'' with $\omega_- \to 0$ for $\kappa \to 1$ can be identified only for large $j \ge 200$.

The various energies in Fig.~\ref{fig:peakposgew} correspond to quantized periodic motion around the one (below the QPT) or two (above the QPT) minima of the classical energy $E(z,\bar \alpha)$ from Eq.~\eqref{EZAlpha}.
In principle it should be possible to obtain these energies, and the corresponding wave functions and the peak weights $w$, with the Wentzel-Krames-Brillouin approximation or other SC quantization schemes~\cite{S03,SJ13}.
A comparison with the numerical data presented here would be most interesting in the vicinity of the phase transition, where deviations from the classical collective mode energies $\omega_\pm$ remain significant also for large $j$.

\section{Classical and quantum chaos}
\label{sec:Chaos}

After our study of the dynamics in the vicinity of the groundstate, we now turn to the general non-equilibrium dynamics for larger energies.
In contrast to the linear response dynamics studied in the previous section,
we can no longer expect a simple relation between the classical dynamics and the time-evolution of quantum-mechanical expectation values.
Additional corrections beyond the leading order of the SC approximation arise,
 e.g., from quantum diffusion in phase space~\cite{AH12,AH12a} that leads to spreading of the wave function.
These corrections manifest themselves in the time-evolution of the wave function, but not in simple expectation values.
Stable or unstable periodic orbits lead to different signatures in the quantum eigenstates~\cite{Gutz90,Haak10},
and require  classification of individual eigenstates  in particular for mixed classical dynamics where regular and chaotic orbits coexist at the same energy.
Conversely, SC quantization schemes can be used to construct stationary or time-dependent wave functions along known classical orbits~\cite{S03,SJ13,SVT12}.
Therefore, we will compare classical orbits with phase space distribution functions of the corresponding quantum orbits and eigenstates rather than the (spin) observables used in the previous section. 
To give a global picture of the dynamics we compare classical and quantum Poincare plots.

\subsection{Classical dynamics}

\begin{figure}
\includegraphics[width=0.32\linewidth]{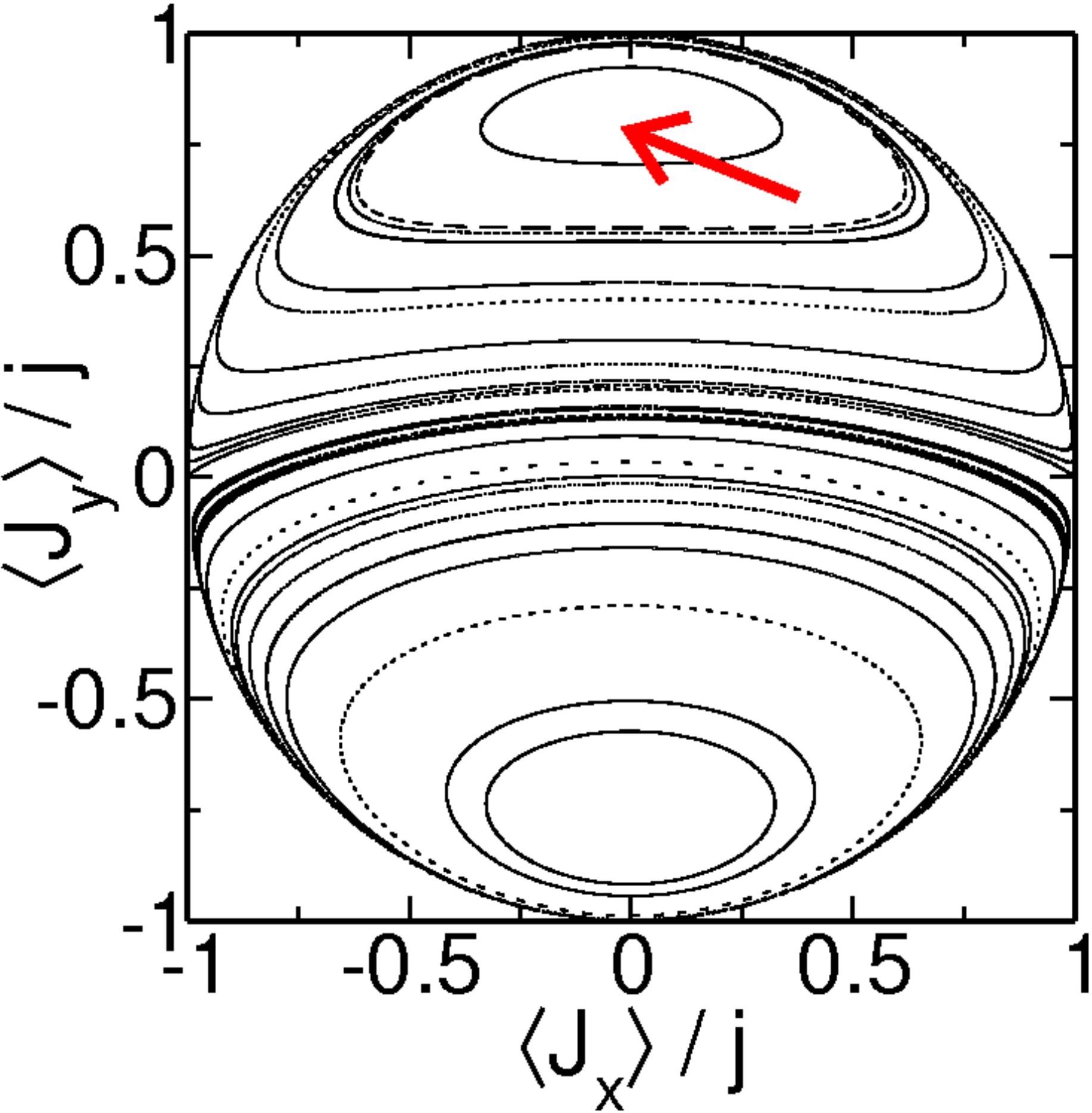}
\includegraphics[width=0.32\linewidth]{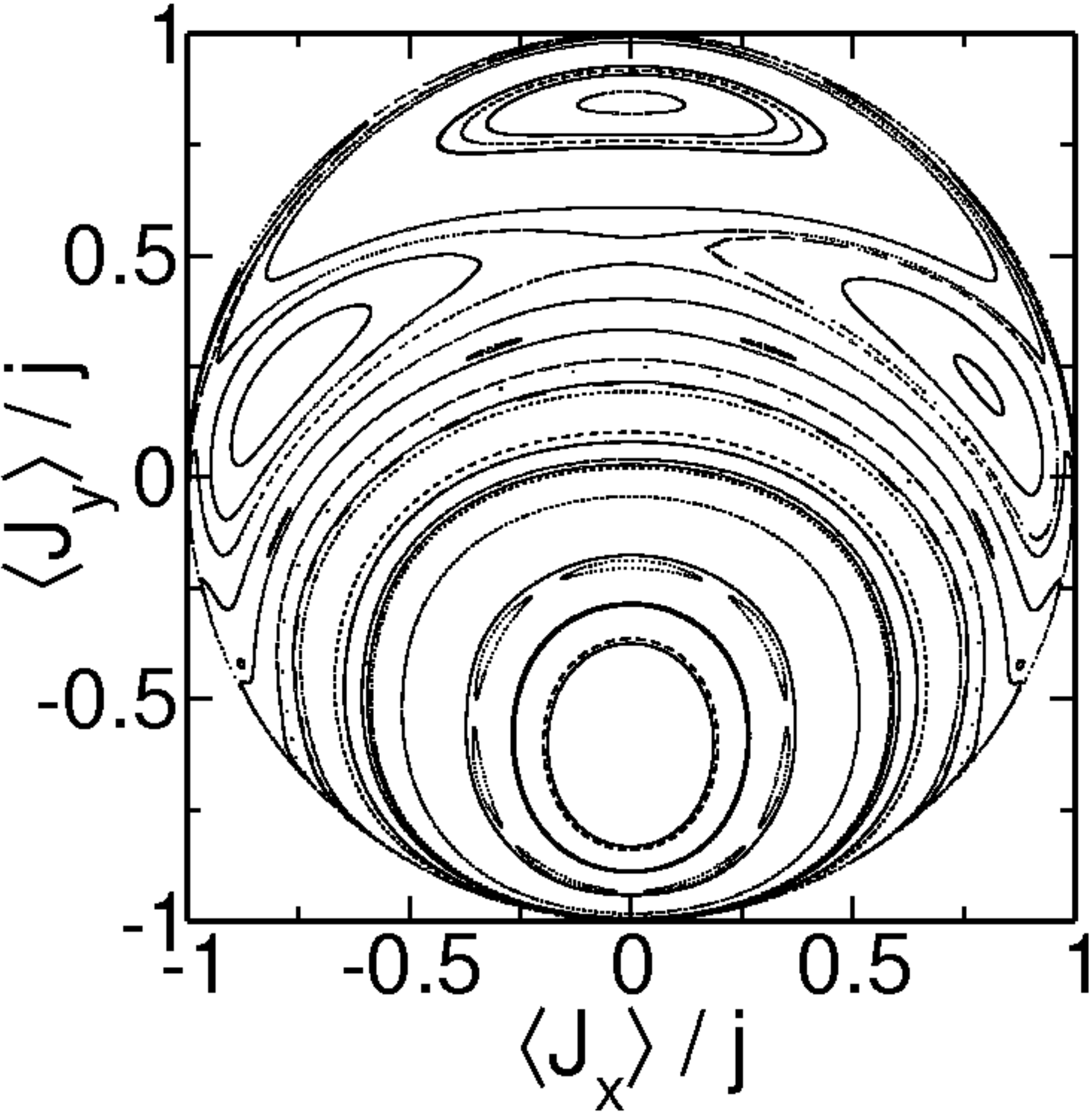}
\includegraphics[width=0.32\linewidth]{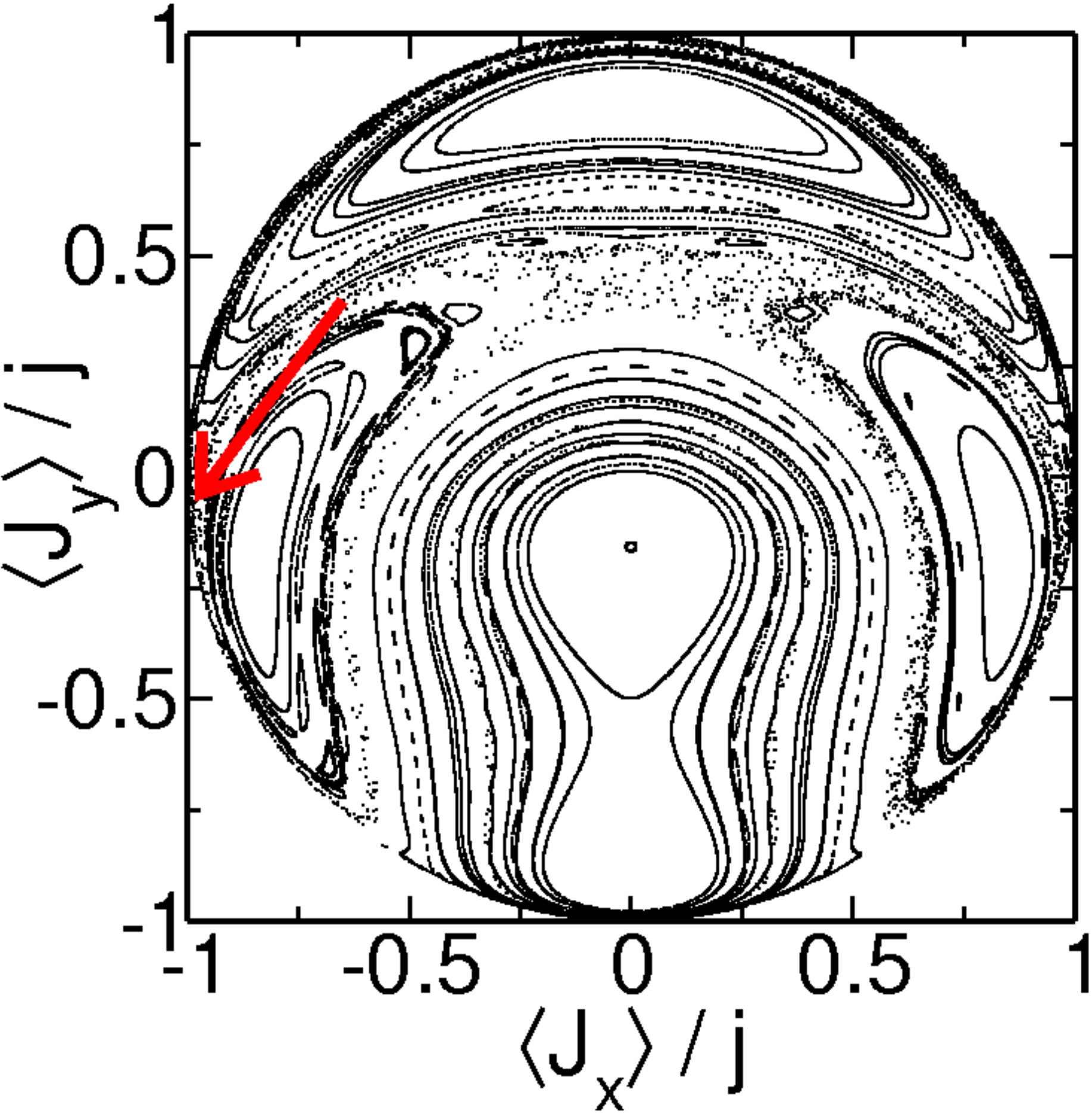} \\
\includegraphics[width=0.32\linewidth]{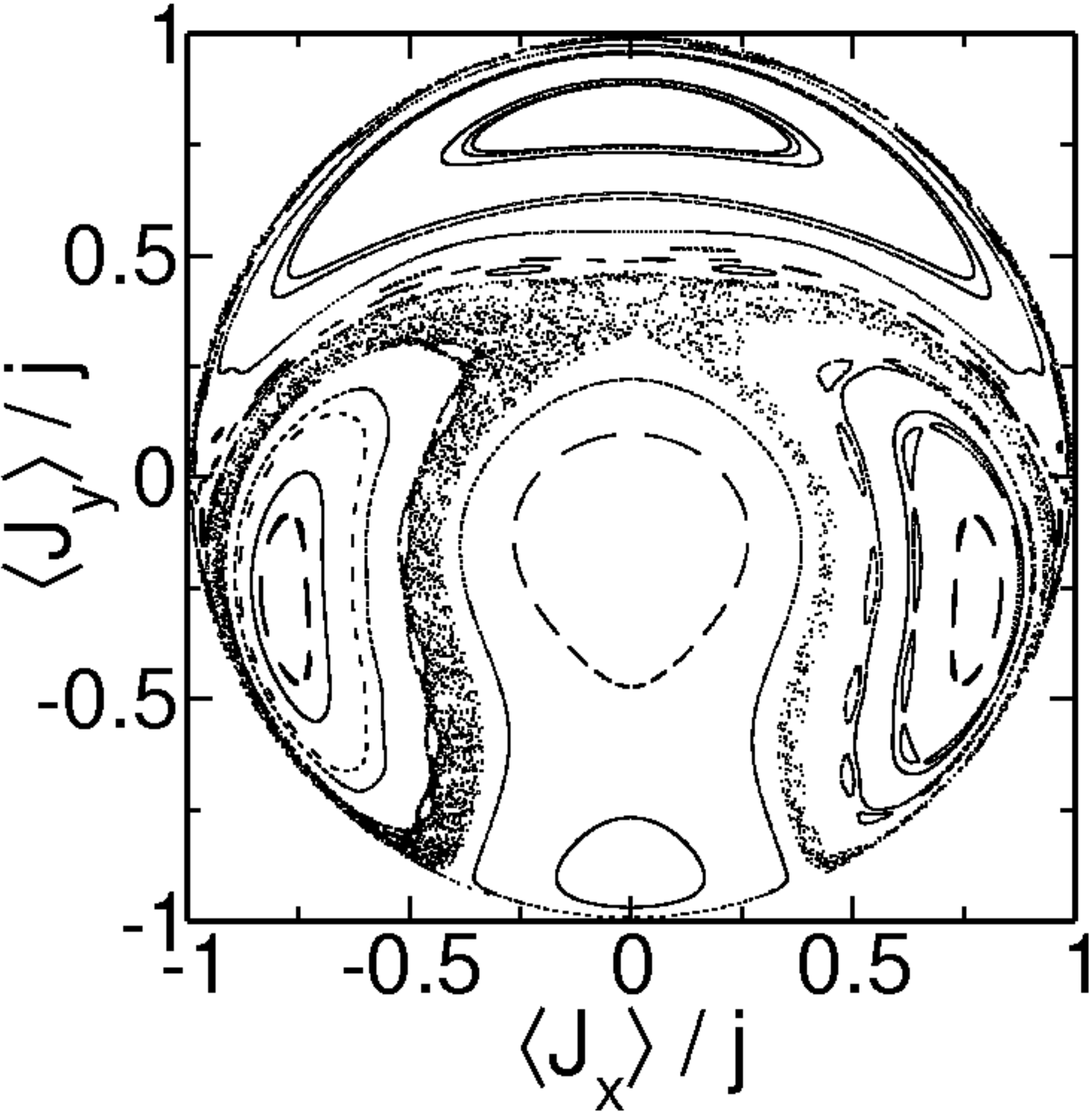}
\includegraphics[width=0.32\linewidth]{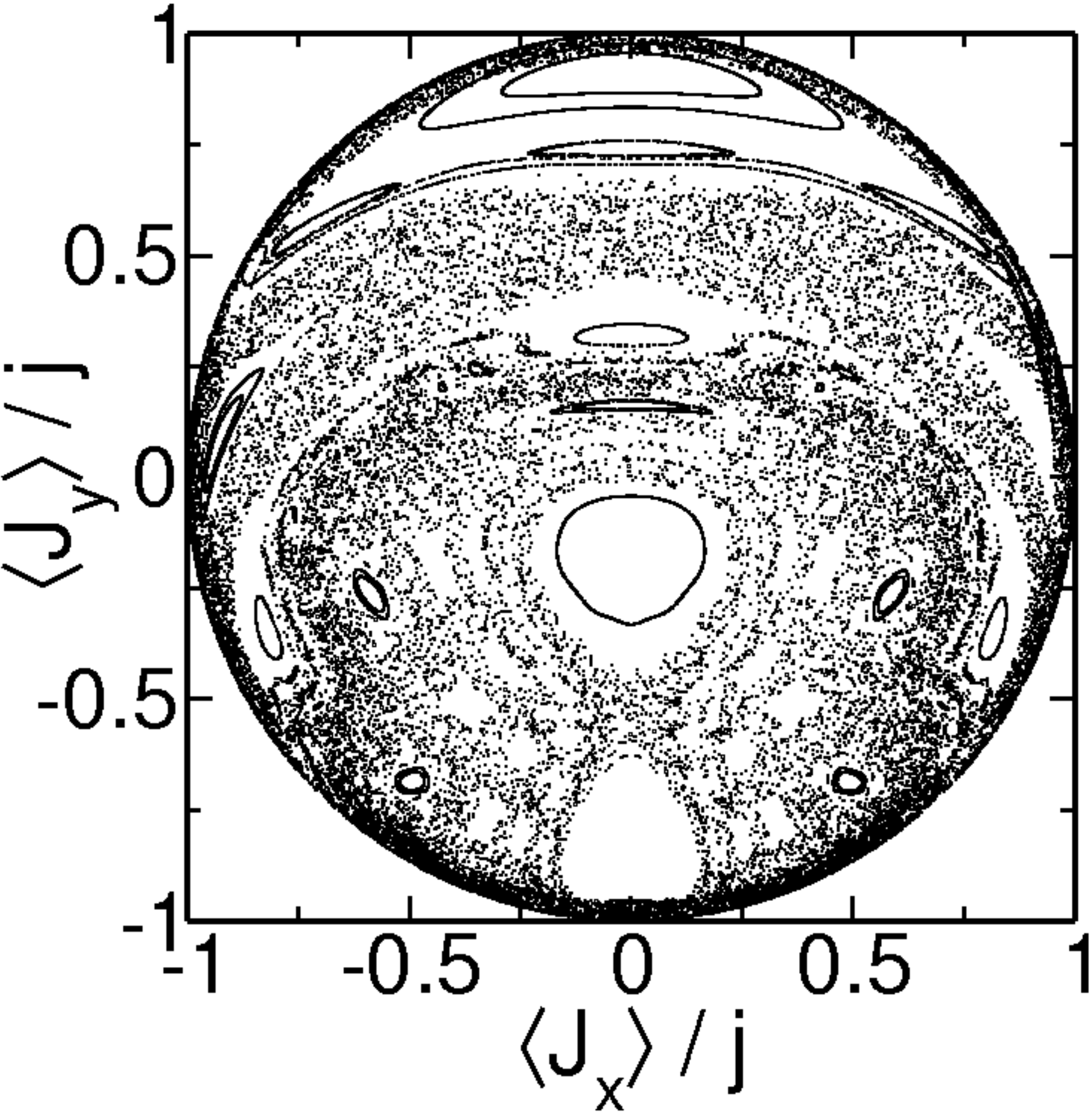}
\includegraphics[width=0.32\linewidth]{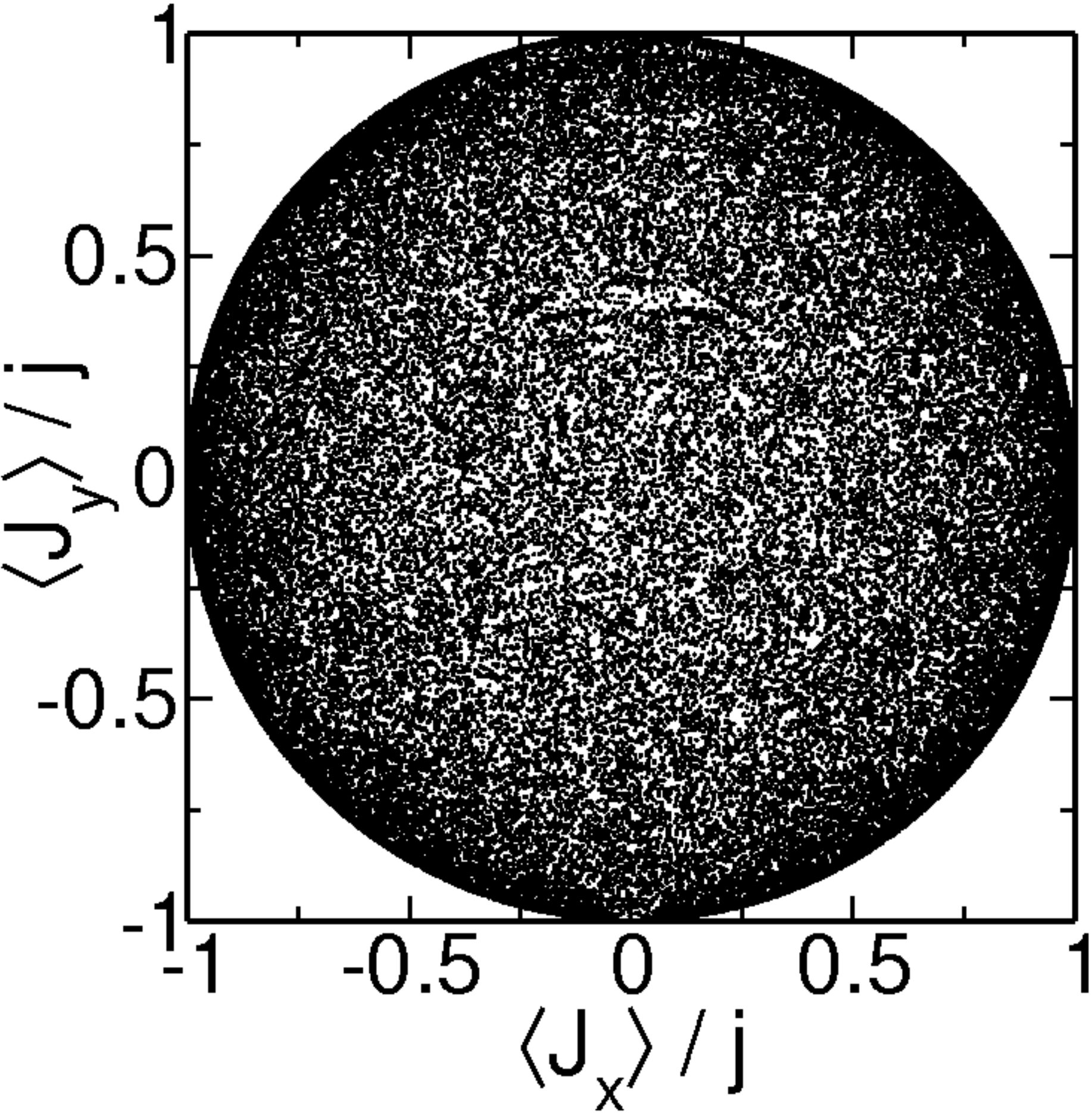}
\caption{Top row: Poincare plots for $E=-0.5$ and
 $\kappa=0.1$ (left), $\kappa=0.5$ (middle),  $\kappa=0.6$ (right).
Bottom row: Poincare plots for $\kappa=0.6$ and $E=-1.0$ (left), $E=1.0$ (middle), $E=9.9$ (right).
Red arrows denote the intitial conditions for the orbits given in Fig.~\ref{fig:TwoClassOrbs} below.
}
\label{fig:ClassPoinc}
\end{figure}

\begin{figure}
\includegraphics[width=0.45\linewidth]{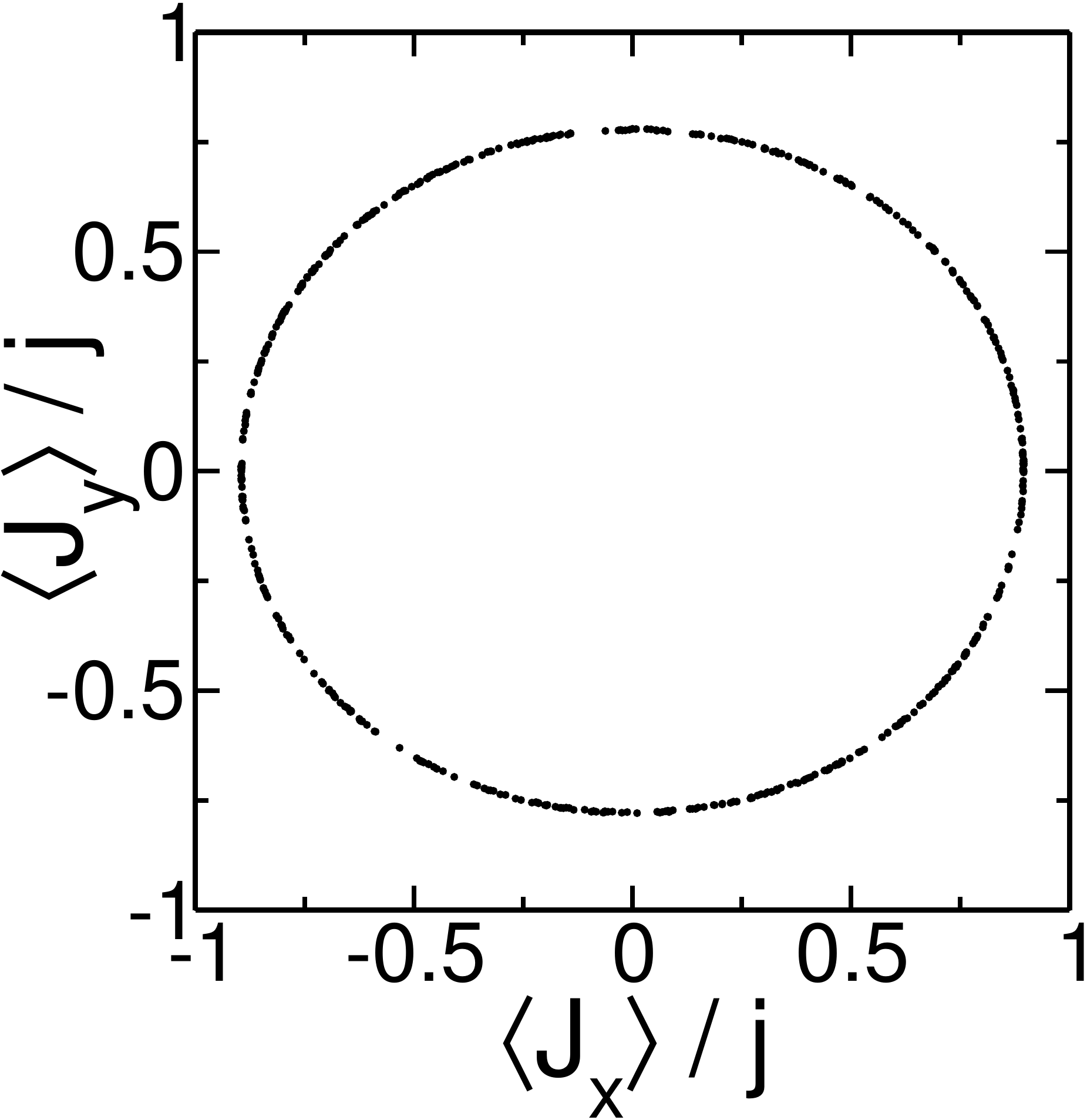}
\hfill
\includegraphics[width=0.45\linewidth]{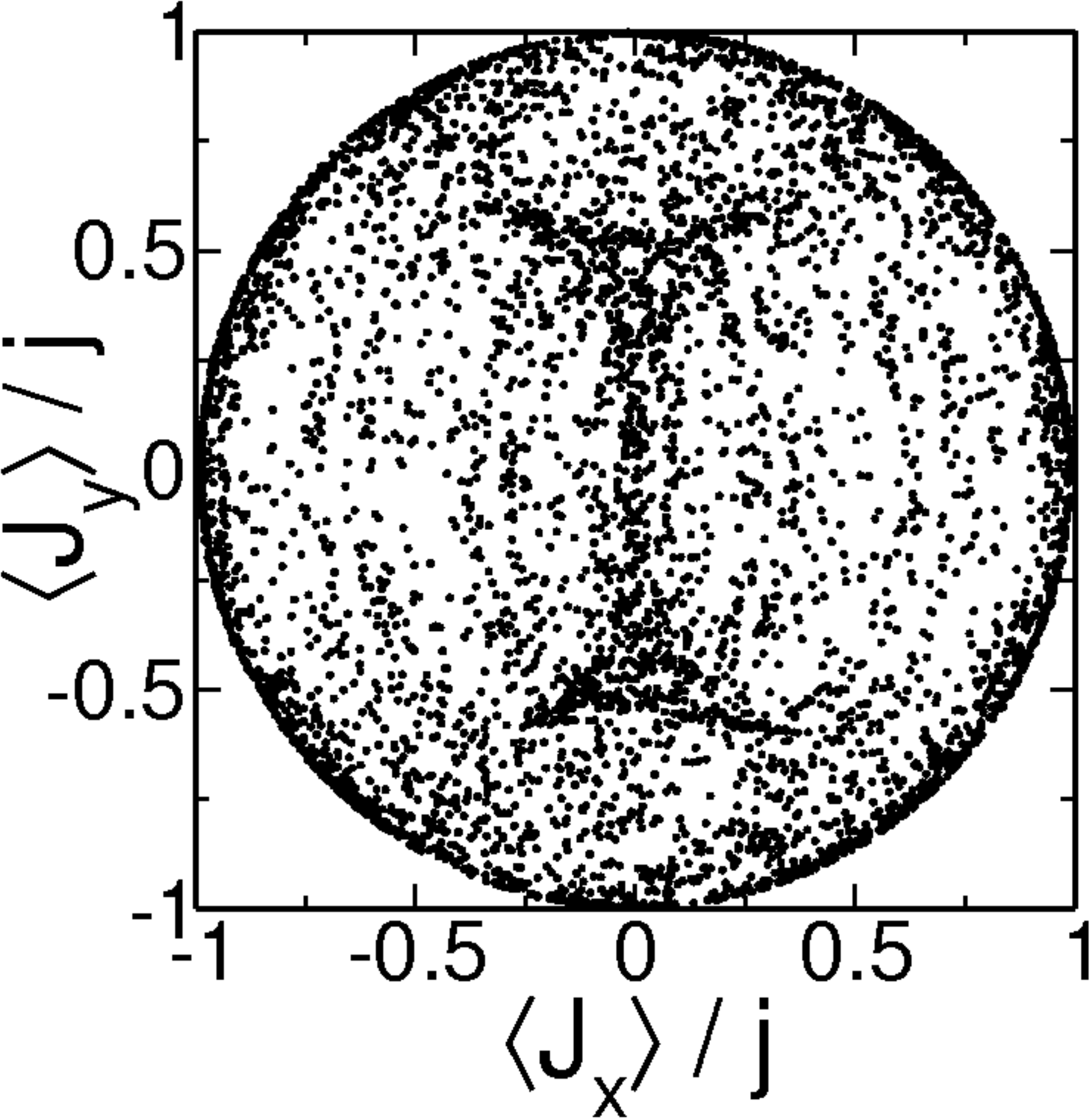}\\
\includegraphics[width=0.45\linewidth]{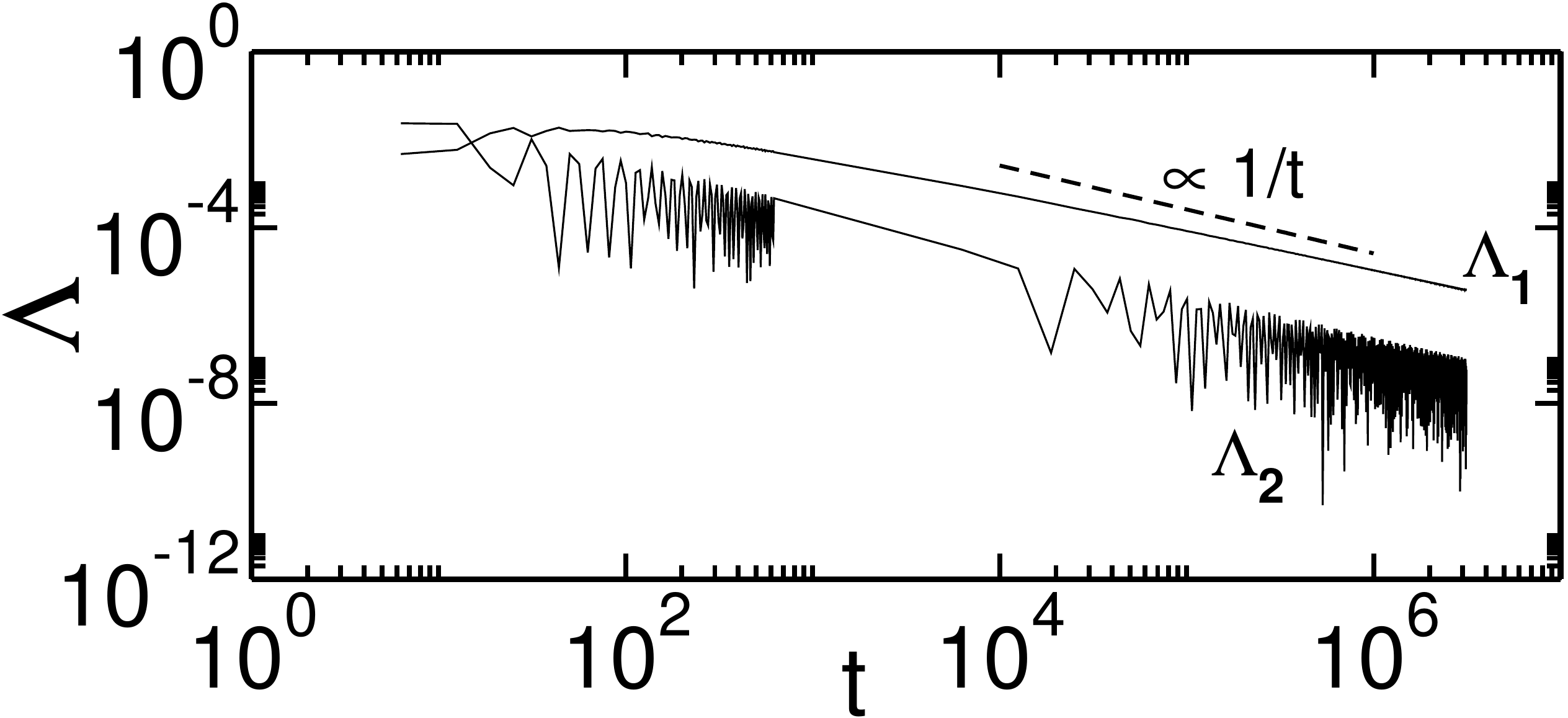}
\hfill
\includegraphics[width=0.45\linewidth]{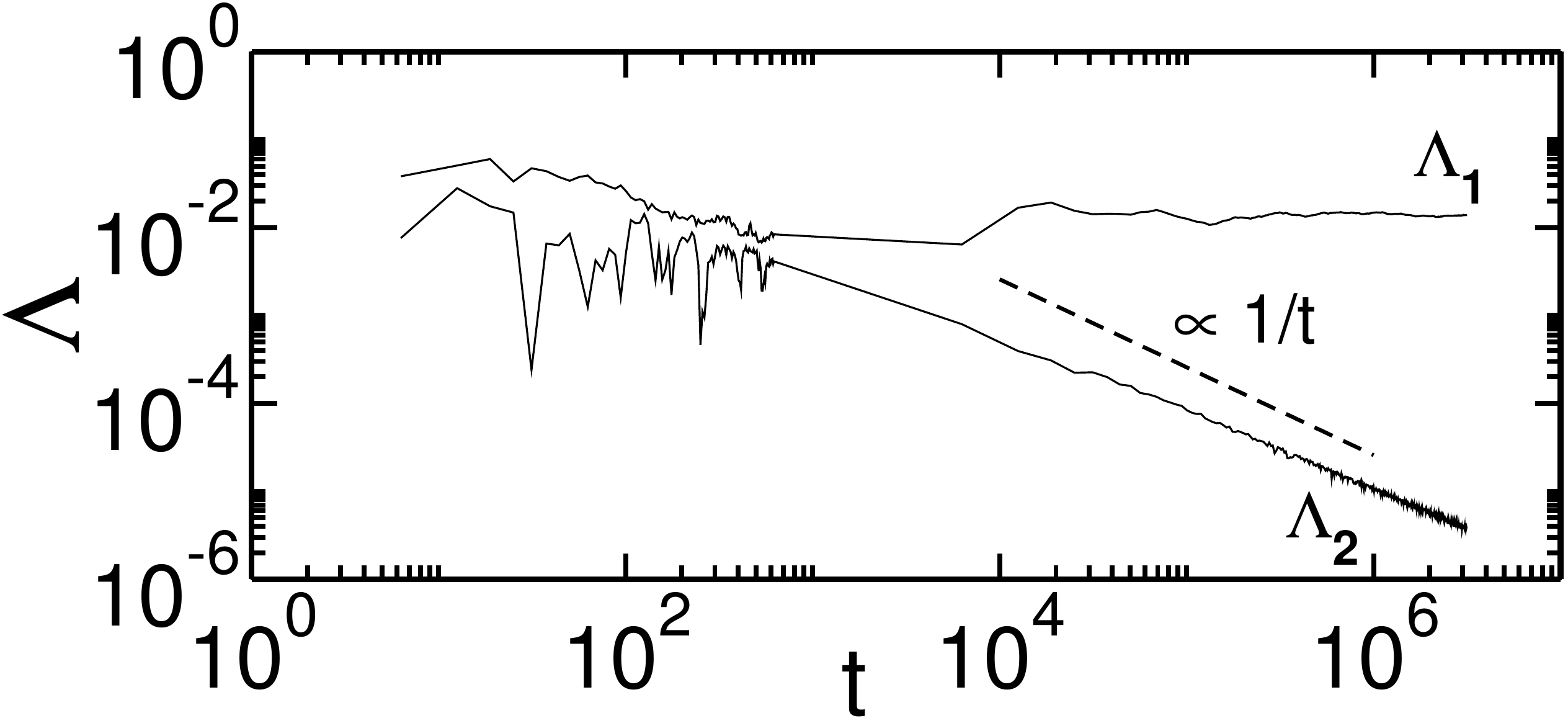}
\caption{Top row: Classical orbits for 
$E=-0.5$, $\kappa=0.1$, $J_x(0)=0.0$, $J_y(0)=0.78$ (left panel) 
and  $E=-0.5$, $\kappa=0.6$, $J_x(0)= -0.987$, $J_y(0)= -0.065$ (right panel),
corresponding to the arrows in the Poincare plots in the previous figure.
Shown is the trajectory in $J_x$--$J_y$ phase space for $0 \le t \le 6000 \times 2\pi/\Delta$.
Bottom row: Positive Lyapunov exponents $\Lambda_{1,2}$ for the two orbits as a function of time.
The left orbit is regular with $\Lambda_{1,2} \to 0$ for $t \to \infty$,
the right orbit is chaotic with $\Lambda_1 \to 0.014 > 0$.
}
\label{fig:TwoClassOrbs}
\end{figure}

Depending on parameters and initial conditions, the SC equations of motion (Eq.~\eqref{SCEom}) predict regular or chaotic dynamics in the limit $j \to \infty$.
This is illustrated by the Poincare plots in Fig.~\ref{fig:ClassPoinc}, which are obtained from classical orbits to fixed energy $E = E(z, \bar \alpha)$. Plotted are the values of $J_x(t)$, $J_y(t)$ at those times $t \ge 0$ when $Q(t) = \Re \bar \alpha(t)=0$.
The knowledge of the four variables $E$, $J_x(t)$, $J_y(t)$, $Q(t)$ fixes the remaining variable $P(t) = \Im \bar \alpha(t)$ because of energy conservation (cf. Eq.~\eqref{EZAlpha}).
The points in the plot are assembled from several orbits at the respective energy.

Regions with regular and chaotic motion can be discerned in the Poincare plots.
For large $E$ all orbits are chaotic, but regular and chaotic dynamics coexist for smaller $E$.
Two different orbits, a stable periodic orbit (left panel) and a chaotic orbit (right panel), are shown in Fig.~\ref{fig:TwoClassOrbs}.
The stability of the classical orbits is characterized by the behavior of the (maximal) Lyapunov exponent $\Lambda(t)$ for $t \to \infty$, 
which we calculate with the ``standard method'' from Refs.~\cite{BGGS80a,BGGS80b}.
In the present case, for a four-dimensional Hamiltonian system, the
Lyapunov exponents appear in two pairs $\pm \Lambda_1(t)$, $\pm \Lambda_2(t)$. Two exponents ($\pm \Lambda_2(t)$) vanish for $t \to \infty$ because motion along the orbit is stable~\cite{BGGS80a}.
For a regular orbit (left panel in Fig.~\ref{fig:TwoClassOrbs}) also $\pm \Lambda_1(t) $ vanish,
while a chaotic orbit (right panel in Fig.~\ref{fig:TwoClassOrbs}) is characterized by a positive Lyapunov exponent $\Lambda_1(t) > 0$ in the limit $t \to \infty$.
Note that the chaotic orbit is ergodic and fills the entire energy shell $E(z,\bar \alpha) = E$ (cf. Eq.~\eqref{EZAlpha}).
We next compare the two classical orbits to their quantum mechanical counterparts for $j < \infty$.

\subsection{Quantum dynamics}

\begin{figure}
\includegraphics[width=0.45\linewidth]{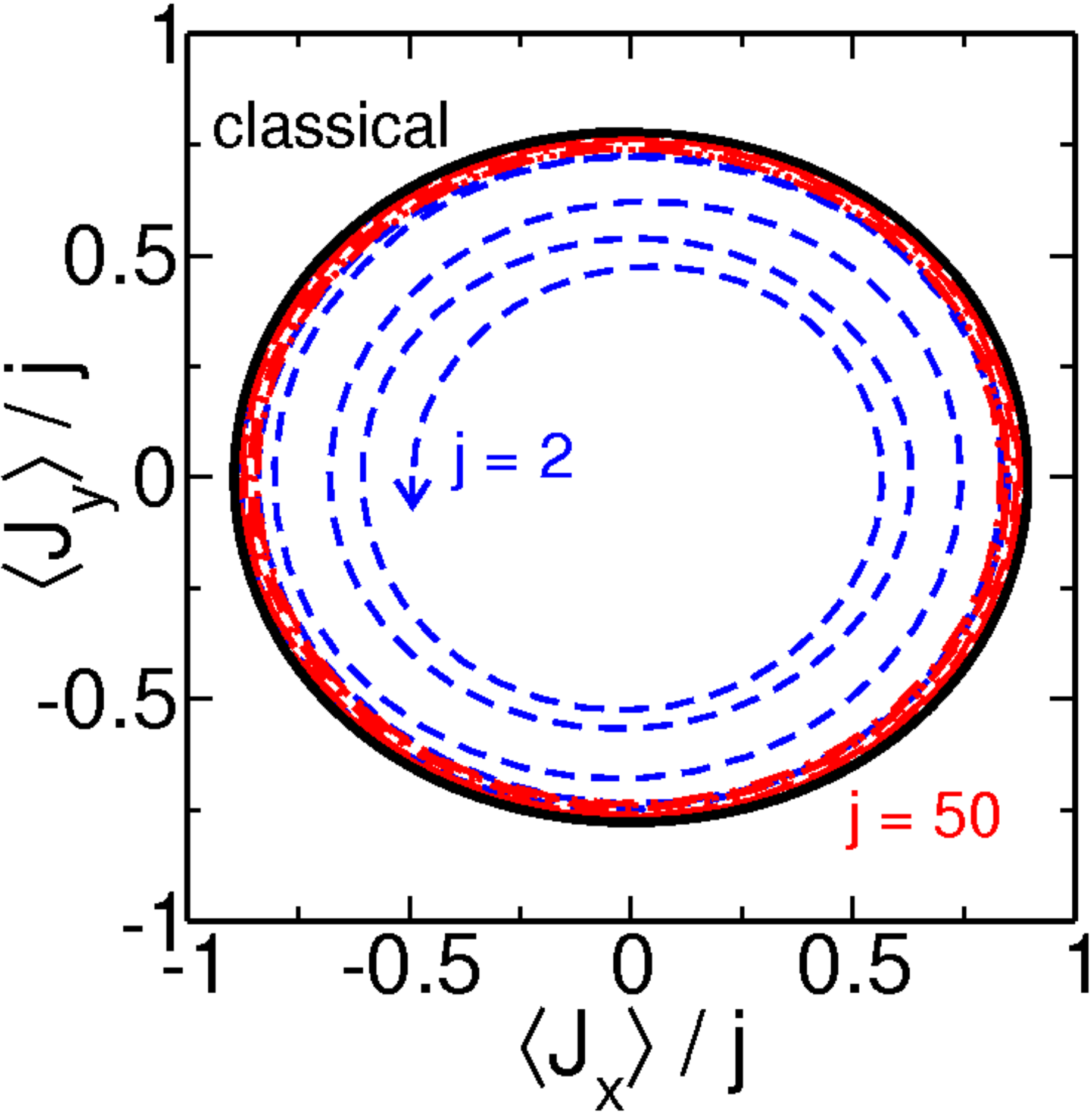}
\hfill
\includegraphics[width=0.45\linewidth]{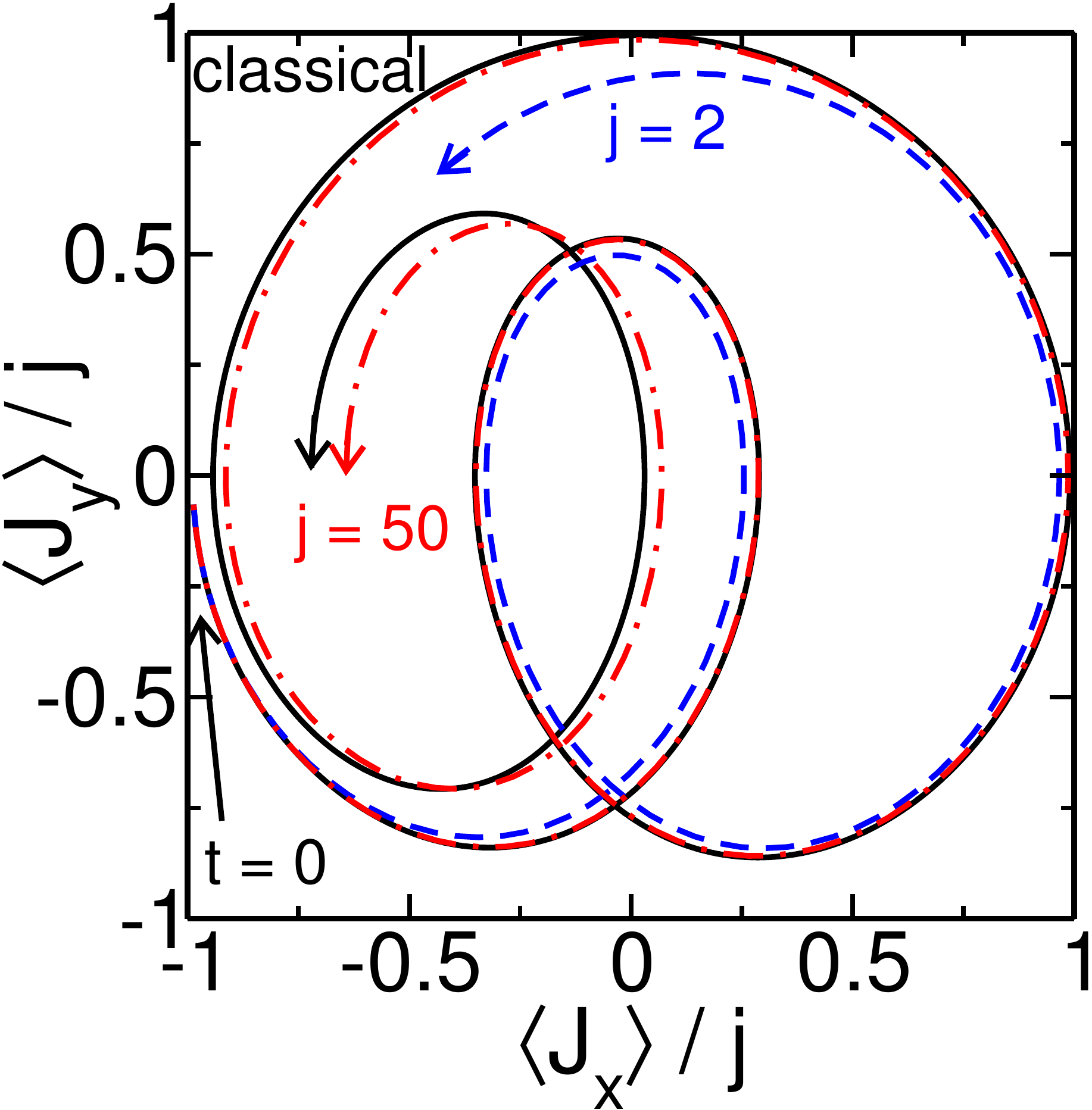}
\\
\includegraphics[width=0.45\linewidth]{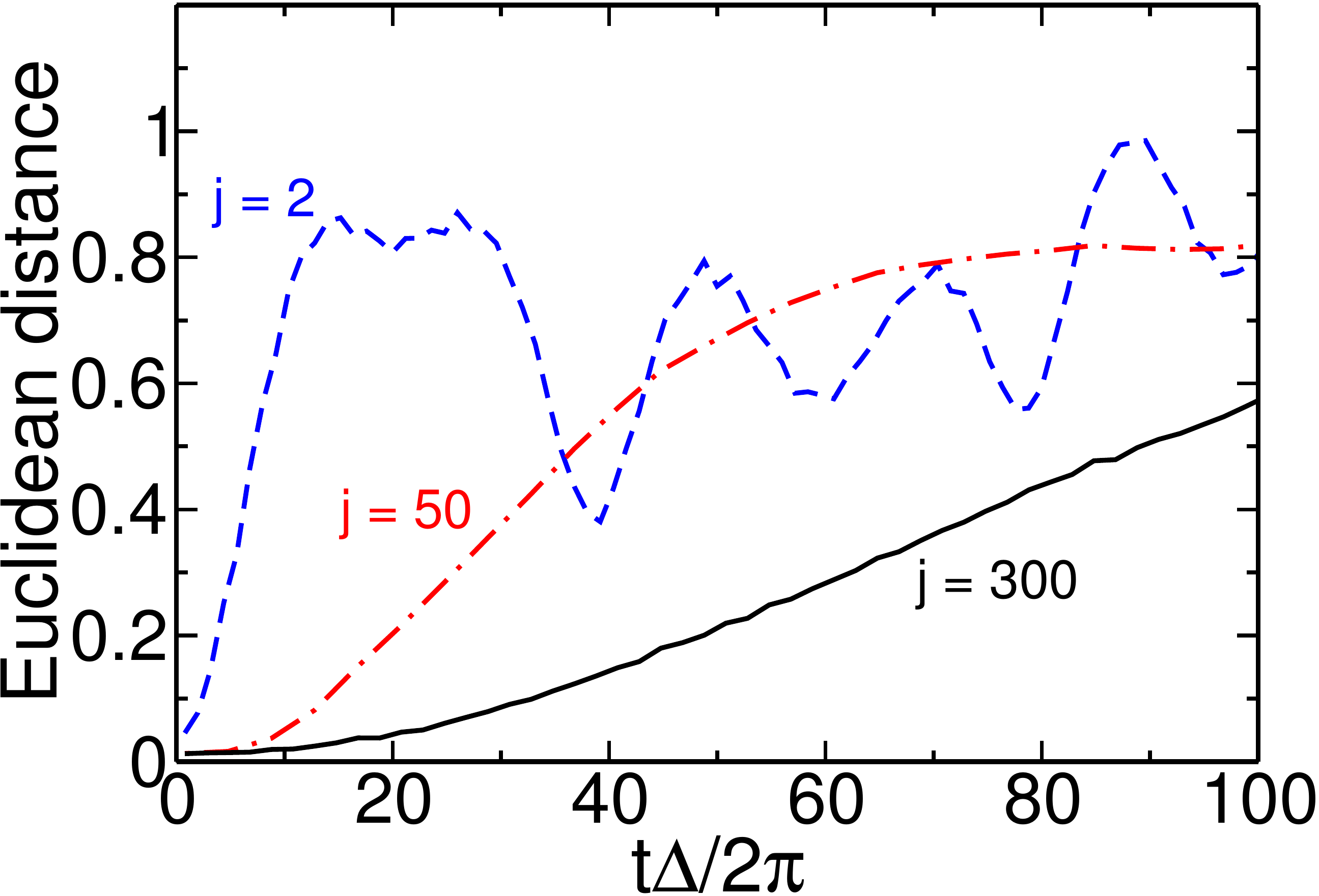}
\hfill
\includegraphics[width=0.45\linewidth]{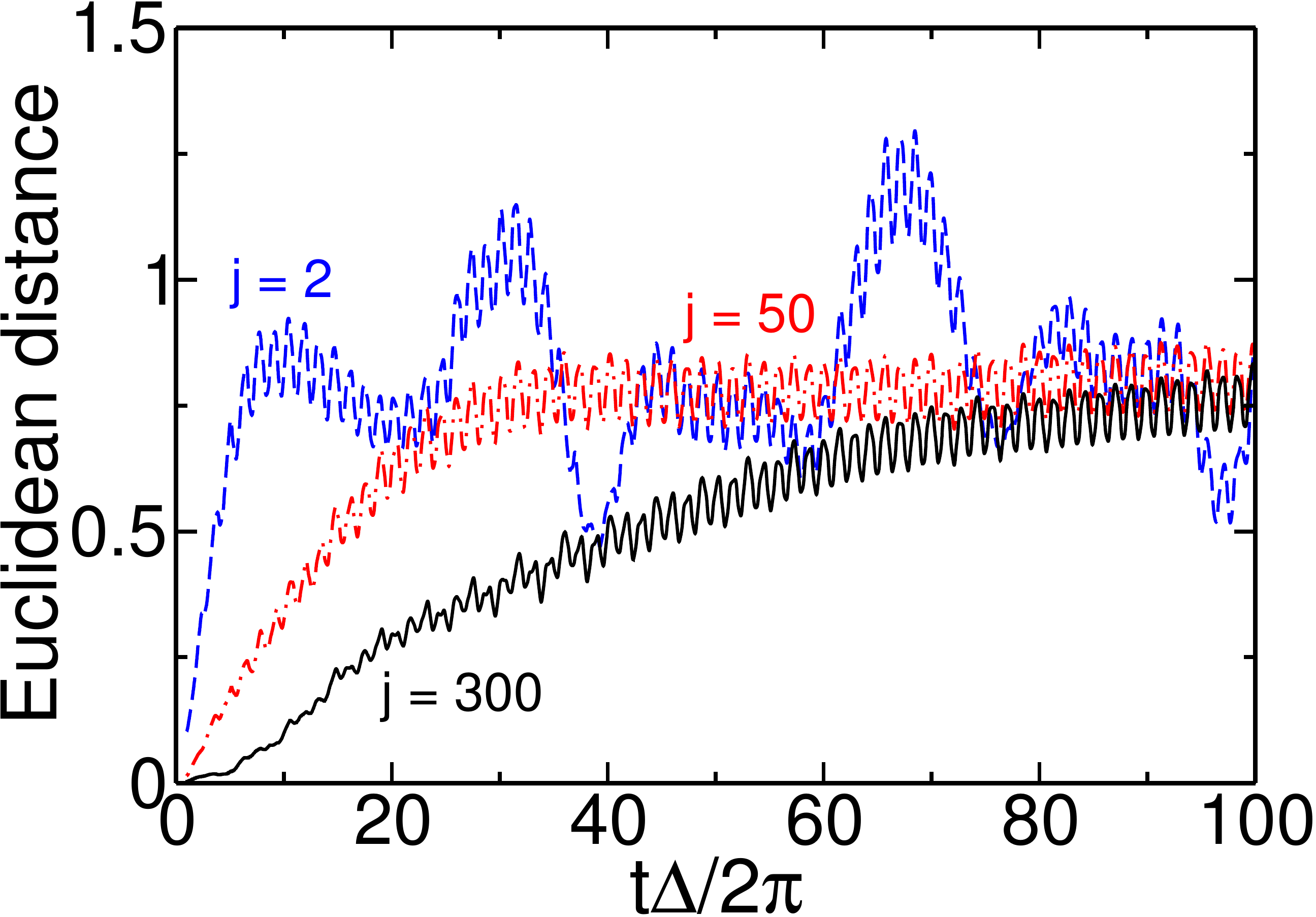}
\caption{Comparison of the classical orbits from Fig.~\ref{fig:TwoClassOrbs}
to the spin expectation values $\langle J_x(t) \rangle$, $\langle J_y(t) \rangle$ from the quantum dynamics for the corresponding initial states as in Eq.~\eqref{eq:cohstate}.
Shown are the classical (solid curves) and quantum (dashed curves) trajectories in $J_x$--$J_y$ phase space (upper row)
and the Euclidean distance between the trajectories as a function of time (lower row), for $j=2, 50, 300$.
}
\label{fig:TwoQuantOrbs}
\end{figure}

For the quantum dynamics, we start from a coherent product state
\begin{equation}
\label{eq:cohstate}
|\psi(0)\rangle = |\alpha(0)\rangle \otimes |z(0)\rangle \;,
\end{equation}
whose parameters are chosen according to the classical initial condition.
The relation to the spin and oscillator expectation values is given by Eqs.~\eqref{ZSpin2}---\eqref{AlphaQP}. 
We obtain the time-evolution of $|\psi(t)\rangle$ numerically with Chebyshev time propagation~\cite{TK84,AF08}.

In Fig.~\ref{fig:TwoQuantOrbs} we show the spin expectation values $\langle J_x(t) \rangle$, $\langle J_y(t) \rangle$ that constitute the quantum trajectory in comparison to the corresponding classical orbits from Fig.~\ref{fig:TwoClassOrbs}.
The classical and quantum trajectory agree only over a short time period,
whose length increases with $j$.
As expected, the agreement is better for the stable orbit than for the chaotic orbit.
Nevertheless, deviations occur even for the stable orbit
already after a few periods (see lower left panel for $j=300$).
In difference to the linear response situation studied in Sec.~\ref{sec:CollModes}, convergence of the quantum to the classical trajectory with increasing $j$ is absent or slow.

\begin{figure}
\includegraphics[width=0.32\linewidth]{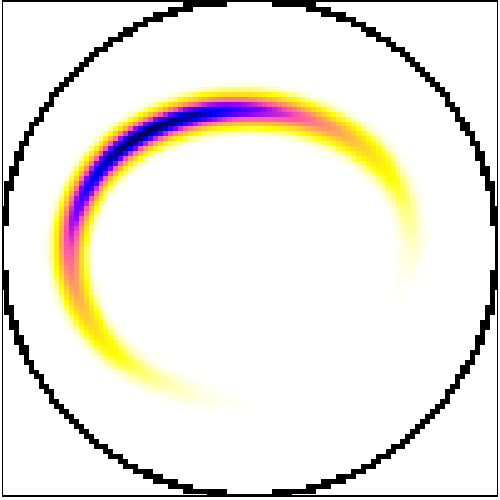} 
\hfill
\includegraphics[width=0.32\linewidth]{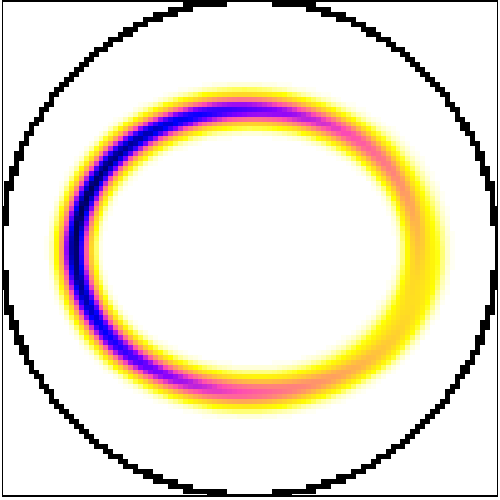} 
\hfill
\includegraphics[width=0.32\linewidth]{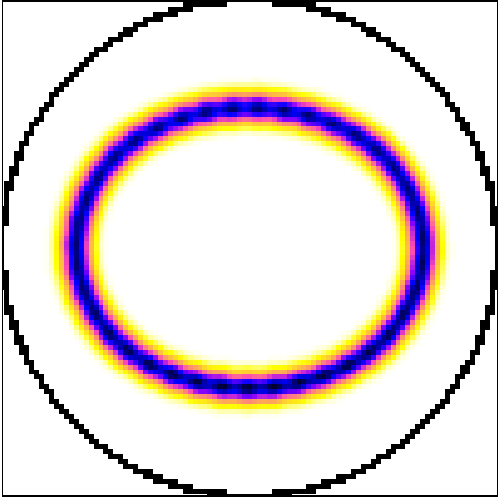} 
\\
\includegraphics[width=0.32\linewidth]{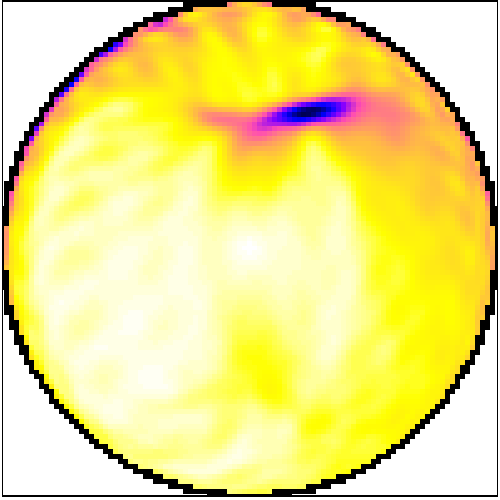}
\hfill
\includegraphics[width=0.32\linewidth]{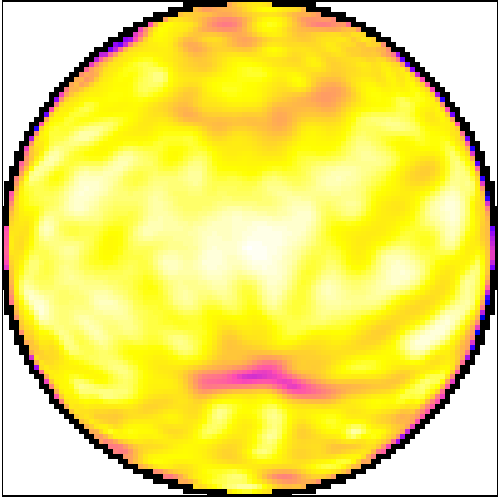}
\hfill
\includegraphics[width=0.32\linewidth]{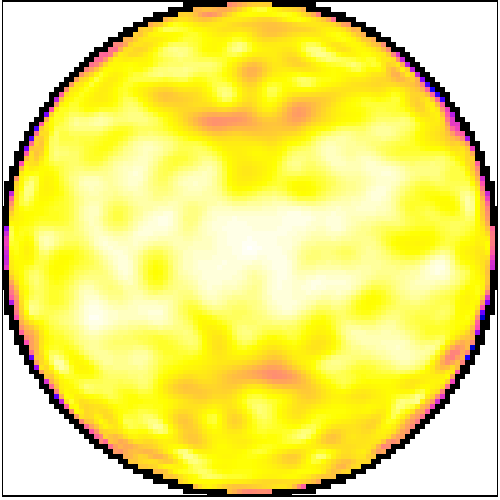}
\caption{Spin Husimi function $Q(\theta,\phi; t)$ 
of the two orbits from Figs.~\ref{fig:TwoClassOrbs} (upper panel),~\ref{fig:TwoQuantOrbs} (lower panel),
for $j=300$ and for $t \Delta/(2\pi) = 50, 100, 200$ from left to right.
Here and in the following figures we show the projection of $Q(\theta,\phi;t)$ onto the $J_x$--$J_y$ plane.
The angle $\theta$ runs from $0$ in the center to $\pi/2$ on the outer circle.
The angle $\phi$ runs counterclockwise from $0$ at the top of the circle to $2 \pi$.
}
\label{fig:QuantHus}
\end{figure}

This behavior can be traced back to the fact that the SC equations of motion coincide with the quantum time-evolution only as long as the quantum state is approximately a coherent product state as in Eq.~\eqref{ProdState}.
Therefore, the classical and quantum trajectories agree only over a finite time $T_E$, the Ehrenfest time, which is of the order of a few spin periods $2\pi/\Delta$ in Fig.~\ref{fig:TwoQuantOrbs}.

A better comparison of the quantum and classical time evolution is possible with phase space functions. We use the spin Husimi function
\begin{equation}
\label{eq:spinhusimi}
Q(\theta, \phi; t) = |\langle \theta, \phi| \psi(t)\rangle|^2 \;,
\end{equation}
which gives the overlap with a coherent spin state $|\theta,\phi\rangle=|z\rangle$ in the spin phase space $\theta, \phi$
(the relation to the complex variable $z$ is as in Eq.~\eqref{ZSpin2}).
For a coherent state in the classical limit $j \to \infty$,
$Q(\theta, \phi)$ shrinks to a point at the respective spin position.
For $j < \infty$, the coherent state covers a phase space volume $\propto 1/j$.

For the present study of the Dicke model we prefer the Husimi function over, e.g., the Wigner function~\cite{Schl01} because it has a well-defined classical limit.
As shown in Refs.~\cite{AH12,AH12a} the exact time-evolution of the joint spin-oscillator Husimi function $Q(z,\alpha,t)$ is determined by a Fokker-Planck equation with a classical drift and a quantum diffusion term. The quantum diffusion term vanishes for $j \to \infty$,
and the Husimi function reduces to a classical probability function on phase space that obeys the Liouville equation.
The equations of motion for the Wigner function contain higher-order derivatives that complicate the classical limit.
Although the Wigner function is successfully used for other systems or the study of other aspects, such as the phase space complexity of quantum dynamics~\cite{SZBC08,BBCG10}, the comparison between the quantum and classical Dicke model is best performed with the Husimi functions.

In Fig.~\ref{fig:QuantHus} we show the spin Husimi function for the two orbits from Fig.~\ref{fig:TwoQuantOrbs}, for large spin length $j=300$.
We now observe convergence of the quantum to the classical dynamics,
in the sense that the spin Husimi function traces out the phase space region accessible to the classical orbits.
However, classical phase space drift and quantum diffusion lead to the spreading of the phase space probability~\cite{AH12}, such that the Husimi function at a single point of time $t$ covers the entire orbit.
Clearly, the values of $t$ in Fig.~\ref{fig:QuantHus} are beyond the Ehrenfest time.
For the stable periodic orbit, the quantum state strictly expands along the one-dimensional classical trajectory in $J_x$--$J_y$ phase space.
The Husimi function remains localized on the classical orbit in spite of the spreading in phase space.
Already at finite (though large) $j$  we thus observe how the classical dynamics constrains the quantum dynamics: The quantum state spreads along, but not perpendicular to the classical orbit.
We note that this behavior, which leads to zero spin expectation values due to the averaging over the entire orbit, 
is related to the collapse of Rabi oscillations for large $j$~\cite{ABF12,BMKP92}.
For the chaotic orbit, the Husimi function fills the entire energy shell also traversed by the classical orbit,
which can be understood as a signature of (microcanonical) thermalization~\cite{AH12,AH12a}.

Spreading of the quantum state along the classical orbit explains why the Ehrenfest time is short
even when convergence to the classical dynamics is observed in the phase space functions.
Because the classical drift term dominates the initial time-evolution of the Husimi function for large spin length~\cite{AH12,AH12a} the Ehrenfest time depends crucially on the associated classical motion~\cite{SVT12}.
For a chaotic orbit classical drift in the unstable directions dominates and the Ehrenfest time scales
as $T_E \sim \Lambda^{-1} \ln (1/V)$, where $\Lambda$ is the maximal Lyapunov exponent and $V$ the initial phase space volume.
For a stable regular orbit the Ehrenfest time is determined by the much slower quantum diffusion along the orbit,
which results in the scaling $T_E \sim 1/\sqrt{V}$.
Indications of this difference between a regular and chaotic orbit can be seen 
already in Fig.~\ref{fig:TwoQuantOrbs}.

To quantify the spreading of a quantum state we use the spin variance $\Delta J_\parallel = \langle J_\parallel^2 \rangle - \langle J_\parallel \rangle^2$
of a rotated spin operator
\begin{equation}\label{eq:RotSpinOp}
J_{||} = \mathbf n \cdot \mathbf J = n_x J_x + n_y J_y + n_z J_z \;,
\end{equation}
which is minimized over all the possible directions $\mathbf n = (n_x,n_y,n_z)^T$ with $|\mathbf n|=1$.
The variance $\Delta J_\parallel$ is the minimum of a quadratic form in $\mathbf n$ and given by the smallest eigenvalue of the $3 \times 3$ matrix
\begin{equation}\label{JVarMat}
  \begin{pmatrix}
  \Delta_{x;x} & \Delta_{x;y} & \Delta_{x;z} \\[1ex]
  \Delta_{x;y} & \Delta_{y;y} & \Delta_{y;z} \\[1ex]
  \Delta_{x;z} & \Delta_{y;z} & \Delta_{z;z}
  \end{pmatrix} \;,
\end{equation}
whose entries are the (mixed) spin operator variances
\begin{equation}
\Delta_{k;l} = \frac{1}{2}(\langle J_k J_l + J_l J_k\rangle - 2\langle J_k\rangle \langle J_l\rangle) \;.
\end{equation}
The spin variance is invariant under rotation. It is $\Delta J_\parallel \ge 0$, and $\Delta J_\parallel = 0$ precisely for a spin coherent state.

\begin{figure}
\includegraphics[width=0.48\linewidth]{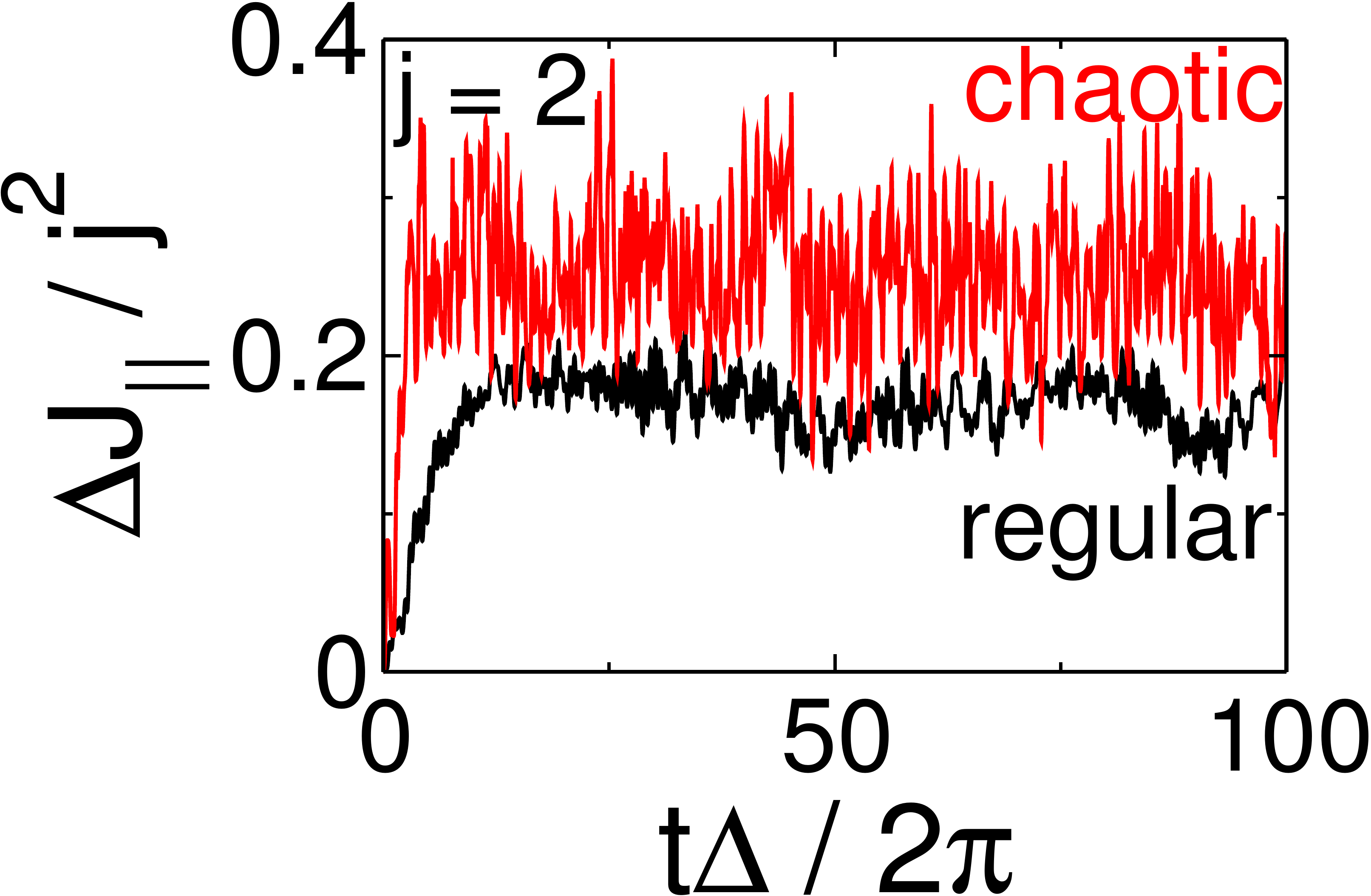}
\hfill
\includegraphics[width=0.48\linewidth]{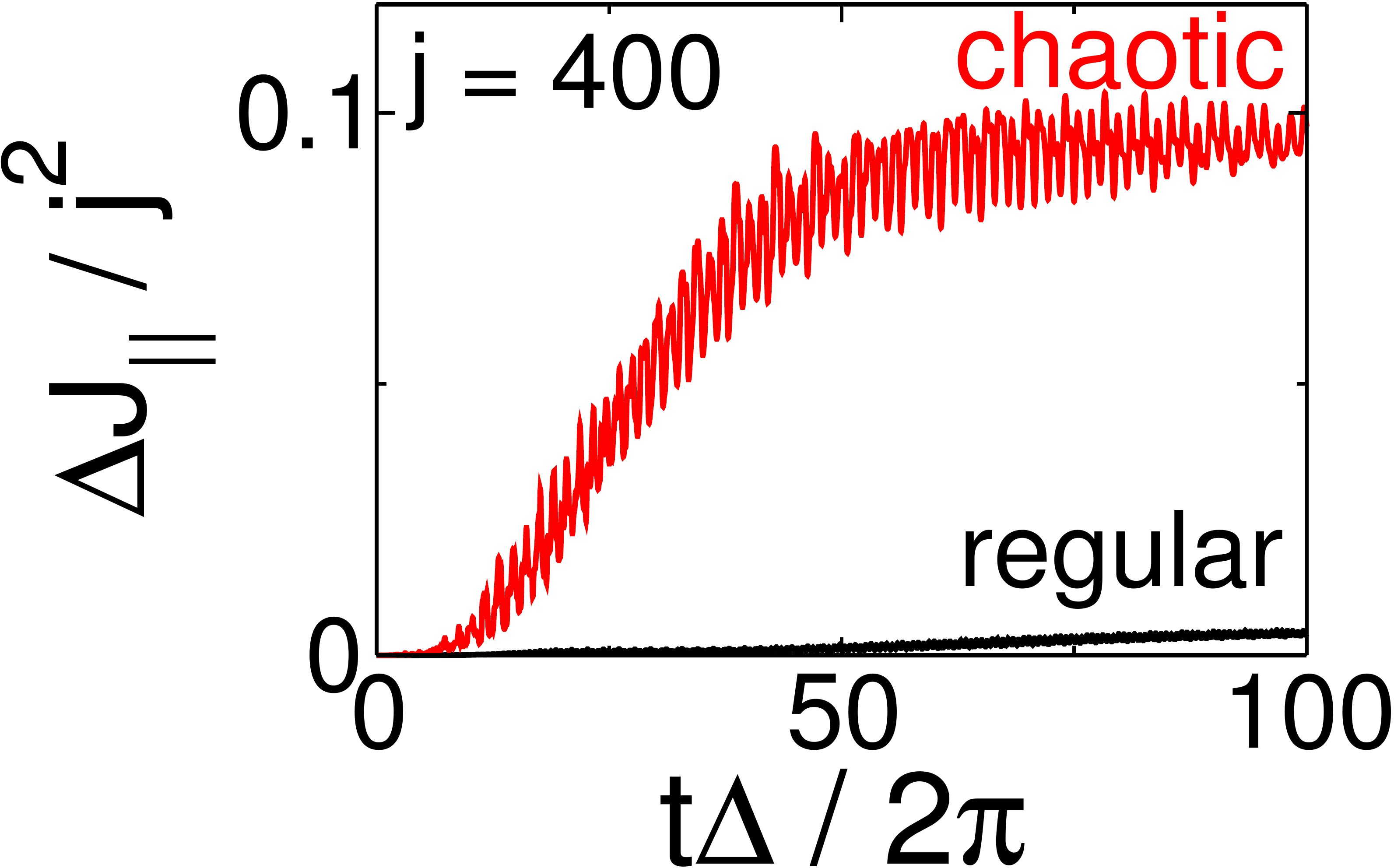}
\caption{Spin variance $\Delta J_{||}$ 
as a function of time, for $j = 2$ (left) and $j = 400$ (right) with initial conditions corresponding to
the regular and chaotic orbit in Fig.~\ref{fig:TwoClassOrbs}.
}
\label{fig:SpinVarRegCha}
\end{figure}

In Fig.~\ref{fig:SpinVarRegCha} we show the spin variance for the quantum dynamics corresponding to the two classical orbits in Fig.~\ref{fig:TwoClassOrbs}. 
For small spin length $j=2$ the spin variance is identical for both orbits, reaching its maximum at about the same time.
Going to large spin length $j=400$ we observe the different scaling of the spin variance and thus the Ehrenfest time.
For the regular orbit the spin variance [at $t=100 (\Delta/2\pi)$] is reduced by a factor $0.02$ and stays small during the plotted time interval.
For the chaotic orbit the spin variance again grows quickly, and is only slightly smaller (by $0.4$) than for $j=2$.
This is a clear sign of the different rates of spreading due to classical drift for chaotic and quantum diffusion for regular orbits.

Note that the quantum diffusion term in the Fokker-Planck equation respects the reversibility of the quantum dynamics~\cite{AH12,AH12a}. Technically, this follows from the invariance under time reversal $t \mapsto -t$ combined with conjugation $z \mapsto z^*$, $\alpha \mapsto \alpha^*$ of the spin and oscillator phase space coordinates, i.e. with $Q(z,\alpha,t)$ also the time-reversed Husimi function $Q(z^*,\alpha^*,-t)$ is a solution of the Fokker-Planck equation.
In spite of this the time evolution shown in Fig.~\ref{fig:QuantHus} and further below
is indicative of irreversible dynamics because it starts from a highly untypical state such as the coherent states used here.
In classical dynamics,
chaotic mixing of trajectories leads to rapid spreading of the initially localized yet somewhat extended phase space distribution at least on times scales smaller than the Poincare recurrence time. 
The perceived irreversibility thus is a consequence of averaging over diverging trajectories starting from nearby phase space points.
For quantum chaotic systems with few degrees of freedom this kind of irreversibility is linked to the complex energy spectrum~\cite{CU10}, as revealed in random matrix theory~\cite{Haak10}.
True irreversibility, involving the approach to a stationary equilibrium state, requires coupling to an infinite number of degrees of freedom provided, e.g., by a bath or the environment~\cite{Wei99,PAF13}.

\subsection{Classical and quantum periodic orbits}
\label{sec:ClassQuantOrbs}

\begin{figure}
\includegraphics[width=0.48\linewidth]{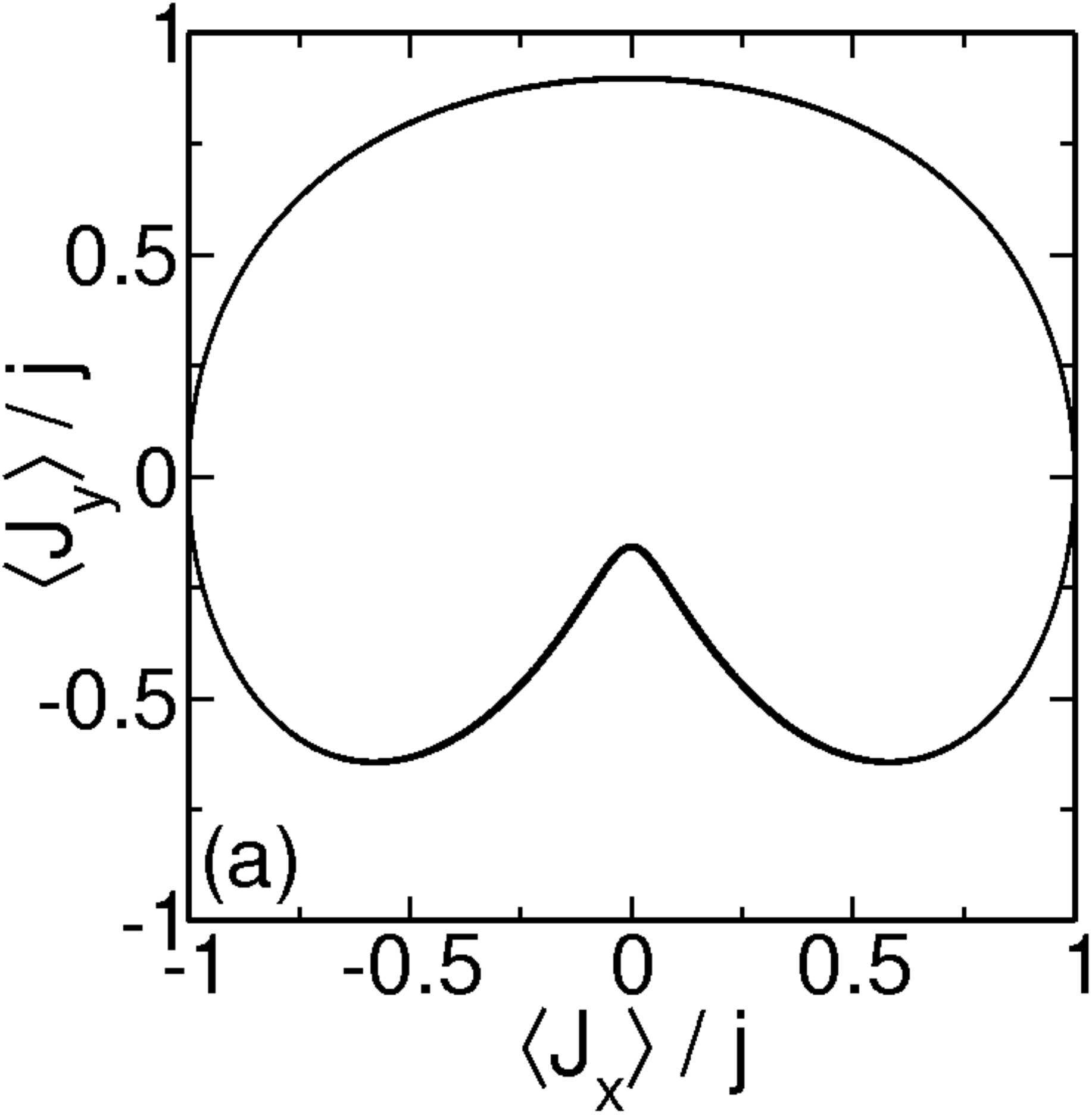}
\hfill
\includegraphics[width=0.48\linewidth]{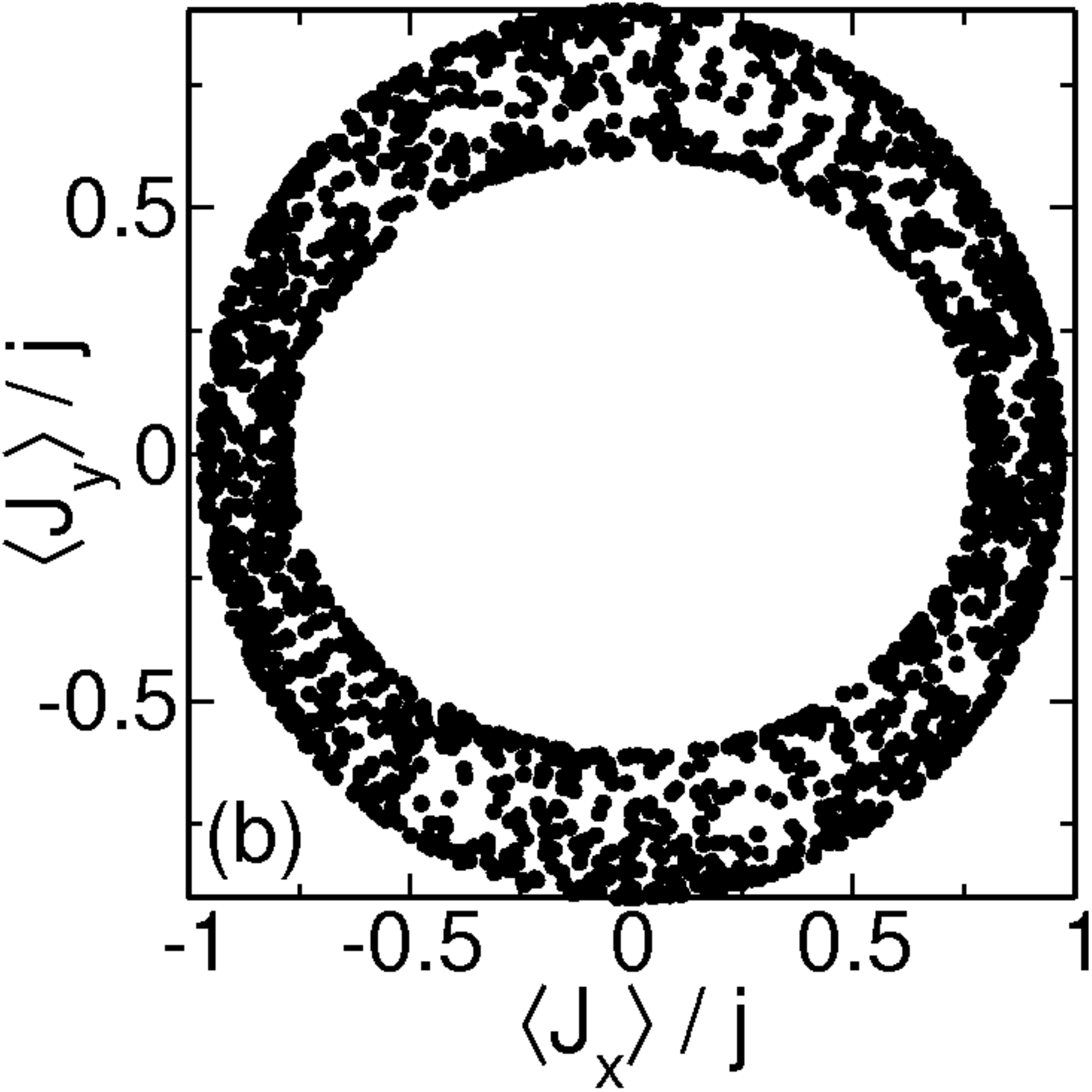}
\\
\includegraphics[width=0.48\linewidth]{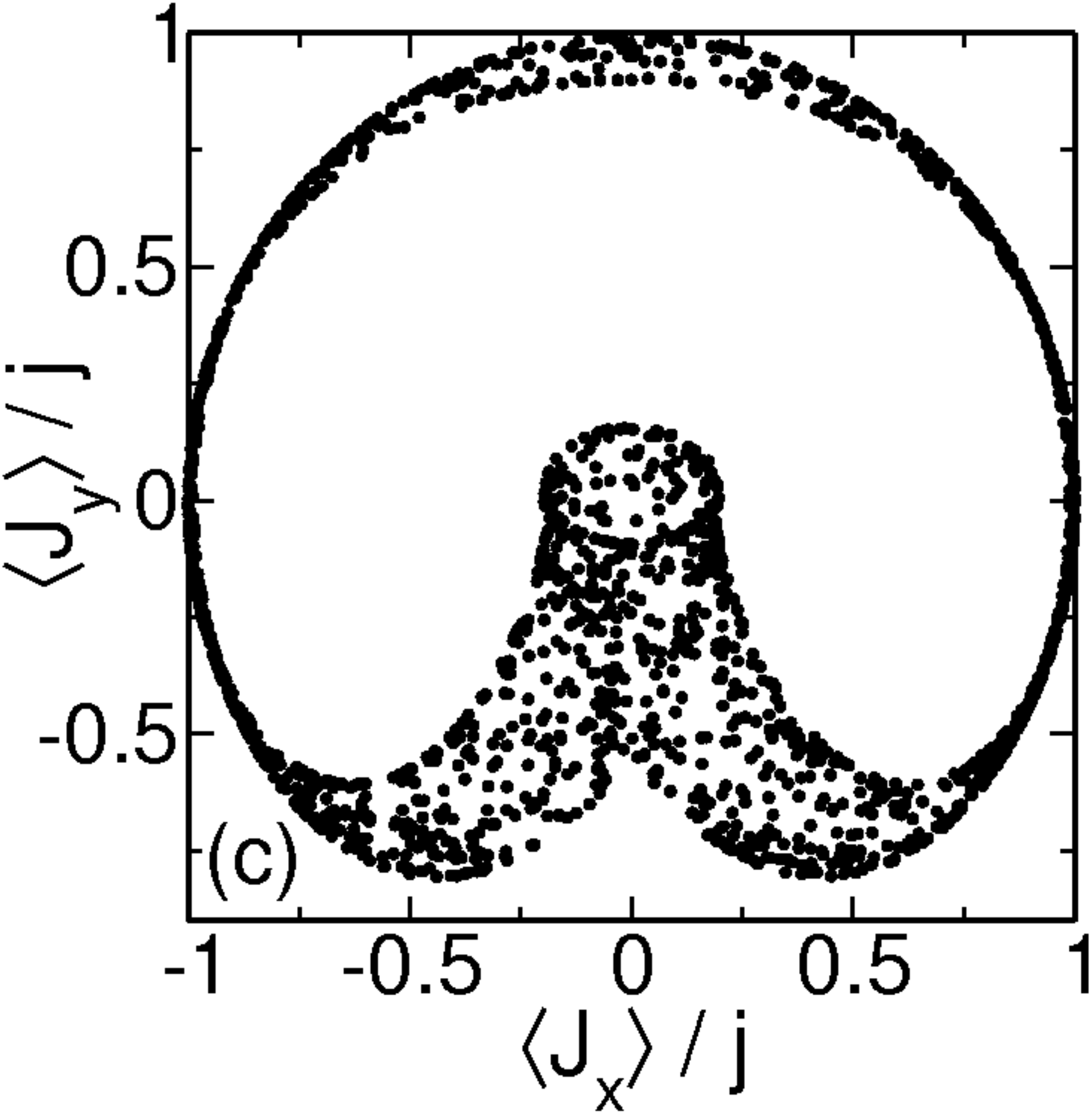}
\hfill
\includegraphics[width=0.48\linewidth]{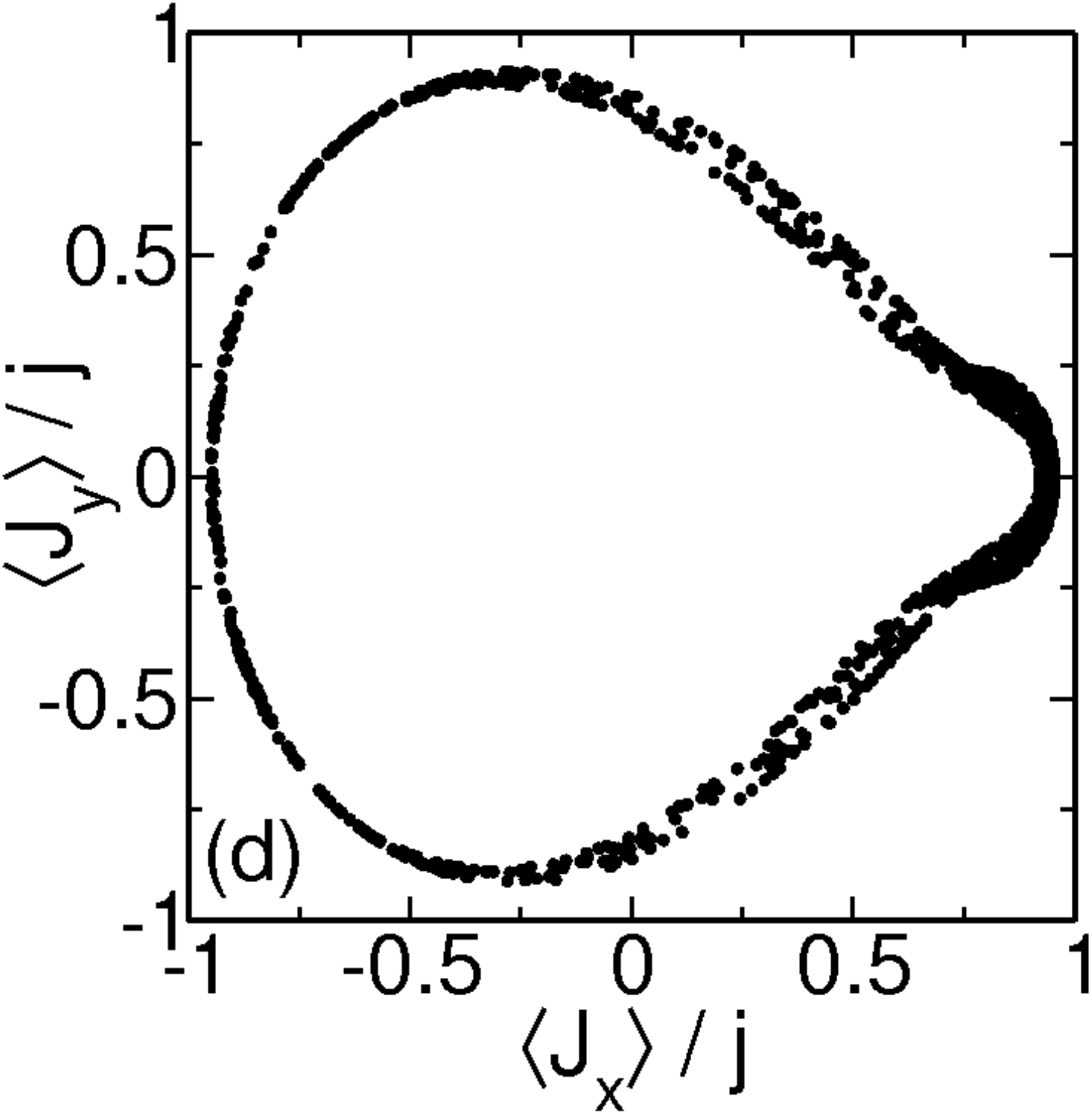}
\caption{Four quasi-periodic classical orbits in $J_x$--$J_y$ phase space,
for $E=-0.5$ and $\kappa = 0.6$, $J_x(0)=0$, $J_y(0)=0.9$ (a),
$\kappa=0.1$, $J_x(0)=0.0$, $J_y(0)=0.9$ (b),
 $\kappa=0.65$, $J_x(0)=0.0$, $J_y(0)=0.9$ (c),
 $\kappa=0.85$, $J_x(0)=0.5$, $J_y(0)=0.5$ (d).
}
\label{fig:ClassOrbits}
\end{figure}

Because the quantum state traces out the classical orbit,
periodic orbits that cover only a low dimensional part of the phase space lead to distinct features in the quantum dynamics.
For further illustration of the relation between classical and quantum dynamics 
we will, therefore, use the four (quasi-)periodic orbits shown in Fig.~\ref{fig:ClassOrbits}.
The quantum signatures of these orbits are identified again with the spin Husimi function.

\begin{figure}
\includegraphics[width=0.99\linewidth]{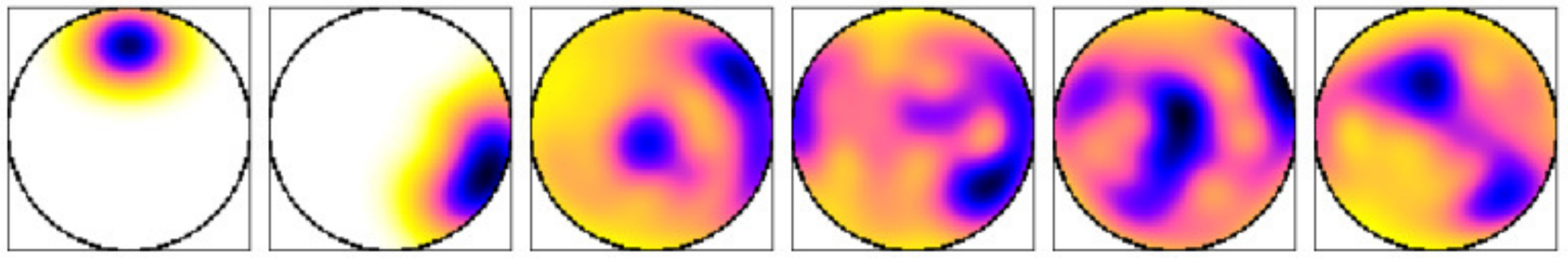}
\includegraphics[width=0.99\linewidth]{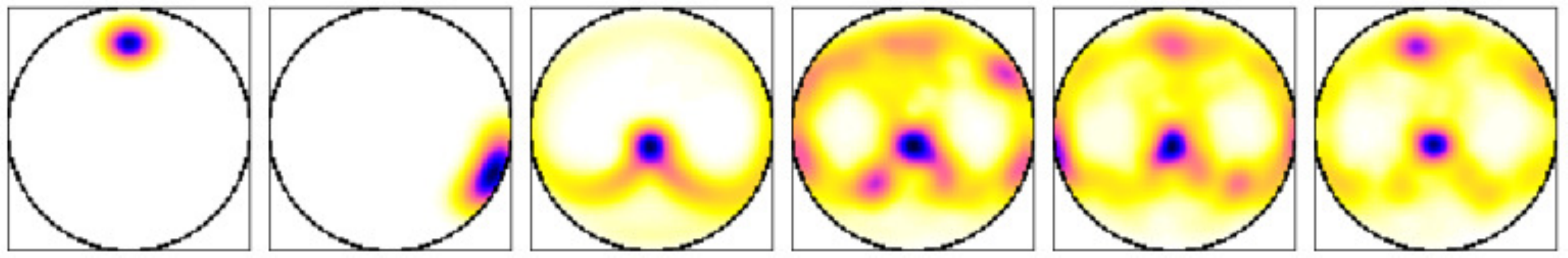}
\includegraphics[width=0.99\linewidth]{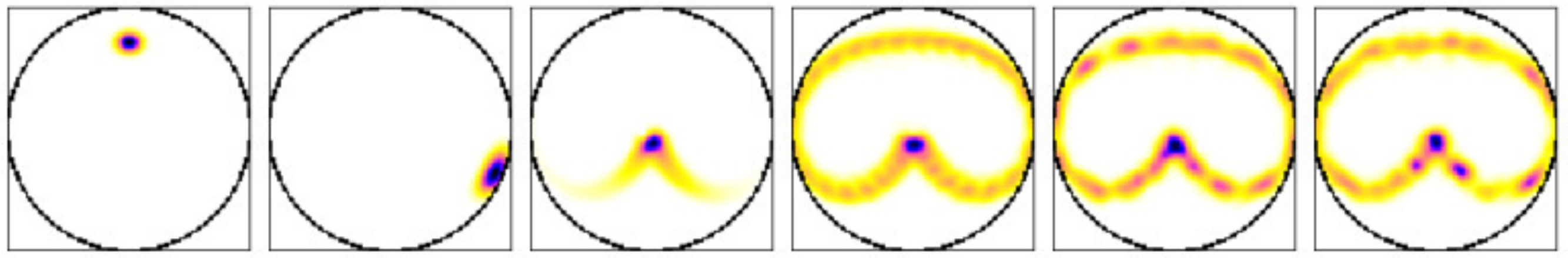}
\includegraphics[width=0.99\linewidth]{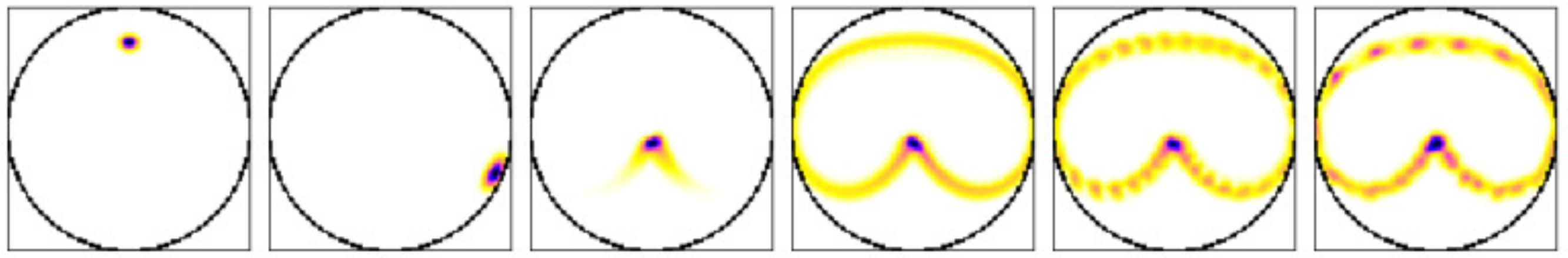}
\caption{(Color online) Spin Husimi function at time steps  $t\Delta/2\pi = 0, 1, 10, 50, 100, 200$ (from left to right) and spin length
(from top to bottom) $j=10, 50, 200, 400$ (from top to bottom).
The initial states correspond to the classical orbit from panel (a) in Fig.~\ref{fig:ClassOrbits}.
}
\label{fig:SpinHusimi1}
\end{figure}

The spin Husimi function for the orbit from panel (a) is shown in Fig.~\ref{fig:SpinHusimi1}, for increasing spin length $j$ and time $t$.
We clearly see the behavior described above, how the spin Husimi function traces out the classical trajectory for larger $j$.
We also observe how the quantum state quickly loses the shape of the initial coherent state after the first few periods (e.g. for $t\Delta/(2\pi)=10$ and $j=400$), while it still follows the classical orbit.

Remainders of the quantum mechanical dynamics are seen for large $t$ (rightmost panels),
where the spin Husimi function fragments into several ``blobs'' located on the classical trajectory~\cite{ABF12}. 
This is a precursor of the revival of the initial state at much larger times,
which occurs because for finite $j$ the quantum dynamics explores only a finite dimensional Hilbert space (the infinite dimensional bosonic part is restricted by energy conservation).

\begin{figure}
\includegraphics[width=0.32\linewidth]{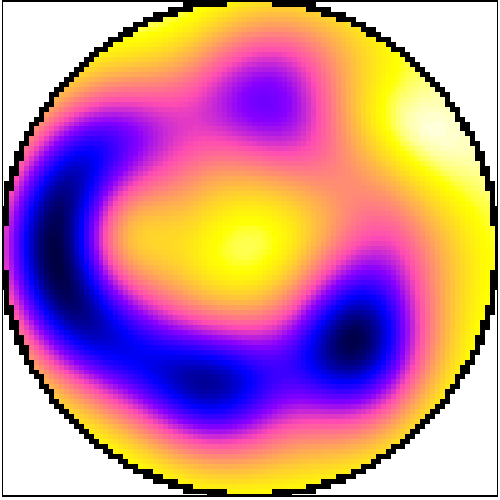}
\hfill
\includegraphics[width=0.32\linewidth]{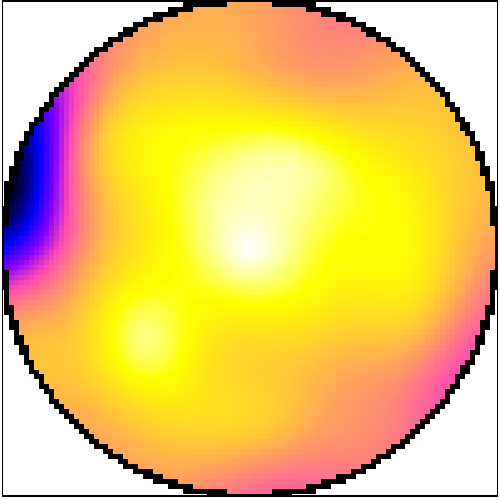}
\hfill
\includegraphics[width=0.32\linewidth]{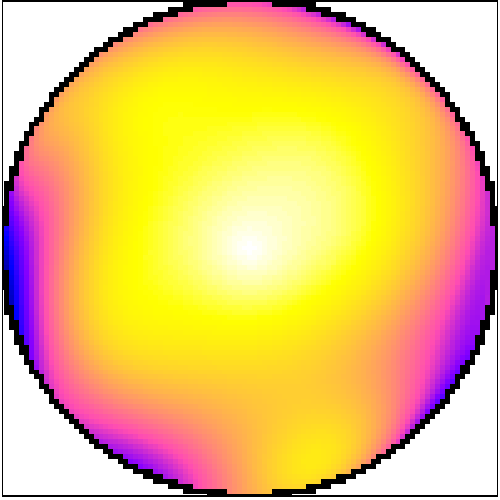}
\\
\includegraphics[width=0.32\linewidth]{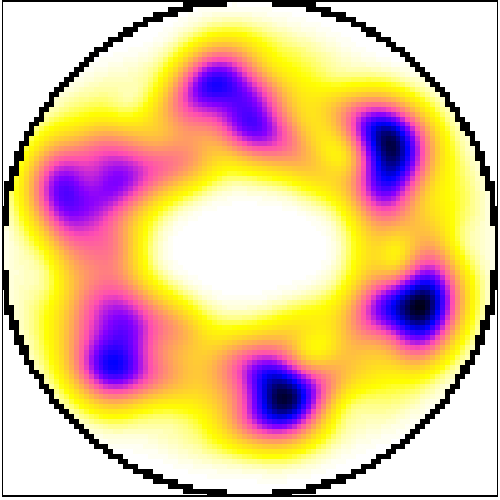}
\hfill
\includegraphics[width=0.32\linewidth]{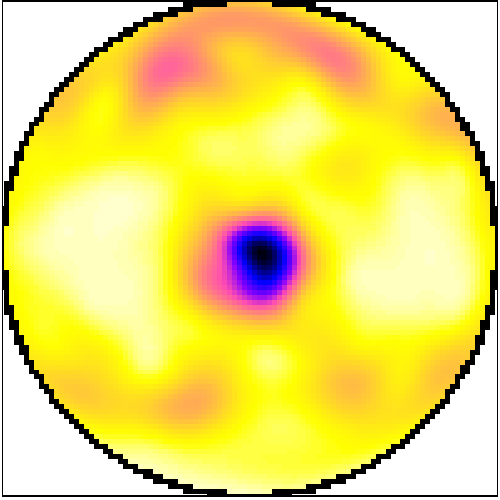}
\hfill
\includegraphics[width=0.32\linewidth]{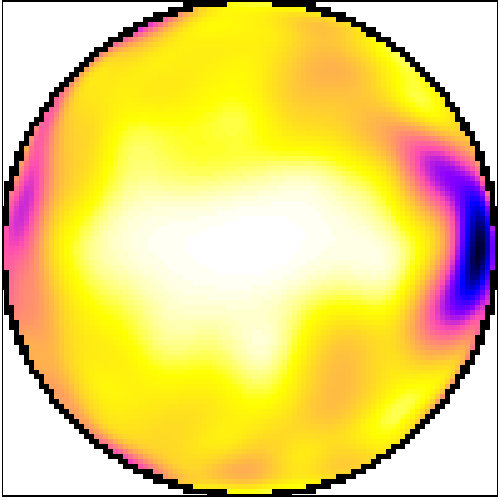}
\\
\includegraphics[width=0.32\linewidth]{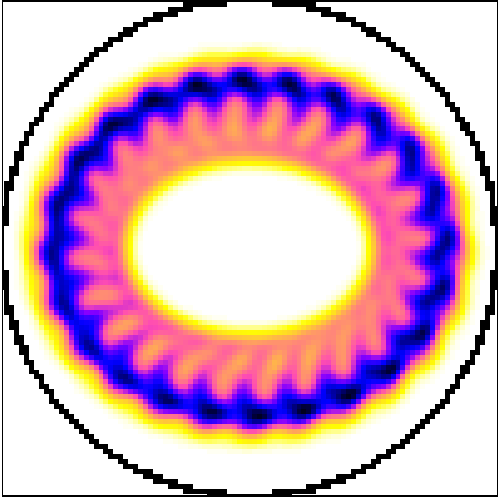}
\hfill
\includegraphics[width=0.32\linewidth]{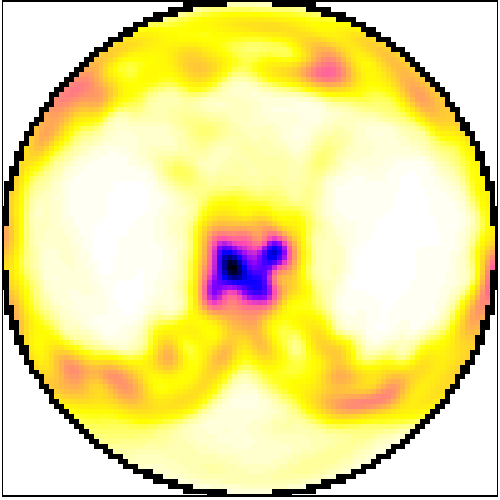}
\hfill
\includegraphics[width=0.32\linewidth]{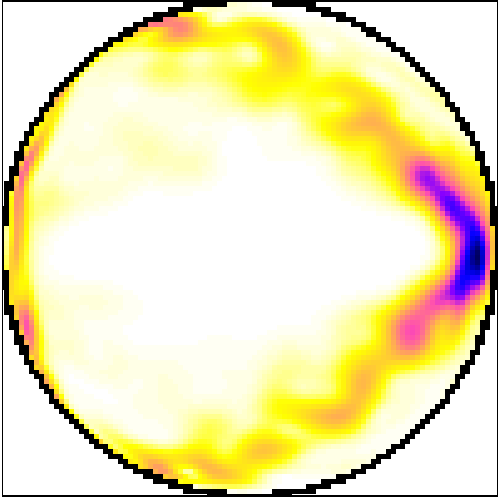}
\\
\includegraphics[width=0.32\linewidth]{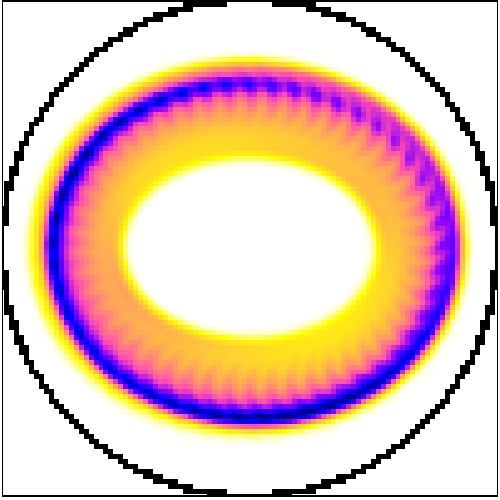}
\hfill
\includegraphics[width=0.32\linewidth]{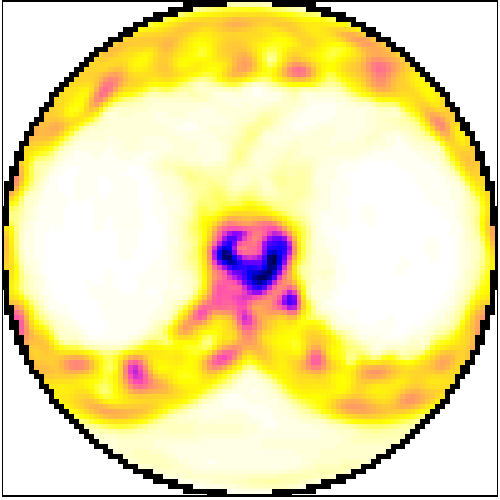}
\hfill
\includegraphics[width=0.32\linewidth]{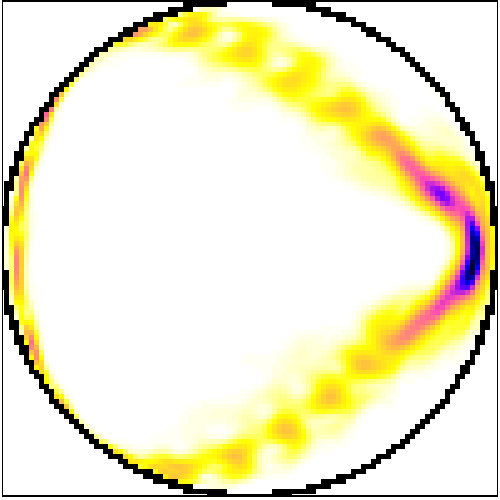}
\caption{(Color online) Spin Husimi functions for the three 
classical orbits from Fig.~\ref{fig:ClassOrbits} (b), (c) and (d),
all at time $t=200 \times 2\pi/\Delta$.
The left, central, right column  corresponds to the respective orbit in panel (b), (c), (d) in Fig.~\ref{fig:ClassOrbits}.
From top to bottom, the spin length grows as $j=10, 50, 200, 400$.
}
\label{fig:SpinHusimi2}
\end{figure}

The scenario of convergence in phase space generally holds for (quasi-) periodic orbits, as the spin Husimi functions in Fig.~\ref{fig:SpinHusimi2} for the remaining three orbits from Fig.~\ref{fig:ClassOrbits} (b), (c) and (d) show.
The required waiting time after which the entire classical orbit can be identified in a ``snapshot'' of the quantum dynamics at time $t$ can become nevertheless large, depending on the rapidity of phase space diffusion.
Therefore, the plots in Fig.~\ref{fig:SpinHusimi2} already show fragmentation of the Husimi function, indicating the later revival of the initial state.

\begin{figure}
\includegraphics[width=0.99\linewidth]{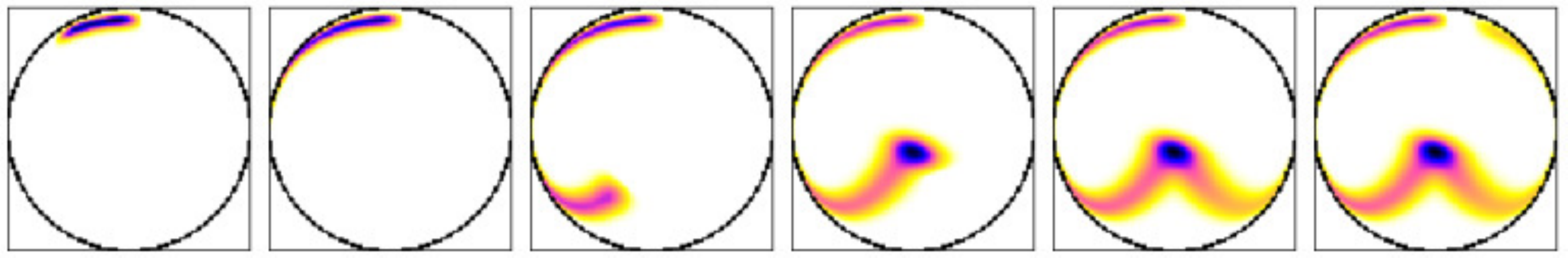} \\
\includegraphics[width=0.99\linewidth]{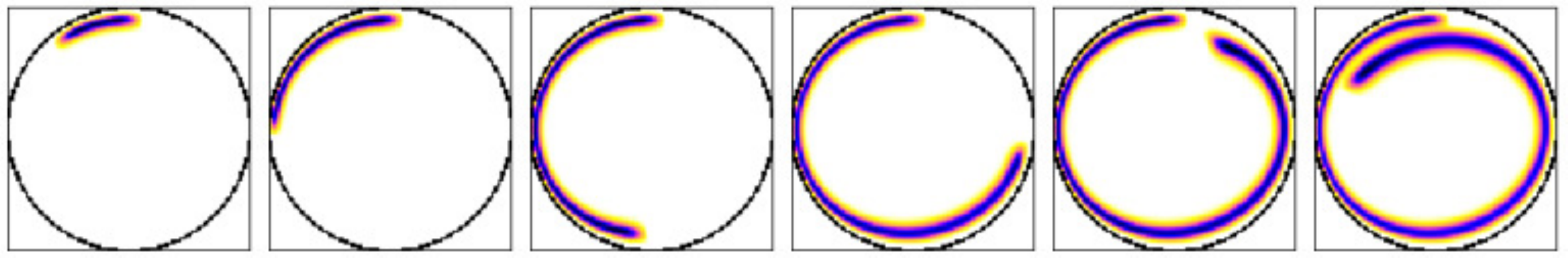} \\
\includegraphics[width=0.99\linewidth]{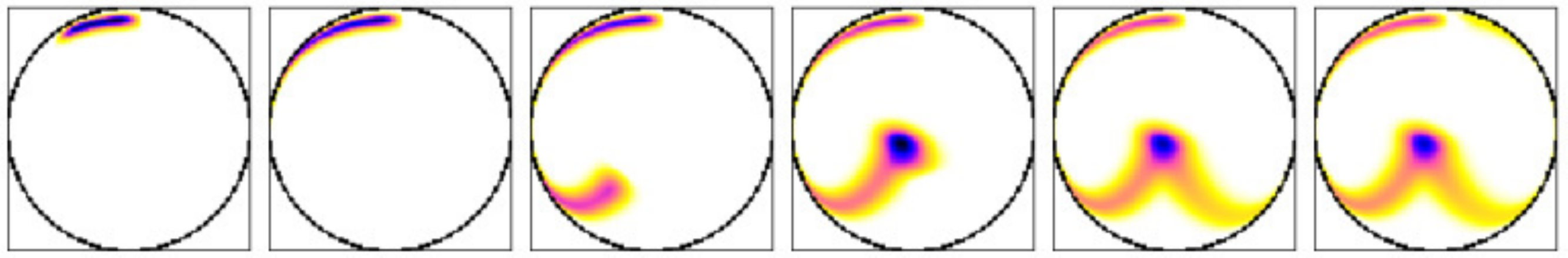} \\
\includegraphics[width=0.99\linewidth]{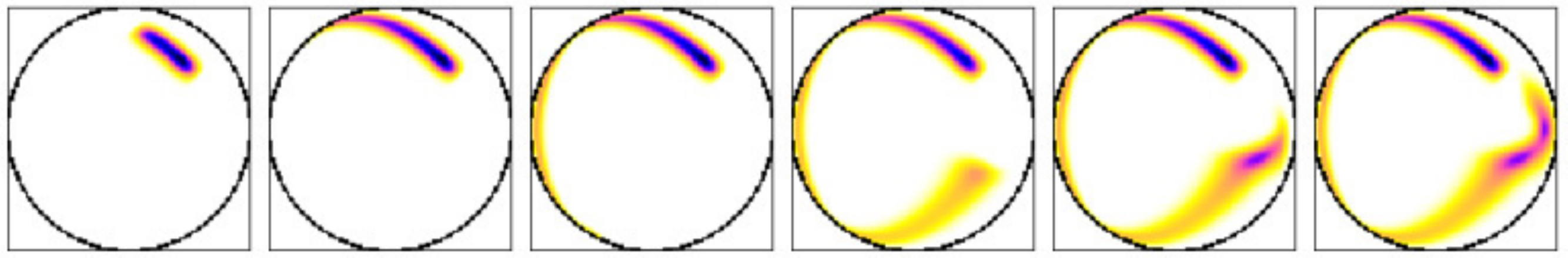}
\caption{(Color online) Time averaged spin Husimi function as defined in Eq.~\eqref{eq:avspinhusimi} for $j=100$ and $T\Delta / (2\pi) = 0.1, 0.25, 0.5, 0.75, 1.0, 1.25$ from left to right.
The rows, from top to bottom, correspond to the classical orbits in panel (a)---(d) in Fig.~\ref{fig:ClassOrbits}.
}
\label{fig:ta_SpinHusimi}
\end{figure}

To identify the classical orbit from the quantum time-evolution already at earlier times we can use the time averaged Husimi function
\begin{equation}
\label{eq:avspinhusimi}
\bar{Q}(\theta,\phi) = \dfrac{1}{2T} \int_{-T}^T \dd t\, Q(\theta,\phi;t) \;,
\end{equation}
where $T$ is of the order of a few periods.
$\bar{Q}(\theta,\phi)$ can be directly computed from the Chebyshev time propagation (see App.~\ref{app:Husimi}), which is a more elegant numerical approach than sampling and averaging of $Q(\theta,\phi;t)$ at many values of $t$.
The time averaged Husimi function as shown in Fig.~\ref{fig:ta_SpinHusimi} now gives a clear picture of the classical trajectory as it is (re-)constructed from the quantum trajectory in $J_x$--$J_y$ phase space.

\subsection{Quantum states close to periodic orbits}
\label{sec:Eigenstates}

Because the time-evolution of a quantum state is directly related to the eigenstates of the Hamiltonian,
the same signatures that appear in the time-dependent Husimi function should show up in the individual eigenstates.
Therefore, we finally try to 
relate the different classical orbits with energy $E$ to the eigenstates
$
H|E_n\rangle = E_n|E_n\rangle
$
with nearby energies $E_n \approx E$.
We use two different spin Husimi functions to characterize the eigenstates,
which give correspondence either to the classical orbits or the Poincare plots.

\begin{figure}
\includegraphics[width=0.23\linewidth]{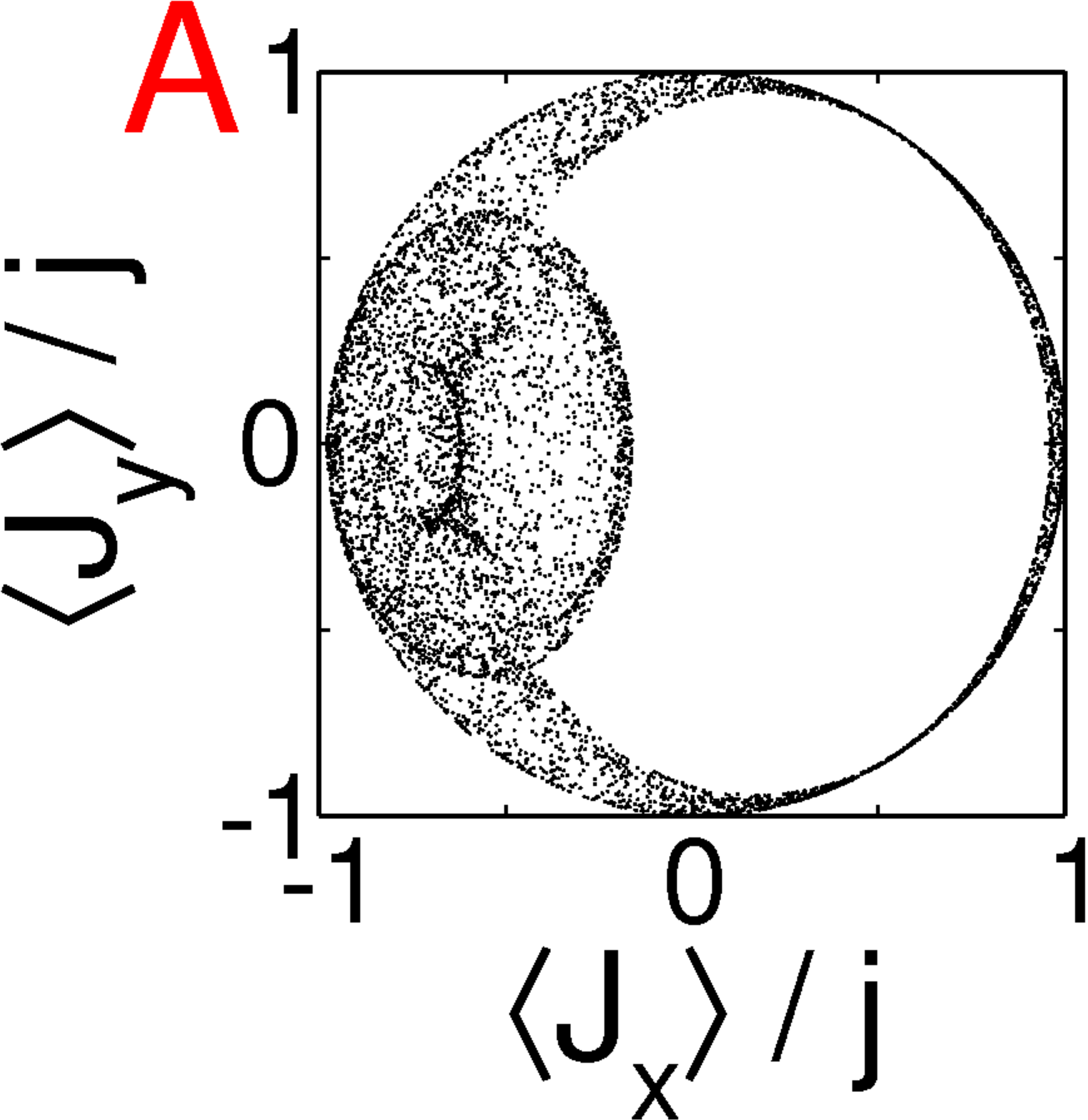}
\raisebox{0.05\linewidth}{\includegraphics[width=0.18\linewidth]{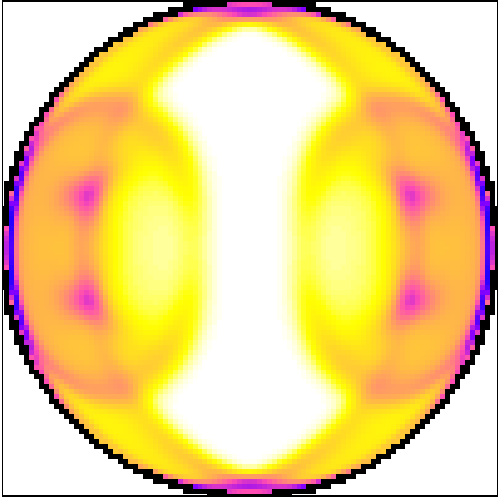}}
\raisebox{0.05\linewidth}{\includegraphics[width=0.18\linewidth]{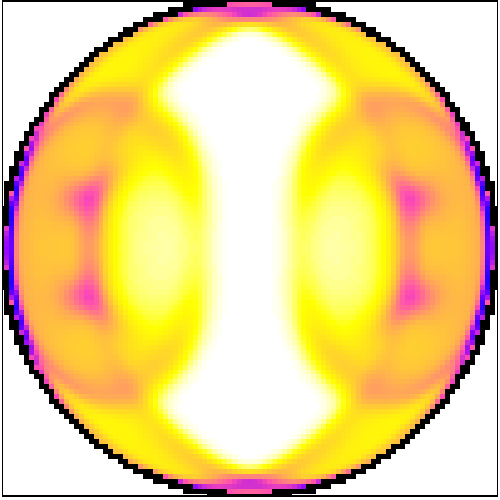}}
\raisebox{0.05\linewidth}{\includegraphics[width=0.18\linewidth]{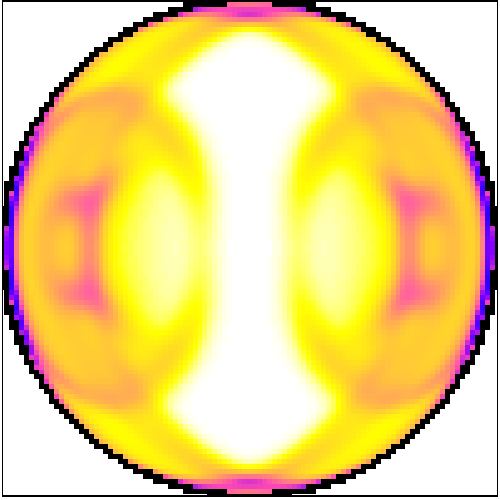}}
\raisebox{0.05\linewidth}{\includegraphics[width=0.18\linewidth]{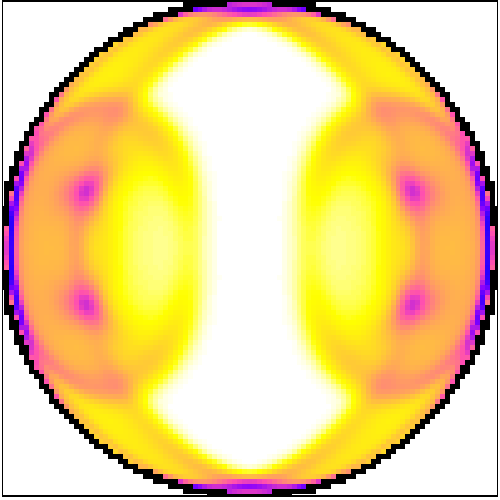}}\\
\includegraphics[width=0.23\linewidth]{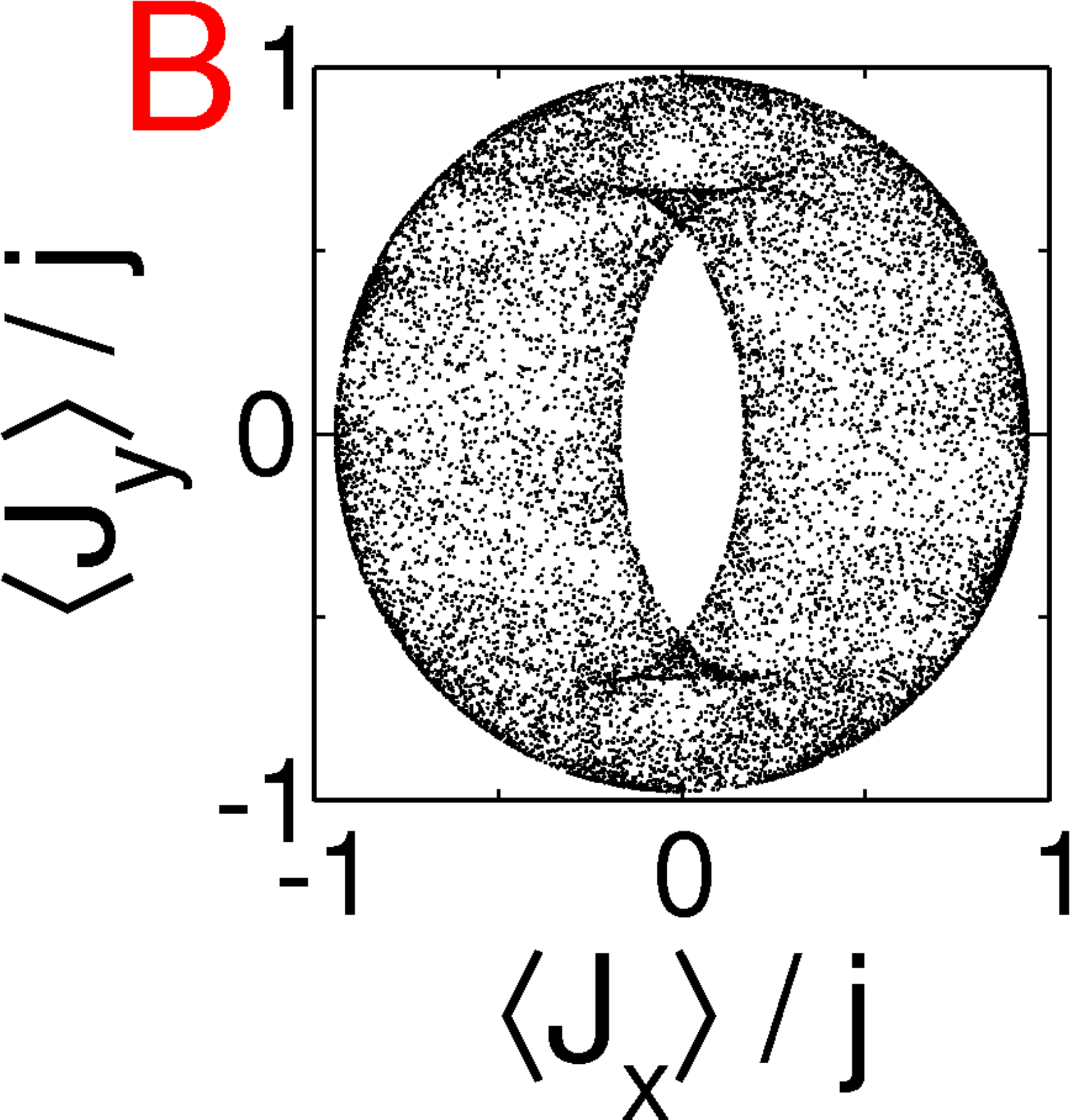}
\raisebox{0.05\linewidth}{\includegraphics[width=0.18\linewidth]{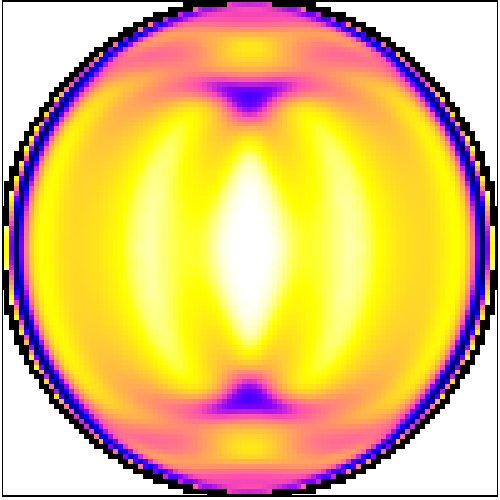}}
\raisebox{0.05\linewidth}{\includegraphics[width=0.18\linewidth]{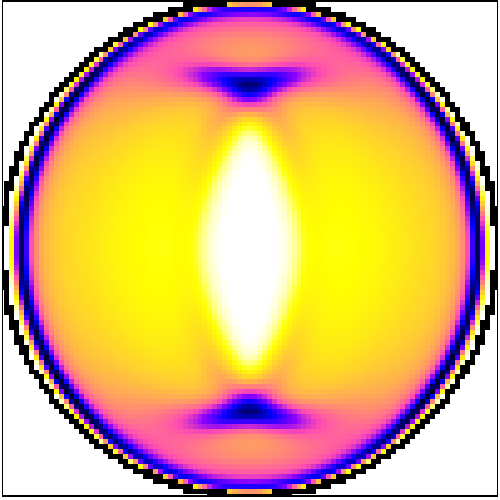}}
\raisebox{0.05\linewidth}{\includegraphics[width=0.18\linewidth]{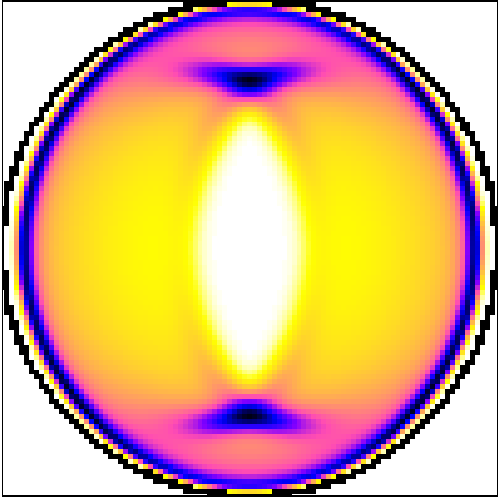}}
\raisebox{0.05\linewidth}{\includegraphics[width=0.18\linewidth]{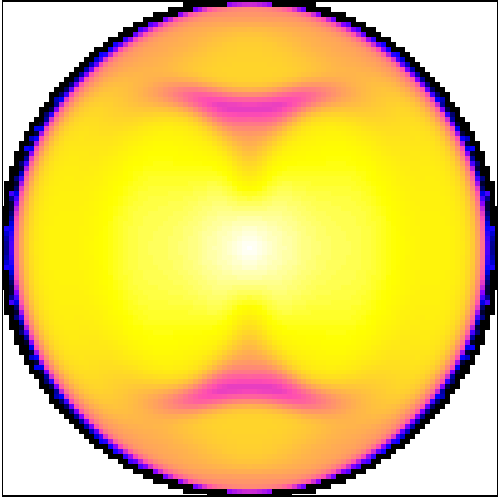}}\\
\includegraphics[width=0.23\linewidth]{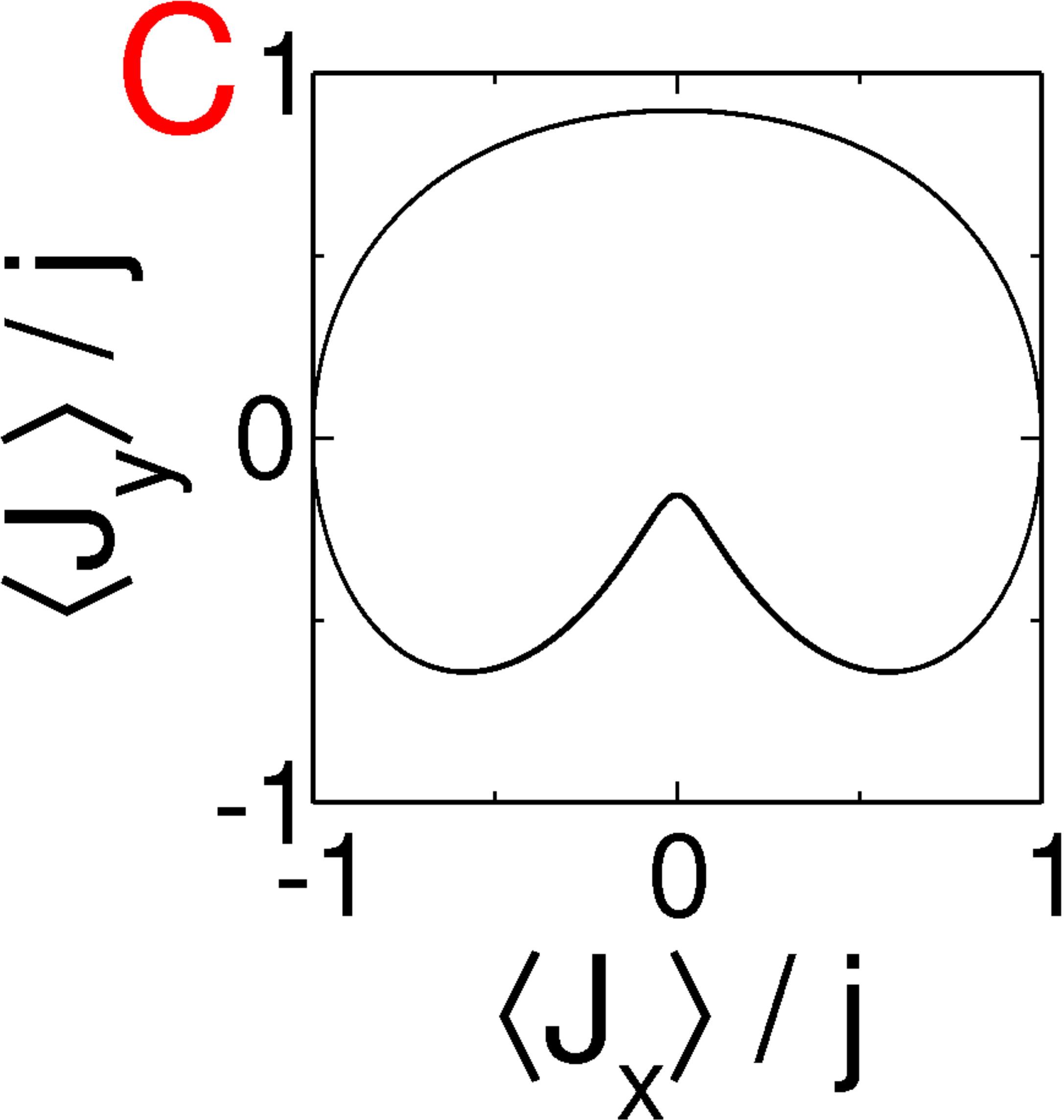}
\raisebox{0.05\linewidth}{\includegraphics[width=0.18\linewidth]{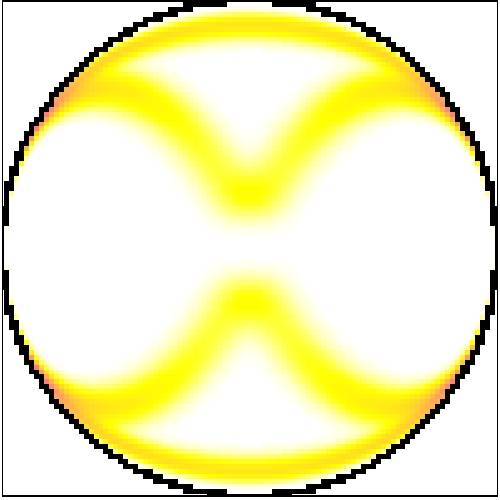}}
\raisebox{0.05\linewidth}{\includegraphics[width=0.18\linewidth]{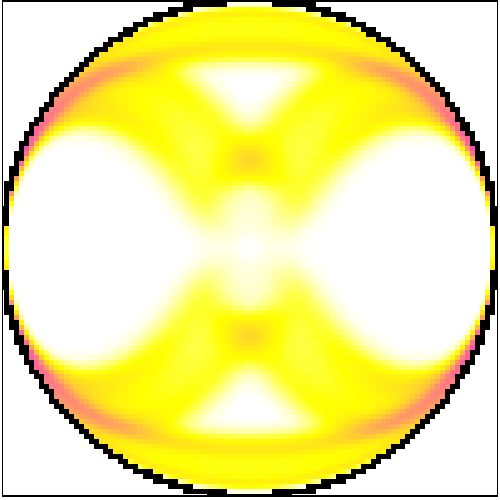}}
\raisebox{0.05\linewidth}{\includegraphics[width=0.18\linewidth]{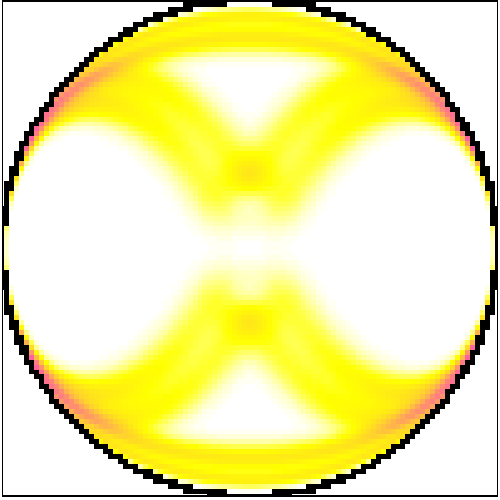}}
\raisebox{0.05\linewidth}{\includegraphics[width=0.18\linewidth]{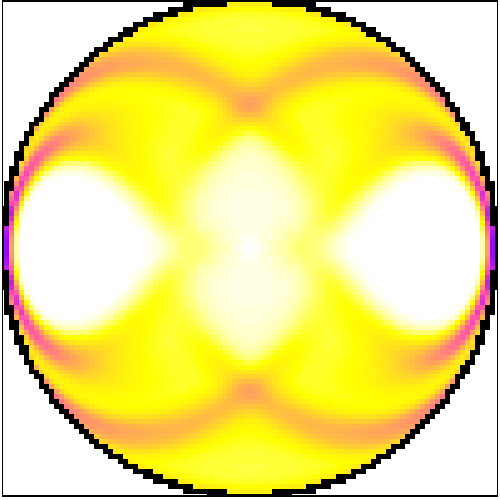}}\\
\includegraphics[width=0.23\linewidth]{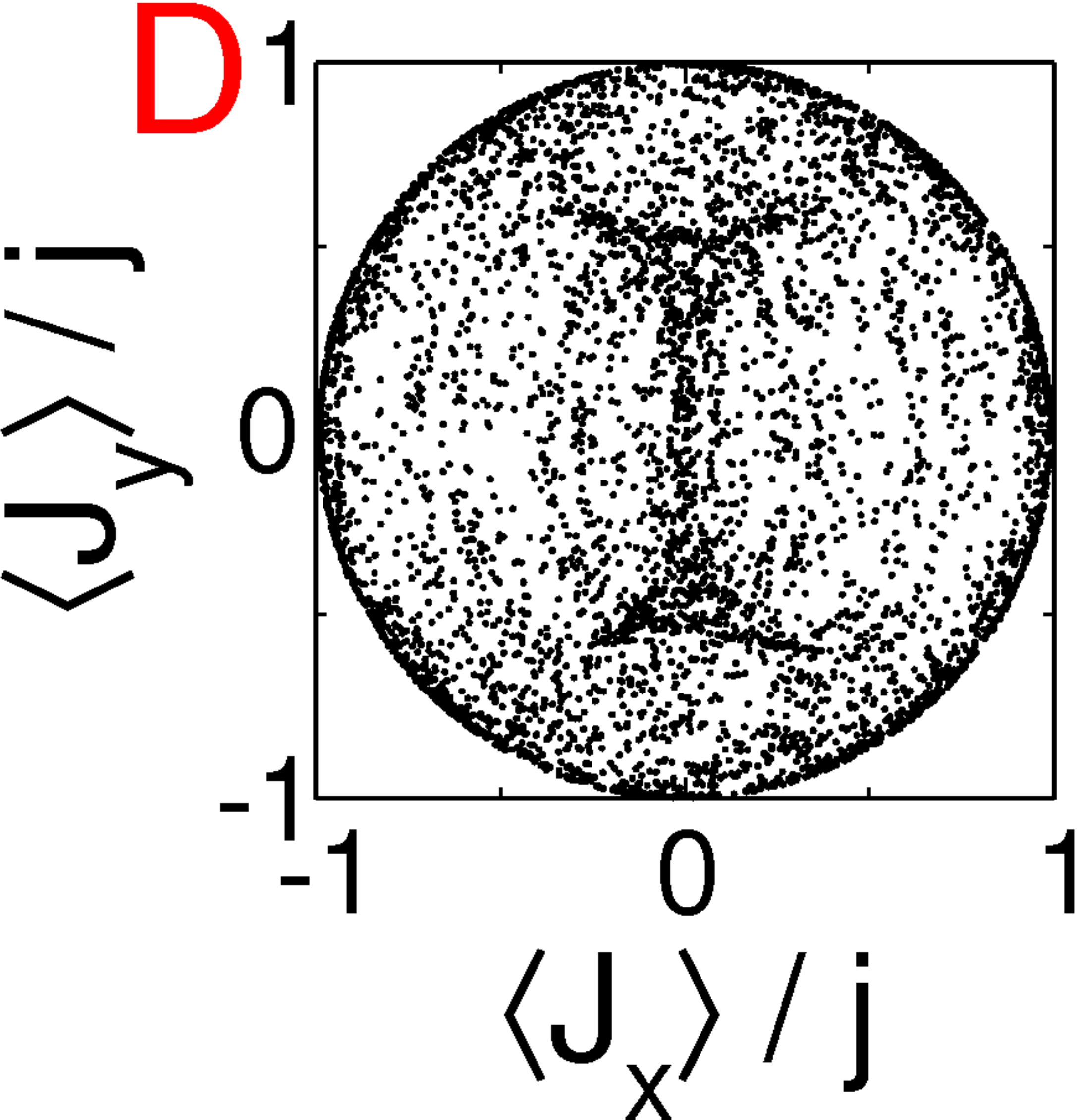}
\raisebox{0.05\linewidth}{\includegraphics[width=0.18\linewidth]{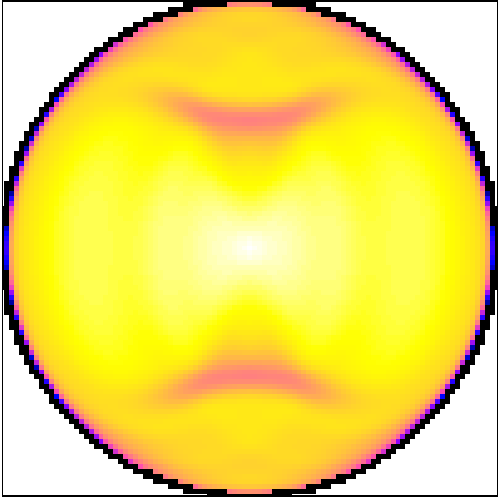}}
\raisebox{0.05\linewidth}{\includegraphics[width=0.18\linewidth]{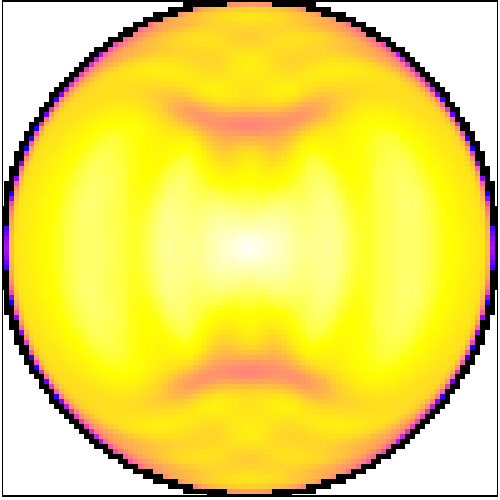}}
\raisebox{0.05\linewidth}{\includegraphics[width=0.18\linewidth]{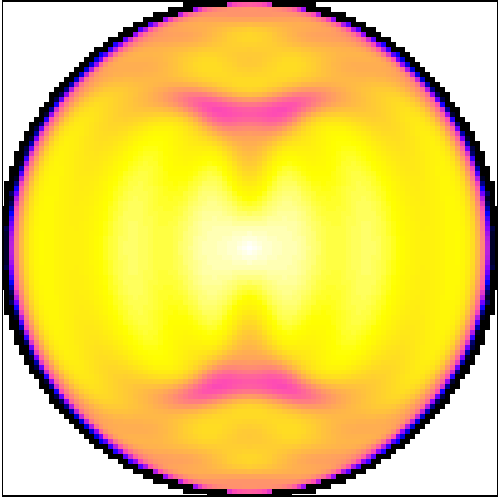}}
\raisebox{0.05\linewidth}{\includegraphics[width=0.18\linewidth]{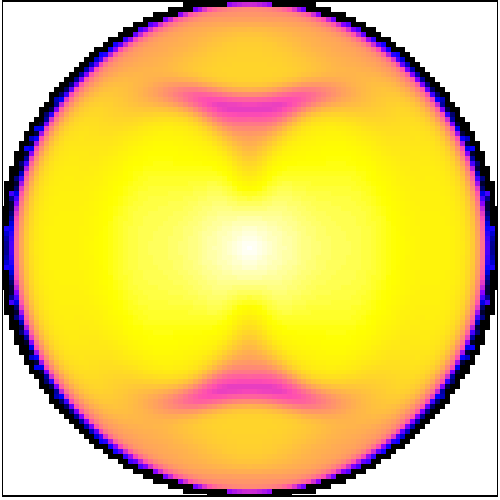}}\\
\includegraphics[width=0.23\linewidth]{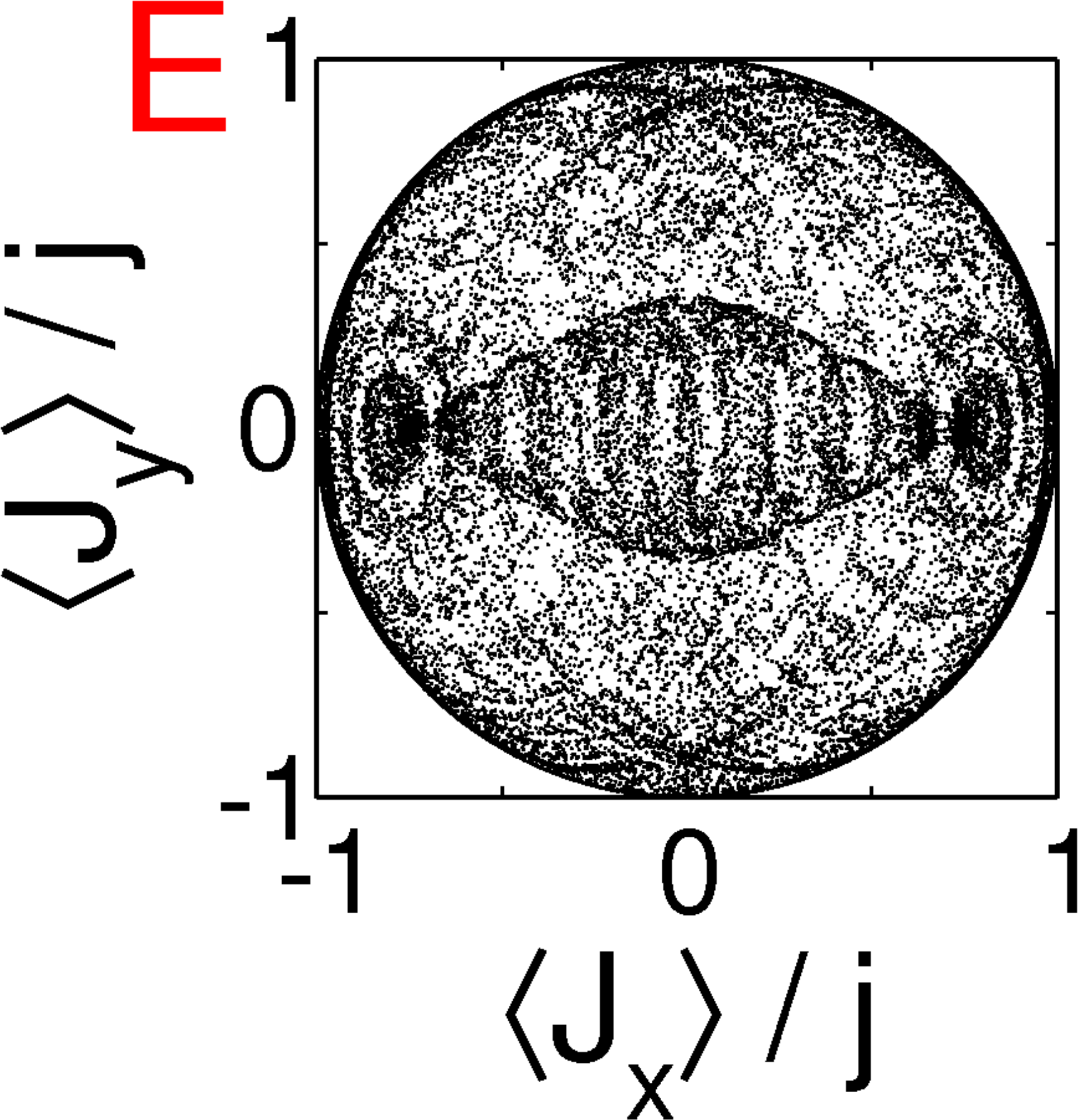}
\raisebox{0.05\linewidth}{\includegraphics[width=0.18\linewidth]{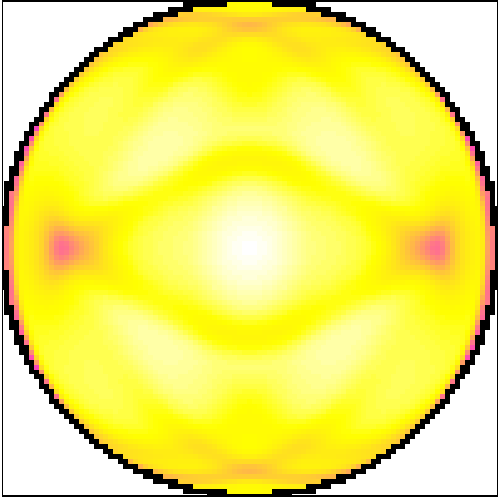}}
\raisebox{0.05\linewidth}{\includegraphics[width=0.18\linewidth]{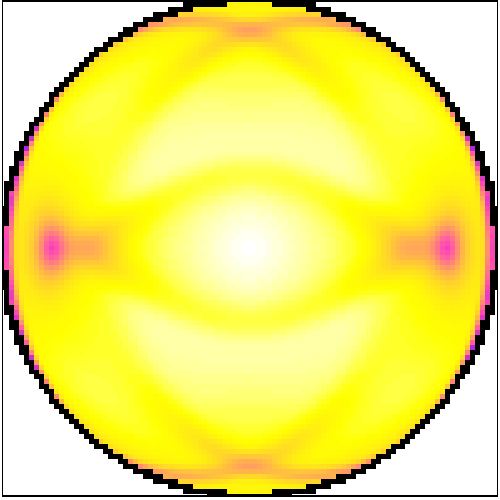}}
\raisebox{0.05\linewidth}{\includegraphics[width=0.18\linewidth]{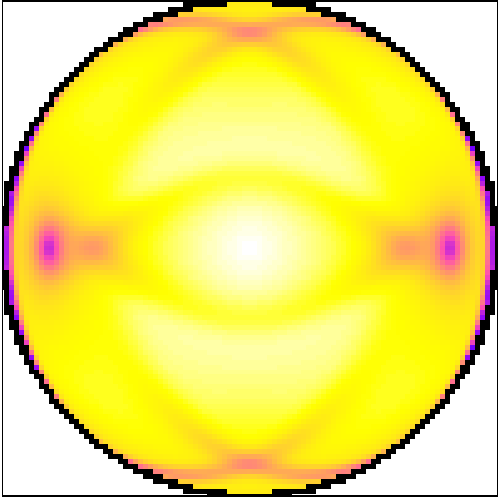}}
\raisebox{0.05\linewidth}{\includegraphics[width=0.18\linewidth]{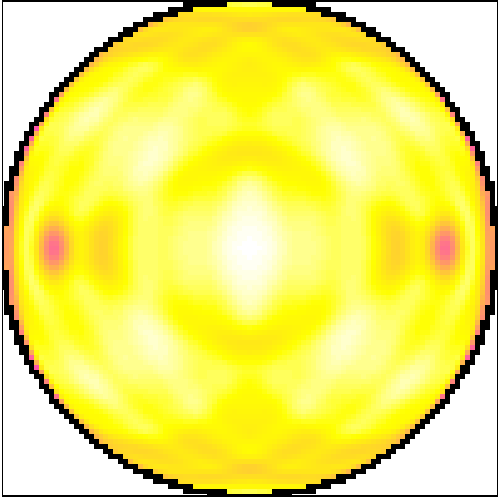}}\\
\caption{(Color online) Five classical orbits at $\kappa = 0.6$ and $E = -0.5$ and eigenstates in the energy range $-0.703 < E_n < -0.249$.
Shown are the respective four eigenstates (to $j=200$) with maximal overlap with initial coherent state
corresponding to the initial conditions of the respective classical orbit.
Orbits (A)--(C) are regular, orbits (D)--(E) are chaotic with $\Lambda_1^{(D)} = 0.014$ and $\Lambda_1^{(E)} = 0.013$.
}
\label{fig:EigenHus_Orbits}
\end{figure}

The spin Husimi function
\begin{equation}\label{SpinHusEig}
Q_n^{\text{Spin}}(\theta, \phi) = |\langle\theta,\phi| E_n \rangle |^2 
\end{equation}
for the eigenstates $|E_n\rangle$ directly corresponds to the time-dependent spin Husimi function from Eq.~\eqref{eq:spinhusimi}.
In Fig.~\ref{fig:EigenHus_Orbits} we show the Husimi functions for different eigenstates in the energy range $-0.703 < E_n < -0.249$.
The eigenstates are arranged according to their overlap with classical orbits to energy $E = -0.5$.
Both regular (A)--(C) and chaotic (D)--(E) orbits appear in the figure because of the classical ``mixed'' dynamics (recall the Poincare plot in Fig.~\ref{fig:ClassPoinc}).
To every orbit, we show the four eigenstates $|E_n\rangle$ with maximal overlap 
$|\langle E_n | \psi(0) \rangle|$ with the initial state $|\psi(0)\rangle$ from Eq.~\eqref{eq:cohstate} that corresponds to the initial conditions of the classical orbit.
The comparison clearly reveals the correspondence between regular classical orbits and the fine structure of the phase space distribution visible in some of the quantum eigenstates.
These eigenstates occupy only part of the admissible phase space.
Classical chaotic orbits, on the other hand, correspond to eigenstates that are spread out over the entire phase space.

The rightmost Husimi functions in the second and fourth row belong to the same eigenstate, which has significant overlap with the two different initial coherent states $|\psi(0)\rangle$ that correspond to the regular (B) or chaotic (D) orbit.
Accordingly, the phase space density of this state shows signatures common to classical orbits of different type.
This effect resembles the ``scars'' of ergodic eigenstates in chaotic systems that arise from (unstable) periodic classical orbits~\cite{Hell84,KH98,WYMGR88,BJ90}.
Note, however, that in the present example with mixed regular and chaotic dynamics 
stable (quasi-) periodic orbits occupy a finite portion of the classical phase space.
Therefore, a finite fraction of the eigenstates shows signatures arising from periodic orbits even in the limit $j \to \infty$,
in contrast to the scars in completely chaotic systems~\cite{Hell84,KH98}.

For small spin length ($j=9/2$) early indications for the localization of the oscillator (but not spin) Husimi function on stable periodic orbits have been observed in Ref.~\cite{AFLN91,FALN92}.
The clear distinction between eigenstates and phase space signatures corresponding to regular or chaotic classical orbits requires the much larger values of $j$ used here.

The Poincare Husimi function is
defined as
\begin{equation}
Q^\text{Poinc}_n(\theta, \phi) = |\langle \bar{\alpha}; \theta, \phi|E_n\rangle|^2 \,,
\end{equation}
where $\bar{\alpha}$ has the value as in the corresponding classical Poincare surface of section, i.e. $Q = \Re \bar{\alpha} = 0$ and $P = \Im \bar{\alpha}$ is determined from the energy constraint $E = E(z,\bar{\alpha})$ (cf. the discussion of Fig.~\ref{fig:ClassPoinc}).
In constrast to the spin Husimi function from Eq.~\eqref{SpinHusEig},
no trace over the bosonic degree of freedom is involved.

\begin{figure}
\includegraphics[width=0.99\linewidth]{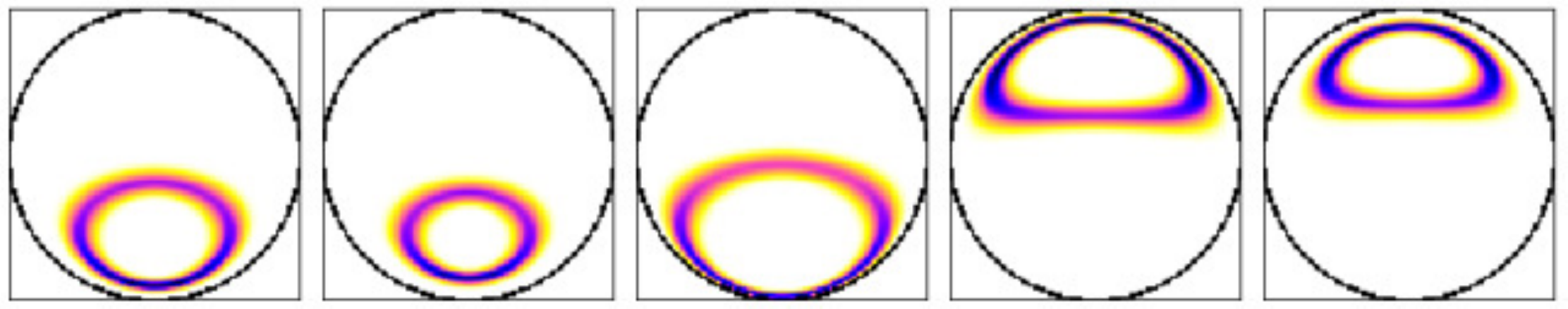}
\includegraphics[width=0.99\linewidth]{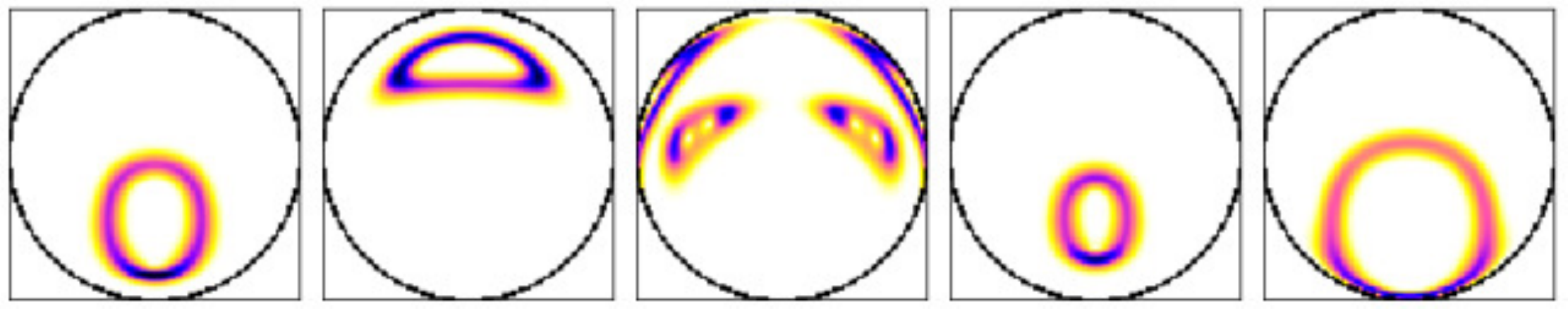}
\includegraphics[width=0.99\linewidth]{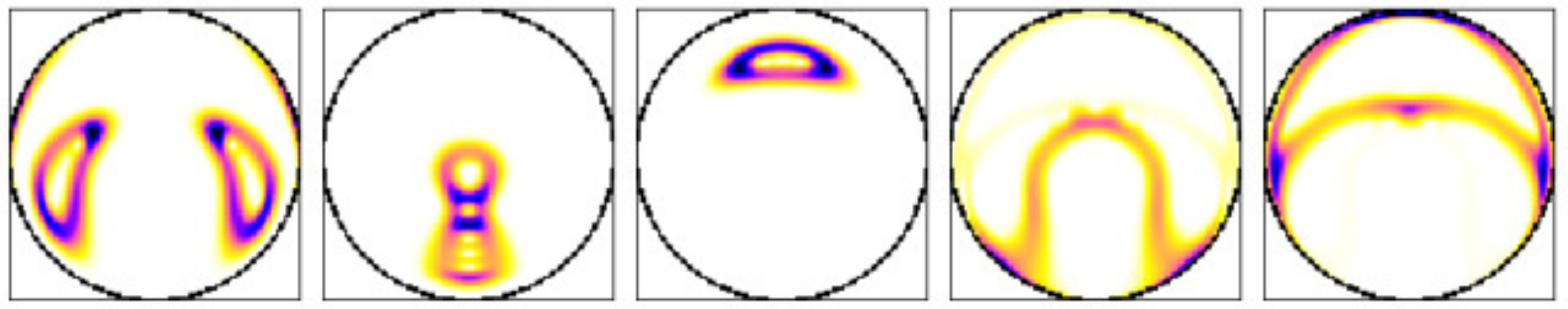}
\caption{(Color online) Poincare Husimi function of the eigenstates for $j=100$ and
$\kappa=0.1$, $-0.570 < E_n < -0.475$ (upper row)
$\kappa=0.5$, $-0.530 < E_n < -0.470$ (middle row)
$\kappa=0.6$, $-0.530 < E_n < -0.460$ (lower row) near the classical energy $E_\text{cl} = -0.5$
of the classical Poincare surface of section in Fig.~\ref{fig:ClassPoinc}.
}
\label{fig:EigenHus_PSOS}
\end{figure}

This function has been considered previously for other models~\cite{LS90, GKS98}.
In Fig.~\ref{fig:EigenHus_PSOS} we show the Poincare Husimi function of several individual eigenstates,
with energies in the vicinity of the energies of the classical Poincare surface of section in Figs.~\ref{fig:ClassPoinc}.
This figure reveals how the eigenstates localize at regular structures in the Poincare plots.

\section{Conclusions}
\label{sec:Conc}

Based on a combination of analytical and numerical data for the quantum dynamics of the Dicke model at large spin length we study the approach towards the classical spin limit $j \to \infty$ in two different situations.

For the low-energy dynamics around the stationary states
linearization of the semi-classical equations of motion gives two classical collective modes. The corresponding quantum observables are Green functions that describe the response of the system to a small perturbation of the ground state.
The quantum mechanical spectrum is dominated by the two classical modes already at small spin length. Convergence is rapid with growing $j$ and allows for clear identification of the ``soft mode'' at the QPT already for $j \simeq 200$.

For the dynamics at higher energies, a direct comparison of quantum and classical trajectories does not show convergence towards the classical dynamics because of rapid spreading in quantum phase space.
Instead, convergence is observed in the Husimi phase space functions only.
They allow us to unambigously identify the signatures of classical (quasi-) periodic orbits and chaotic orbits in the quantum dynamics and in individual eigenstates.

In conclusion, our results give a direct picture how the classical dynamics determines the quantum dynamics at larger $j$.
In short, the quantum dynamics is a combination of motion along a classical orbit, and spreading of the phase space probability along, but not perpendicular to, the classical orbit.
The spreading can be attributed to classical phase space drift and quantum diffusion. This behaviour is most naturally observed for classical \mbox{(quasi-)}periodic orbits, which lead to distinct signature in the quantum dynamics and the eigenstates.
Our results thus corroborate the general scenario developed for the Dicke model in, e.g., Ref.~\cite{AH12}.
For very long times, fragmentation of phase space functions indicates the 
revival of the initial state, which poses a natural limit to the almost classical phase space dynamics at large but finite $j$.

\begin{acknowledgments}
We thank B. Bruhn for helpful discussions.
This work was supported by Deutsche Forschungsgemeinschaft through Sonderforschungsbereich 652 (B5).
\end{acknowledgments}

\appendix

\section{Derivation of the equations of motion from the Dirac-Frenkel variational principle}
\label{app:DF}

From the derivative of the product state in Eq.~(\ref{ProdState}) with respect to the parameters $\alpha$, $z$, one obtains the three linearly independent states
\begin{equation}
\left\{ |\alpha,z\rangle; \ad |\alpha,z\rangle; J_+ |\alpha,z\rangle \right\} \;,
\end{equation}
which span the tangent space of the manifold of variational states. 
To apply the Dirac-Frenkel variational principle~\cite{Di30,Fren34}, we have to build 
an orthonormal basis in the tangent space. This is given by
\begin{equation}
\label{eq:OrthSys}
\left\{ |\alpha,z\rangle; |\tilde{\alpha},z\rangle; |\alpha,\tilde{z}\rangle \right\} \;,
\end{equation}
where
\begin{equation}
|\tilde{\alpha}\rangle=\ad |\alpha\rangle - \alpha^*|\alpha\rangle
\end{equation}
and
\begin{equation}
|\tilde{z}\rangle = \dfrac{1+|z|^2}{\sqrt{2j}}\left(J_+|z\rangle - \dfrac{2j z^*}{1+|z|^2}|z\rangle\right) \,.
\end{equation}
Projection of $H|\psi\rangle$ onto the basis set~\eqref{eq:OrthSys} results in
\begin{equation}
\mathcal P H|\psi\rangle =   \xi_1 |\alpha,z\rangle + \xi_2 |\tilde{\alpha},z\rangle + \xi_3 |\alpha,\tilde{z}\rangle \,,
\end{equation}
with
\begin{equation}
\xi_1 = j\Delta\left(\dfrac{|z|^2-1}{|z|^2 + 1} + \dfrac{\kappa}{2}|\bar \alpha|^2 + 2\kappa\dfrac{\Re(\bar \alpha)\Re(z)}{1+|z|^2}\right) \;,
\end{equation} 
\begin{equation}
\xi_2 = \sqrt{\dfrac{j\Delta\Omega\kappa}{2}}\left(\bar \alpha + \dfrac{2\Re(z)}{1+|z|^2}\right) \;,
\end{equation}
and
\begin{equation}
\xi_3 = \sqrt{2j}\Delta\left(\dfrac{z}{1+|z|^2} + \dfrac{\kappa}{2}\dfrac{1-z^2}{1+|z|^2}\Re(\bar \alpha)\right) \;.
\end{equation}

On the other hand, it is
\begin{equation}
  \ii \frac{d}{dt} |\psi_\text{SC}\rangle = \chi_1 |\alpha,z\rangle  + \chi_2 |\tilde{\alpha},z\rangle + \chi_3 |\alpha,\tilde{z}\rangle
\end{equation}
with
\begin{equation}
\begin{split}
 \chi_1 = & \dfrac{j\Delta\kappa}{2\Omega}\left(\Re \dot{\bar{\alpha}} \Im \bar \alpha - \Re \bar \alpha \Im \dot{\bar{\alpha}}\right) \\
          & + \dfrac{2j}{1+|z|^2}\left( \Re \dot z \Im z  - \Re z \Im \dot z \right) \;,
\end{split}
\end{equation}
and 
\begin{equation}
\chi_2 =  \ii \sqrt{\dfrac{j\Delta\kappa}{2\Omega}}\dot{\bar \alpha} \;, \quad \chi_3 = \ii \dfrac{\sqrt{2j}}{1+|z|^2} \dot{z} \;.
\end{equation}
From comparison of the coefficients $\xi_1, \xi_2, \xi_3$ and $\chi_1, \chi_2, \chi_3$ one directly obtains the SC equations of motion for $\alpha$, $z$ (Eq.~\eqref{SCEom}).

\section{Equation of motion for the classical collective modes}
\label{app:EOM}

Eq.~\eqref{LinEOM} is a linear equation of motion, which can be written as
\begin{equation}
\label{app:LinEOM}
\ii \dfrac{d}{dt} \begin{pmatrix} \Re \delta \bar\alpha \\ \ii \Im \delta \bar\alpha \\ \Re \delta z \\ \ii \Im \delta z^*\end{pmatrix} =
 \mathbf g_\text{lin}  \begin{pmatrix} \Re \delta \bar\alpha \\ \ii \Im \delta \bar\alpha \\ \Re \delta z \\ \ii \Im \delta z^*\end{pmatrix} \;,
\end{equation}
with a $4 \times 4$ matrix of the form
\begin{equation}
 \mathbf g_\text{lin} = \begin{pmatrix} 0 & g_1 & 0 & 0 \\ g_1 & 0 & g_2 & 0 \\ 0 & 0 & 0 & g_4 \\ g_3 & 0 & g_4 & 0 \end{pmatrix} \;,
\end{equation}
whose parameters are read off Eq.~\eqref{LinEOM} as
\begin{equation}
\begin{split}
 g_1 & = \Omega \;, \;\; g_2 = \Omega \frac{2(1-z_s^2)}{(1+z_s^2)^2} \;, \\
 g_3 & = \Delta \frac{\kappa}{2}(1-z_s^2) \;, \;\;
 g_4 =  \Delta (1-\kappa \bar\alpha_s z_s) \;.
\end{split}
\end{equation}
For $\kappa < 1$, it is
\begin{equation}\label{gBelow}
 g_1 = \Omega \;, \;
 g_2 = 2 \Omega \;, \;
 g_3 = \frac{\Delta \kappa}{2} \;, \;
 g_4 = \Delta \;, \;
\end{equation}
and for $\kappa >1$,
\begin{equation}\label{gAbove}
 g_1 = \Omega \;, \;
 g_2 =  \frac{\Omega(\kappa+1)}{\kappa^2} \;, \;
 g_3 =  \frac{\Delta \kappa}{\kappa+1} \;, \;
 g_4 =  \Delta \kappa \;.
\end{equation}

Eq.~\eqref{app:LinEOM} is the equation of motion of two coupled oscillators and can be solved as such.
The eigenvalues of $\mathbf g_\text{lin}$ are
\begin{equation}
  \omega^2_\pm = \frac{g_1^2+g_4^2}{2} \pm \sqrt{\frac{(g_1^2-g_4^2)^2}{4} + g_1 g_2 g_3 g_4} \;. 
\end{equation}
For oscillatory motion, it must $\omega^2 > 0$, which gives the criterion
\begin{equation}
 g_1 g_4 > g_2 g_3 \;.
\end{equation}
Then, four different real eigenvalues $\pm \omega_+$, $\pm \omega_-$ exist.

Let us now assume that $g_1 \ge g_4$, and swap $g_1$ and $g_4$ otherwise.
Then, $\omega_+^2 \to g_1^2$ and $\omega_-^2 \to g_4^2$ for $g_2,g_3 \to 0$.
The eigenvectors of $\mathbf g_\text{lin}$ are
\begin{equation}
 x_{1/2,+} = 
 \begin{pmatrix} 
  1 \\
  \frac{\omega}{g_1} \\
  \frac{\omega^2-g_1^2}{g_1 g_2} \\
  \frac{\omega(\omega^2-g_1^2)}{g_1 g_2 g_4} 
 \end{pmatrix}
\end{equation}
for the eigenvalues $\pm \omega_+$ with $\omega = \omega_+$ for $x_{1+}$, and $\omega= -\omega_+$ for $x_{2+}$, and
\begin{equation}
 x_{1/2,-} = 
 \begin{pmatrix} 
   \frac{\omega^2-g_4^2}{g_3 g_4}\\
  \frac{\omega(\omega^2-g_4^2)}{g_1 g_3 g_4}  \\
  1 \\
  \frac{\omega}{g_4}
 \end{pmatrix}
\end{equation}
for the eigenvalues $\pm \omega_-^2$ with $\omega = \omega_-$ for $x_{1-}$, and $\omega= -\omega_-$ for $x_{2-}$.
These expressions converge to the eigenvectors of the uncoupled oscillators for $g_2, g_3 \to 0$.

For the computation of $\delta J_x(t)$ in Eq.~\eqref{JxSC1}, we make the ansatz
\begin{equation}
 \begin{pmatrix} 0 \\ 0 \\ 1 \\ 0 \end{pmatrix} =  
a \frac{1}{2} (x_{1-}+x_{2-})  + b \frac{\sqrt{g_1 g_2}}{\sqrt{g_3 g_4}} (x_{1+} + x_{2+}) 
\end{equation}
such that
\begin{equation}
a + b \xi = 1 \;,   a \xi - b = 0 \;, 
\end{equation}
with
\begin{equation}
 \xi = \frac{\omega_+^2 - g_1^2}{\sqrt{g_1 g_2 g_3 g_4}} = - \frac{\omega_-^2 - g_4^2}{\sqrt{g_1 g_2 g_3 g_4}} \;.
\end{equation}
Here we have $\xi \ge 0$.
We can alternatively write
\begin{equation}
 \begin{pmatrix} a & b \\ -b & a  \end{pmatrix} 
\begin{pmatrix} 1 \\ \xi  \end{pmatrix} = \begin{pmatrix} 1  \\ 0  \end{pmatrix} \;,
\end{equation}
which is the characteristic equation for a Givens rotation.
For $\xi \ge 0$, this can be solved as
\begin{equation}
 a = \cos^2 \beta \;, \quad b \xi = \sin^2 \beta 
\end{equation}
with
\begin{equation}
  \cos 2 \beta = \frac{1 - \xi^2}{1 + \xi^2}
\end{equation}
or
\begin{equation}\label{app:beta}
 \tan 2 \beta = \pm \frac{\sqrt{1 - \cos^2 2 \beta}}{\cos \beta} = \pm \frac{2 |\xi|}{1 - \xi^2} = \pm \frac{2 \sqrt{g_1 g_2 g_3 g_4}}{g_1^2 - g_4^2} \,.
\end{equation}
Insertion of $g_1, \dots, g_4$ from Eqs.~\eqref{gBelow},~\eqref{gAbove} 
gives Eqs.~\eqref{WeightBelow},~\eqref{WeightAbove}.
Note that the angle $\beta$ in Eq.~\eqref{app:beta} has to be chosen from the correct branch of the $\arctan$ function.
For $g_1 \ge g_4$, we take $|\beta| < \pi/2$ from the principal branch.
In the opposite case $g_1 < g_4$, 
we use $\pi/2 < |\beta| < \pi$ (or similar) which coincides with the result after swapping $g_1$ and $g_4$ in the equations.

The third component $x_3(t)$ of the solution vector $x(t)$ of Eq.~\eqref{app:LinEOM}, to initial condition $x(0)=\mathbf e_3$, then is
\begin{equation}
\begin{split}
 x_3(t) & =  a \cos \omega_- t +  b \, \xi \cos \omega_+ t \\
 & = \cos^2 \beta \, \cos \omega_- t \; + \; \sin^2 \beta \, \cos \omega_+ t \;,
\end{split}
\end{equation}
yielding Eq.~\eqref{JxSC1}.

\section{Calculation of the time averaged Husimi function}
\label{app:Husimi}

We give here the deviation of Eq. (\ref{eq:avspinhusimi}).
We start with the definition of the time averaged Husimi function,
\begin{equation}
\bar{Q}(\theta, \phi) = \dfrac{1}{2T}\int_{-T}^T \mathrm{d}t\, |\langle \theta, \phi|\psi(t)\rangle|^2 \,.
\end{equation}
The time evolved state $|\psi(t)\rangle$ is calculated by means of the Chebyshev expansion~\cite{TK84,AF08}
\begin{equation}
|\psi(t)\rangle = \sum_{n=0}^N c_n(t) T_n(H) |\psi(0)\rangle \,,
\end{equation}
with the Chebyshev polynomials $T_n(x)$ and the expansion coefficients $c_n(t) = (-\ii)^n J_n(at)$, where $a$
is a scaling factor chosen such that the spectrum of $(1/a) H$ lies in the interval $[-1,1]$.
$J_n(x)$ is the Bessel function
\begin{equation}
J_n(x) = \dfrac{1}{2\pi} \int_{-\pi}^\pi \dd\tau\, e^{-\ii(n\tau - x\sin\tau)}\,.
\end{equation}
The absolut-squared overlap of $|\psi(t)\rangle$ with the coherent state $|\theta, \phi\rangle$
is given by
\begin{equation}
|\langle \theta, \phi|\psi(t)\rangle|^2 = \sum_{m,n=0}^N c^*_m(t) c_n(t) \mu^*_m(\theta, \phi) \mu_n(\theta, \phi) \,,
\end{equation}
with $\mu_n(\theta, \phi) = \langle \theta, \phi | T_n(H) | \psi(0) \rangle$.
This allows us to write the time average as a matrix-vector product according to
\begin{equation}
\begin{split}
\bar{Q}(\theta, \phi) &= \dfrac{1}{2T}\sum_{m,n=0} \int_{-T}^T dt \, c_m^*(t) c_n(t) \mu_m^*(\theta, \phi) \mu_n(\theta, \phi) \\
           &= \vec{\mu}^*(\theta, \phi) \mathbf{C} \vec{\mu}(\theta, \phi) \,,
\end{split}
\end{equation}
where the matrix coefficients are
\begin{equation}
C_{mn} = \dfrac{1}{2T} \int_{-T}^T \dd t\, c_m^*(t) c_n(t) \;.
\end{equation}
Since the integrand is given by
\begin{equation}
c_m^*(t) c_n(t) = \dfrac{\ii^{(m-n)}}{(2\pi)^2} \int_{-\pi}^\pi \dd x\, \dd y\,
                     e^{-\ii(nx - my)} e^{\ii at(\sin x - \sin y)}\,,
\end{equation}
we obtain
\begin{equation}
\begin{split}
C_{mn} &= \dfrac{\ii^{(m-n)}}{2T(2\pi)^2} \int_{-\pi}^\pi \dd x\, \dd y\, e^{-\ii(nx - my)}
            \int_{-T}^T \dd t\, e^{\ii at(\sin x - \sin y)} \\
        &= \dfrac{\ii^{(m-n)}}{(2\pi)^2} \int_{-\pi}^\pi \dd x\, \dd y\,
               e^{-\ii(nx-ny)} \text{sinc}[aT(\sin x - \sin y)] \,.
\end{split}
\end{equation}
The remaining integral can be evaluated numerically, e.g. by means of a discrete Fourier transformation in the form
\begin{equation}
\begin{split}
C_{mn} &= \dfrac{\ii^{(m-n)} (-1)^{(m+n)}}{N^2} \times \\
        &\times \sum_{\nu=0}^{N-1} \sum_{\mu=0}^{N-1}
               \text{sinc}[aT(\sin x_\nu + \sin y_\nu)] e^{-\ii n x_\nu} e^{-\ii m y_\mu}
\end{split}
\end{equation}
with $x_\nu = \dfrac{2\pi\nu}{N}$, $y_\nu = \dfrac{2\pi\mu}{N}$, $\nu, \mu = 0,1, ..., N-1$.

\end{document}